\documentclass[aip,cha,reprint]{revtex4-1}

\usepackage{amsmath}
\usepackage{amssymb}
\usepackage{graphicx}
\usepackage{hyperref}

\begin{document}

\newcommand{\dd}{\,\text{d}} 
\newcommand{\ee}{\text{e}} 

\title{Characterizing mixed mode oscillations shaped by noise and bifurcation structure}

\author{Peter Borowski}
\email{peterphysik@gmail.com}
\author{Rachel Kuske}
\email{rachel@math.ubc.ca}
\author{Yue-Xian Li}

\affiliation{Department of Mathematics, University of British Columbia, Vancouver, Canada}
\author{Juan Luis Cabrera}
\affiliation{Centro de F\'isica, Instituto Venezolano de Investigaciones Cienti\'ificas, Caracas, Venezuela}

\pacs{87.18.Tt  Noise in biological systems; 05.45.-a       Nonlinear dynamics and chaos; 05.45.Tp         Time series analysis
     }

\begin{abstract}
Many neuronal systems and models display a certain class of mixed mode oscillations (MMOs) consisting of periods of small amplitude oscillations interspersed with spikes. Various models with different underlying mechanisms have been proposed to generate this type of behavior. Stochastic versions of these models can produce similarly looking time series, often with noise-driven mechanisms different from those of the deterministic models. We present a suite of measures which, when applied to the time series, serves to distinguish models and classify routes to producing MMOs, such as noise-induced oscillations or delay bifurcation. By focusing on the subthreshold oscillations, we analyze the interspike interval density, trends in the amplitude and a coherence measure. We develop these measures on a biophysical model for stellate cells and a phenomenological FitzHugh-Nagumo-type model and apply them on related models. The analysis highlights the influence of model parameters and reset and return mechanisms in the context of a novel approach using noise level to distinguish model types and MMO mechanisms. Ultimately, we indicate how the suite of measures can be applied to experimental time series to reveal the underlying dynamical structure,  while exploiting either the intrinsic noise of the system or tunable extrinsic noise.
\end{abstract}

\maketitle

\section{Introduction}

Mixed mode oscillations (MMOs), composed of well-defined periods of subthreshold oscillations (STOs) separating spikes, appear in biological rhythms, neural dynamics, chemical oscillations, and network oscillators, with many models proposed to capture this phenomenon in these and other applications~\cite{ChaosFocus2008}. One question in applications is the significance of the STOs -- that is, do the interspike intervals (ISIs) and STOs encode important information for the system (e.g., Refs.~\onlinecite{Epsztein2010} or~\onlinecite{Longtin1996}). An element in the answer is identifying the mechanisms that drive the STOs. A challenge in this identification is that noisy time series generated by structurally different models can appear hardly distinguishable from each other, even  quantitatively. Experimentally observed behavior can be captured with different types of models with different noise levels, so that model identification and calibration  requires consideration of several classes of models over wide parameter ranges.

Previous studies indicate that a minimum of three degrees of freedom is necessary to produce MMOs in deterministic models (see examples in Ref.~\onlinecite{ChaosFocus2008}), while  stochastic van der Pol and FitzHugh-Nagumo-type models illustrate how noise can drive MMOs in  simple 2D models~\cite{Muratov2008,Makarov2001}. Similarities in  MMO signals often stem from an underlying Hopf bifurcation and canard structure in the system~\cite{Rotstein2006,Makarov2001,Desroches2008}. In periods where the system is in some sense close to the Hopf bifurcation, STOs can be both supported and disrupted by noise. In these models it is possible to choose parameters so that the MMOs are driven by the deterministic dynamics, while for other parameter choices the appearance of MMOs relies on stochastic fluctuations.

 While on the one hand  noise can make it difficult to distinguish between different  models with various routes for producing stochastic MMOs,  it is nevertheless possible to use noise as a way to identify underlying mechanisms. In this paper we compare a suite of measures at different noise levels to extract the structure of the model, and thus the mechanism for the STOs and MMOs.

The choice of measures is partly motivated by the presence of multiple time scales, an important feature that allows both well-defined periods of STOs and more than one mechanism for MMO generation. As described in more detail in the next section, the common substructure of the models we consider corresponds to a subsystem with a slowly varying control parameter. Depending on the combination of the other model parameters, that control parameter can slowly vary through a Hopf bifurcation, in which case  the STO-spike transition exhibits  the dynamics of bifurcation delay (Ref.~\onlinecite{DCDSSSpecial2009} and references therein). For other parameter combinations corresponding to quiescence in the deterministic system, noise excites subthreshold coherent oscillations with a frequency  near that of the Hopf bifurcation.  The weak damping of these oscillations  is due to the presence of multiple time scales,  observed for parameters near a  Hopf bifurcation. These noise-induced STOs are just one type of oscillation appearing in the context of coherence resonance (CR).  In its broadest sense CR is observed when a range of noise levels excites coherent oscillations in a system that is quiescent without noise, with a maximum coherence at an optimal noise level.  This phenomenon is most commonly cited for transitions between steady equilibria with large excursions, such as relaxation oscillations~\cite{Pikovsky1997,Lindner2004}, but in this paper it refers to the noise-induced STOs near a HB, also exhibiting CR-type behavior.

Whether the STOs appear via slow passage through a Hopf bifurcation (SPHB) or via a CR-type mechanism, the impact of the noise on the MMOs is concentrated in the interspike interval (ISI) where the STOs arise and eventually transition to a spike. It is no accident that this is also the interval where  the slow dynamics are most prominent, and it is well known that noise can have a significant impact in such intervals~\cite{Lythe1993,Celet1998,Georgiou1992}. The suite of measures used in this paper focuses on the STOs and the ISI behavior, capitalizing on previous analytical and computational studies of noise sensitivity of SPHB~\cite{Celet1998} and CR-driven oscillations~\cite{Klosek2005,Yu2006}. These previous analyses show how noise can enhance or suppress the STOs through interactions with the fast-slow dynamics inherent in the MMOs, and these characteristics are captured by the behavior of the suite of measures. Focused on the noise sensitive STOs in the ISI, these measures can isolate differences between models and identify mechanisms for the MMOs as noise levels and  bifurcation and control parameters vary.

An additional advantage of these  measures is that they are based on features that are easy to analyze in  time series data,  making them amenable for use on experimental and simulated data. Furthermore, recognition of  the multiple time scales of the MMOs suggests ways to experimentally introduce fluctuations into the system as a tunable extrinsic noise. When scaled appropriately by considering the ISI dynamics, this extrinsic noise can be used to mimic the effects of other intrinsic noise sources. Then this introduction of noise, combined with the suite of measures, can be applied both to  identify the likely mechanism for MMOs and  limit the variety of models one must consider for calibration.

{\bf Summary of results}

We use a combination of amplitude trend, ISI density, and coherence measure based on  power spectral density (PSD) to classify and differentiate  mechanisms for MMOs. We compare phenomenological and physiological models to explore the types of STO and MMO behaviors that can  be captured by simplified or reduced stochastic models over a range of noise levels. We then analyze these results within intra-model comparisons to get a deeper understanding of those MMO characteristics related to model structure, those dictated by deterministic dynamics, and those driven by noise. We focus particularly on cases where  CR and/or SPHB are key mechanisms. In the conclusion we give a detailed list of identifying characteristics of MMO mechanisms obtained from this suite of measures. Understanding the dynamical features reveals  the sensitivity in model calibration on bifurcation structure and on model choice,  and illustrates ways that noise can be used to help to differentiate between models.

For STOs of the CR-type   we observe longer tails in the ISI density together with stronger  peaks in the coherence measure vs. noise, and no trend in the average STO amplitude leading up to the spike. In contrast, SPHB-dominated STOs have different behaviors in all three measures.

 Simplified low dimensional integrate and fire (IF) models can be used to capture some STO and ISI behavior over a range of noise levels through adjustment of the speed of the control variable and reset. However, they can not mimic differences observed from underlying bifurcation structure, particularly in the case where a combination of CR and SPHB are at play.

Models  with voltage-dependent control variables can have a rich variety of underlying deterministic behavior, allowing possibilities for both CR- and SPHB-driven MMOs. These differences  can drive significantly different stochastic behavior in the ISI and PSD behavior as compared to simple IF models.

Related to this last observation, our analysis also illustrates how refractory dynamics, dynamics following the spike at the initial stage of the ISI, can influence ISI density for increasing noise, and disrupt the coherence  of the STOs in MMOs. The influence of the reset or return mechanism, as well as the underlying deterministic behavior, observed in the measures considered here is consistent with the influence of these modeling aspects observed for noise-driven clusters of spikes, that is, repeated spikes without STOs in the ISI~\cite{Kuske2009}.

The article is organized as follows: In Section~\ref{sec:models} we first give a brief overview of two mathematical models in the literature -- one physiological, the other phenomenological -- that have been used to explain MMOs in neural systems, and discuss different mechanisms for MMOs in these stochastic systems. In the same section we present a number of different computational measures for analyzing important characteristics of the time series in order to identify the underlying model structure. In Section~\ref{sec:analysis_inter} we generate and classify MMOs with the physiological model and find matching MMOs generated by the phenomenological model. We then apply the measures to these time series and provide a set of features that -- in combination -- make the time series distinguishable. Secs.~\ref{sec:analysis_intra} and~\ref{sec:intra_FHN} focus on the comparison of time series generated by the same model with varying parameters, to further highlight differences in the computational measures that appear with different model or bifurcation structure. We also highlight the dramatic effect of weak noise on different families of MMOs that are close in parameter space. In Sec.~\ref{sec:other_models} we outline how  tunable extrinsic noise can be introduced to mimic the effect of intrinsic noise, based on the inherent multiple time scales and the fact that the noise impact is concentrated in the ISI. We also illustrate the applicability of our analysis to another MMO-generating model.



\section{Models and Measures}
\label{sec:models}

We review the setup and dynamics of two models whose STO dynamics will be analyzed in this paper: A detailed biophysical model for MMOs in stellate cells (`SC' model) as well as a modified FitzHugh-Nagumo model (`MFN'). The two main MMO-generating mechanisms in these noisy models are slow passage through a Hopf bifurcation (SPHB) and coherence resonance (CR).

\subsection{Biophysical model -- SC}
\label{ssec:scmodel}

We consider a biophysical model for the MMOs observed experimentally in the layer II stellate cells of the medial entorhinal cortex introduced in Ref.~\onlinecite{Acker2003}. It consists of a voltage equation that includes the three usual Hodgkin-Huxley currents for sodium, potassium and a leak current together with a persistent sodium current and a hyperpolarisation-activated current. Together with the dynamical equations for six gating variables, this model is a system of seven coupled ordinary differential equations (`7DSC'), given in App.~\ref{app:sc_eq}.

Since we focus on the STO dynamics, for most of this paper we analyze a reduced, three-dimensional version (3DSC) of the full model, introduced in Ref.~\onlinecite{Rotstein2006},  consisting of only the equations for the voltage and two gating variables $r_f$ and $r_s$ (In the equation for $r_s$, we use the term from the original work~\cite{Acker2003} (see footnote in Ref.~\onlinecite{Rotstein2006}).). This 3DSC model provides a good approximation of the ISI dynamics of the full 7D model~\cite{Rotstein2006}. Throughout this article, we refrain from giving units in the equations or for parameters (see App.~\ref{app:sc_eq} for the original units and parameter values). The three equations considered for the reduced model are:
\vspace{0.5cm}

\begin{widetext}
\begin{align}
\dot{V} & = \frac{1}{C}\Big[ I_{\rm{app}} - G_L(V-E_L) - G_p \left( \frac{1}{1+\exp\left(-\frac{V+38}{6.5}\right)} + 0.15\sqrt{2D}\eta(t)\right) (V-E_{\rm Na}) - G_h(0.65 r_f + 0.35 r_s)(V-E_h)\Big] \label{eq:sc_V} \\
\dot{r}_f & = \left[1/[1+\exp\left((V+79.2)/9.78\right)]-r_f\right] / \left[0.51/[\exp\left((V-1.7)/10\right)+\exp\left(-(V+340)/52\right)]+1\right] \label{eq:sc_rf} \\
\dot{r}_s & = \left[1/[1+\exp\left((V+71.3)/7.9\right)]-r_s\right] / \left[5.6/[\exp\left((V-1.7)/14\right)+\exp\left(-(V+260)/43\right)]+1\right] . \label{eq:sc_rs}
\end{align}
\end{widetext}

Eq.~\ref{eq:sc_V} contains a noise term $\eta(t)$ representing white noise, corresponding to fluctuations in the persistent sodium current. According to Ref.~\onlinecite{White1998}, this current gives the main stochastic contribution  in stellate cells, due to the low number of channels. The model (Eqs.~\ref{eq:sc_V}--\ref{eq:sc_rs}) is treated as an integrate and fire (IF) model, where spiking and refractory dynamics of the full (7D) model are replaced by a reset value once the trajectory of the system crosses a threshold defined in terms of $V$. The value of reset is chosen such that the trajectory is placed close to a post-spike value of the full deterministic system ($V=-80$, $r_f=r_s=0$ as in Ref.~\onlinecite{Rotstein2006}). The deterministic ($D=0$) version of this reduced model has been analyzed in detail~\cite{Rotstein2006,Rotstein2008,Wechselberger2009}, and there have been also studies of the noisy ($D\ne 0$) case~\cite{Rotstein2006,Kuske2009}.

Both the full 7D and the 3DSC deterministic system  produce attracting states of either a steady state, MMOs, or tonic spiking, depending on  $I_{\rm{app}}$. In the 3DSC model these different long time patterns are observed for $I_{\rm app} \lesssim I_{\rm app,H} \approx -2.575$, $-2.575\lesssim I_{\rm app} \lesssim -2.241$, and $I_{\rm app} \gtrsim -2.241$, respectively. These values are slightly shifted in the 7D version of the model, e.g., the Hopf bifurcation value $I_{\rm{app,H}}$ separating steady state solutions from MMOs lies at $I_{\rm{app,H}}^{\rm 7D}\approx -2.702$ when using the alternative Eq.~\ref{eq:sc_rs_58} for the dynamics of $r_s$ (see App.~\ref{app:sc_eq}).

As discussed in Ref.~\onlinecite{Rotstein2006}, the 3DSC system can be viewed as a 2D subsystem in $V$-$r_f$ with slowly varying control parameter $r_s$. In the nondimensionalized version of the 3DSC model analyzed in Ref.~\onlinecite{Rotstein2008}, the ratio of time scales for $V$ and $r_f$ is approximately 0.02, with the time scale for $r_s$ even slower than that of $r_f$ by a factor of $0.3$.

\begin{figure}
\centering
\includegraphics[width=.32\textwidth]{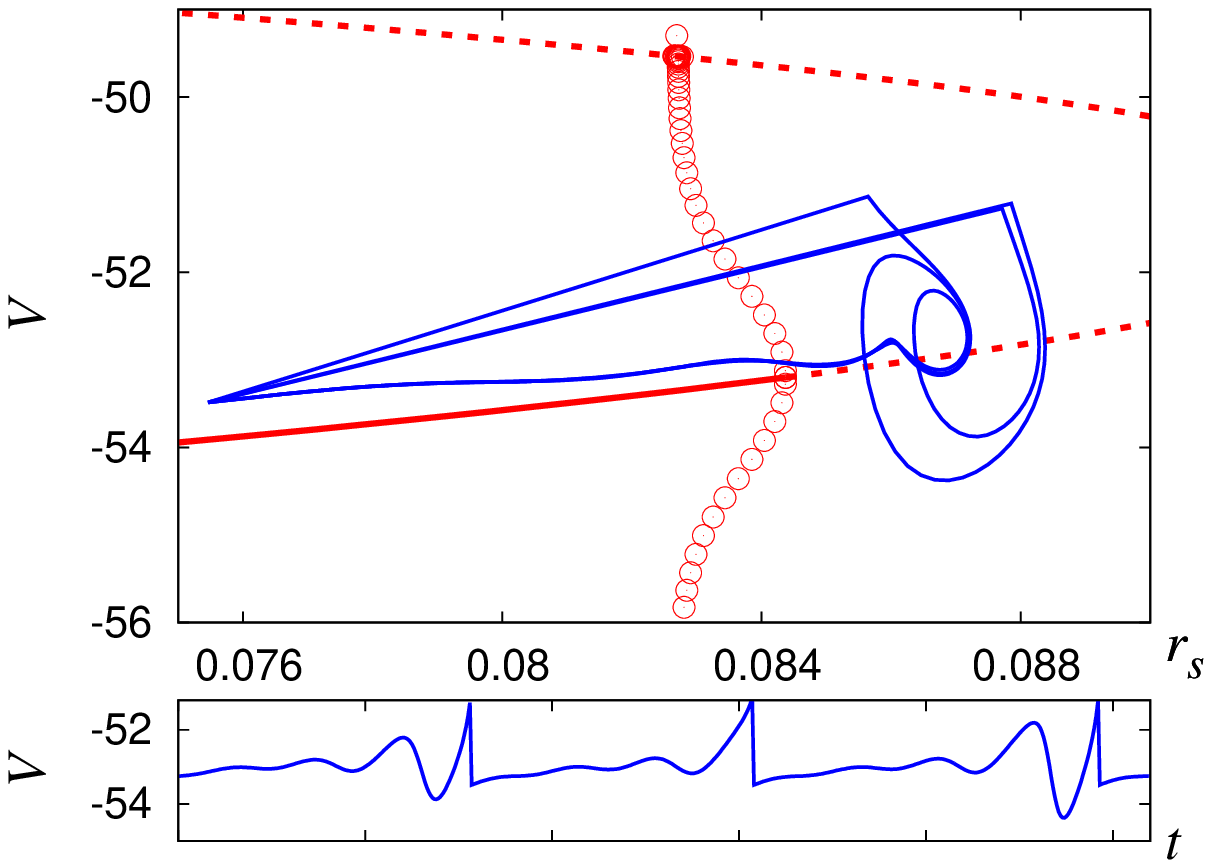}
\includegraphics[width=.32\textwidth]{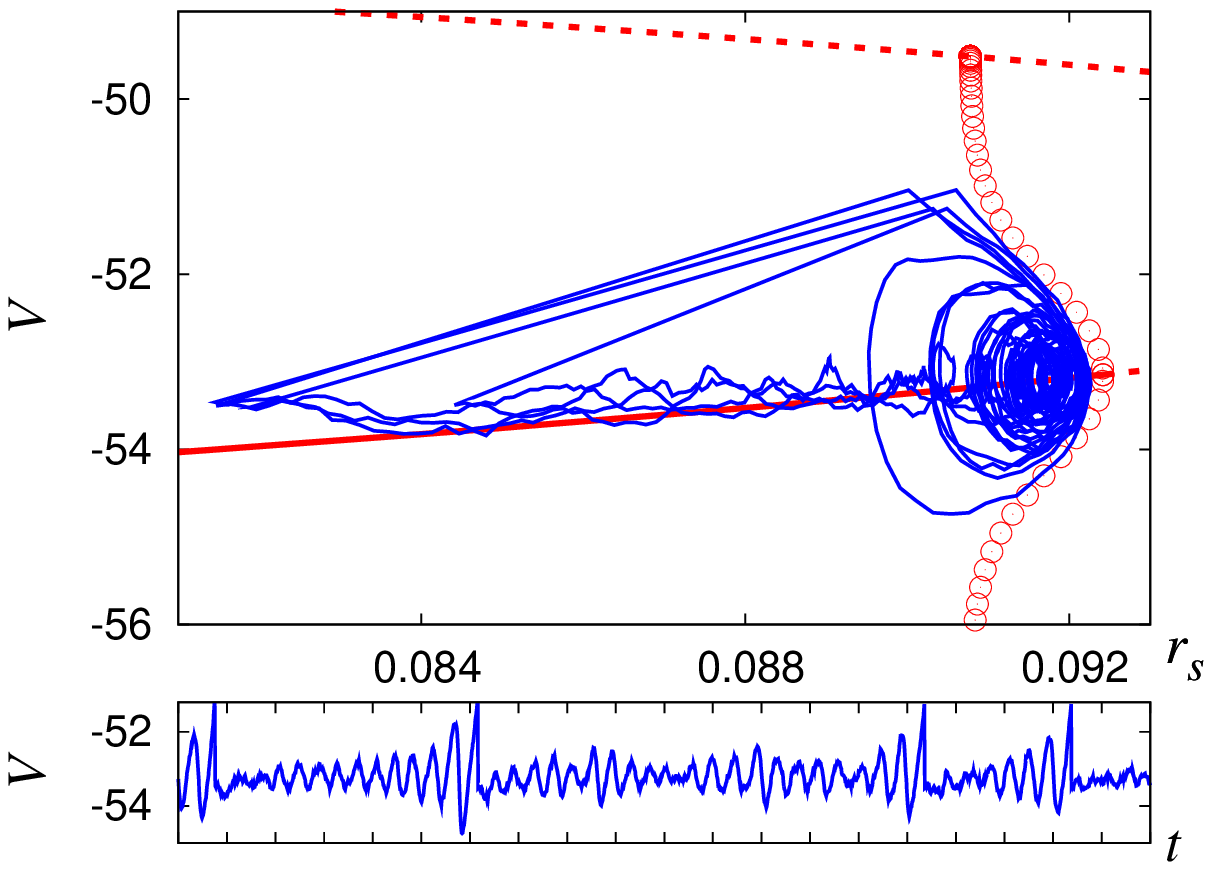}
\caption{Examples of the dynamics of the 3DSC model (for units see App.~\ref{app:sc_eq}): trajectory in the underlying 2D bifurcation diagram and time series. {\bf Top:} slow passage through a HB (see Class 1 later; $I_{\rm app}=-2.45$, $D=10^{-9}$); {\bf bottom:} STOs of the CR-type (see Class 2 later; $I_{\rm app}=-2.58$, $D=5\times 10^{-6}$). The scale in the time series is 250. A solid/dashed red line represents a stable/unstable fixed point. Solid/open circles represent stable/unstable limit cycles. There is another stable steady state at $V\approx -8$.}
\label{fig:bifurcation_sc}
\end{figure}

From that viewpoint, Fig.~\ref{fig:bifurcation_sc} shows the relevant part of the bifurcation diagram (with fixed $I_{\rm{app}}$) of the underlying 2D subsystem with $r_s$ treated as the bifurcation parameter. For the parameters considered, there is a subcritical Hopf bifurcation at $r_{s,\rm H}$ and a canard transition at $r_{s,c}$ ($r_{s,\rm H}\approx 0.08437$, $r_{s,c}\approx 0.08271$ for $I_{\rm{app}}=-2.45$ and $r_{s,\rm H}\approx 0.09241$, $r_{s,c}\approx 0.09078$ for $I_{\rm{app}}=-2.58$ -- obtained with XPPAUT~\cite{XPPAUT}).

In the stochastic 3DSC model ($D>0$), MMOs are observed for a broader range of  $I_{\rm app}$ values than listed above, generated by two mechanisms: a slow passage through a Hopf bifurcation and coherence resonance. For intermediate values of $I_{\rm app}$ and noise, mixtures of both of these mechanisms are observed. One of the purposes of this paper is to provide measures that characterize and identify these different mechanisms in various settings.
The top panel of Fig.~\ref{fig:bifurcation_sc} shows the stochastic dynamics of the 3DSC model when the deterministic dynamics corresponds to a SPHB, or delay bifurcation. The dependent variable $V$ is attracted to a steady state for $r_s<r_{s,\rm H}$. As the variable $r_s$ slowly ramps through $r_{s,\rm H}$, the full system does not immediately make the transition to spiking. Instead, there is a  delayed transition, as STOs increase gradually about the unstable steady state. It is well known from previous studies that for delay bifurcations, or more generally for certain systems with alternating slow/fast dynamics,  certain characteristics of the slow  dynamics are exponentially sensitive to noise. That is, for noise levels above those exponentially small in terms of the parameter of the slow time scale, the behavior of the slow dynamics is qualitatively changed by the noise, as has been discussed for examples of delay bifurcations and resonance~\cite{Baer1989,Berglund2002,Celet1998,Muratov2008,Su2004} and in other slow systems~\cite{Georgiou1992, Kuske1998, Lythe1993}. In the example in Fig.~\ref{fig:bifurcation_sc} the stochastic model typically has shorter intervals of STOs as compared with the deterministic model, as even small noise typically drives faster transitions to spiking.

The example in the bottom panel of Fig.~\ref{fig:bifurcation_sc} corresponds to STOs of the CR-type. The underlying deterministic system has a steady state corresponding to $r_s = r_{s,0}<r_{s,\rm H}$, so STOs are damped. In the stochastic system  coherence resonance (CR) occurs when  noise amplifies the STOs in the presence of weak damping (see Refs.~\onlinecite{Gang1993}, \onlinecite{Klosek2005}, \onlinecite{Kuske2007}, \onlinecite{Yu2006} and references therein; note that this type of CR is different to what is described, e.g., in Refs.~\onlinecite{Lindner2004} or~\onlinecite{Pikovsky1997}). These STOs have a frequency corresponding to that of the Hopf bifurcation at $r_{s,\rm H}$ of the 2D subsystem. Results in Ref.~\onlinecite{Yu2006} indicate that  CR drives STOs in the range of noise levels where coherence is optimal, that is, where the PSD is relatively narrow with significant power. The analysis in Ref.~\onlinecite{Yu2006} shows that the amplification factor of the STOs is inversely proportional to the proximity to the Hopf bifurcation, so that the STOs are a prominent and prolonged feature in the ISI for $r_s$ near $r_{s,\rm H}$. Variability in the amplitude of these STOs yields a nontrivial probability of spiking, leading to frequent appearance of MMOs, even for $I_{\rm app}<I_{\rm{app,H}}$ if noise is strong enough~\cite{Rotstein2006}.

\subsection{Modified FitzHugh-Nagumo model -- MFN}
\label{ssec:fhn}

In Ref.~\onlinecite{Makarov2001}, a FitzHugh-Nagumo-type (FN) model is given by the following pair of coupled nonlinear stochastic differential equations: 
\begin{align}
\epsilon \dot{u} & = u(u-a)(1-u)-v , \label{eq:Makarov_u}
\\
\dot{v} & = g(u-b) + \sqrt{2D}\xi(t) .
\label{eq:Makarov_v}
\end{align}

We follow Ref.~\onlinecite{Makarov2001} and use $\epsilon=0.005$, $a=0.9$ and $g(x)=7x^2+0.08(1-\ee^{-x/0.08})$, chosen to give a ratio of the time scales of the STOs and the spikes (relaxation oscillations) that is close to that typically observed experimentally and in the SC model. The parameter $b$ is a control or bifurcation parameter. We refer to this model as the modified FN model (MFN). Taking $a=-1$, $D=0$, adding a constant term in Eq.~\ref{eq:Makarov_u} and a $v$-dependent term in Eq.~\ref{eq:Makarov_v} yields the general FN model~\cite{Scholarpedia_FHN}. With $a=-1$, $g(x)=x$, $b=D=0$ and a simple rescaling Eqs.~\ref{eq:Makarov_u} and~\ref{eq:Makarov_v} are the original van der Pol (vdP) model~\cite{Scholarpedia_vdP}.

\begin{figure}
\centering
\includegraphics[width=.32\textwidth]{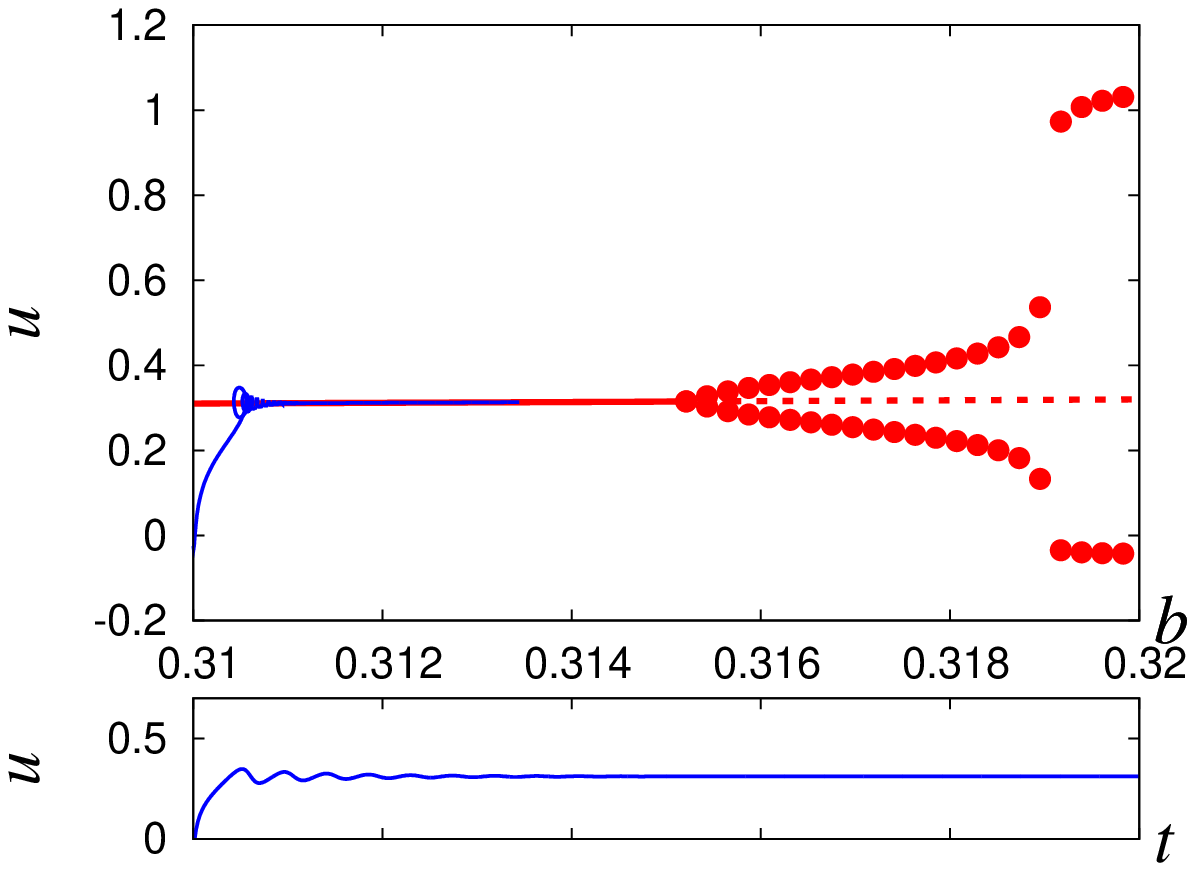}
\includegraphics[width=.32\textwidth]{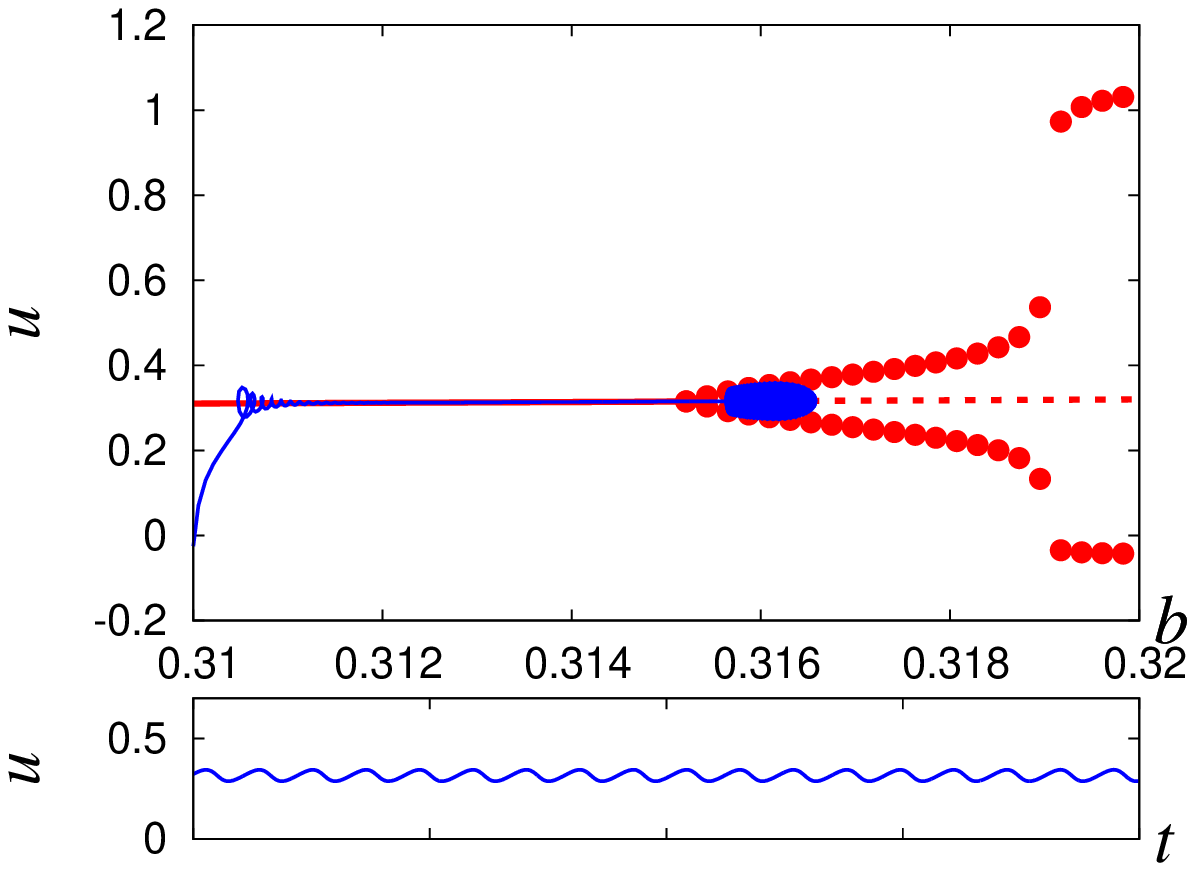}
\includegraphics[width=.32\textwidth]{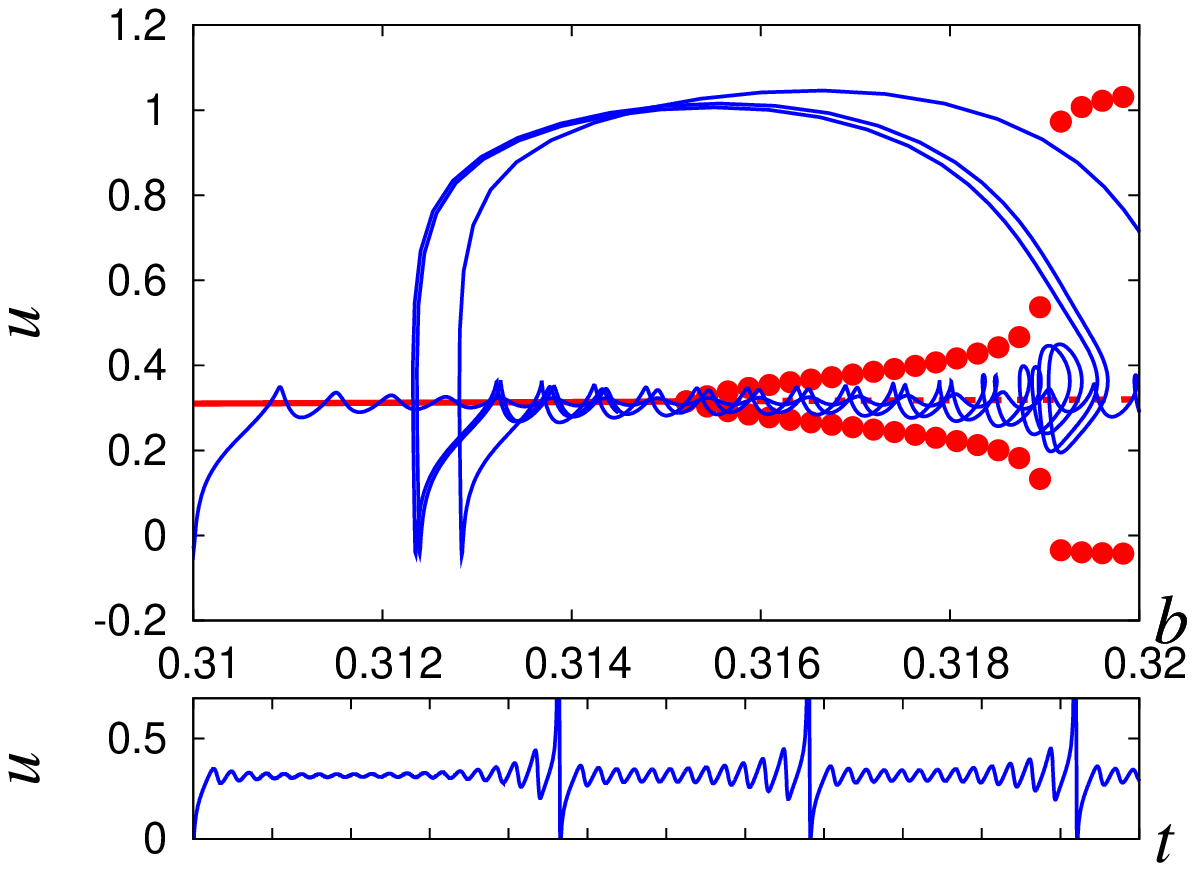}
\caption{The bifurcation diagram of the MFN model with trajectories of the deterministic NLMFN model superimposed. The three figures show the three different dynamics of the deterministic model depending on the value of $c_b$ (see text): {\bf top:} $c_b=1.54$, stable steady state; {\bf middle:} $c_b=1.5$, stable STOs; {\bf bottom:} $c_b=1.03$, stable MMOs. The timescale in the time series is 2. Lines and symbols as of Fig.~\ref{fig:bifurcation_sc}.}
\label{fig:bifurcation_Ma}
\end{figure}

Fig.~\ref{fig:bifurcation_Ma} shows the corresponding bifurcation diagram: a supercritical Hopf bifurcation with a canard transition to large amplitude relaxation oscillations. Depending on the value of the control parameter $b$, the deterministic system ($D=0$) takes one of three solutions: a stable steady state for $b<b_{\rm H}\approx 0.31535$ where $b_{\rm H}$ is the Hopf bifurcation value, stable small amplitude oscillations (STOs) for $b_{\rm H}<b<b_c\approx 0.31854$ with $b_c$ the canard value (see, e.g., Ref.~\onlinecite{Krupa2001}), and large amplitude oscillations for values above the canard value $b>b_c$. As discussed in Ref.~\onlinecite{Makarov2001}, applying noise to this system ($D\ne 0$) can produce MMOs as noise drives the system between small and large amplitude oscillations.

The influence of noise on the MMO dynamics of the vdP model has been studied recently in detail~\cite{Muratov2008}, providing scaling relationships between the proximity to the Hopf value and the noise levels that lead to different types of behavior. Here we use the model given by Eqs.~\ref{eq:Makarov_u} and~\ref{eq:Makarov_v} as a basis for comparison with the biophysical model for two reasons. First, the use of $g(x)$ gives control over the time scale and shape of the spike (see above). The other important effect that the nonlinearity has on the system is a shift of the canard point, enlarging the parameter region of stable STOs (i.e., $b_c-b_{\rm H}$) by a factor of roughly 1.5. This leads to MMOs with well-defined STOs over significant parameter ranges even in the presence of stronger noise values, in contrast to the vdP model studied in Ref.~\onlinecite{Muratov2008}. One other difference, compared to the vdP model, is that the choice of $g(x)$ yields a less abrupt transition from STOs to relaxation oscillations, allowing for a greater probability of observing spikes with reduced amplitude, an effect rarely observed in the SC model.

As in Ref.~\onlinecite{Kuske2009} we augment the MFN model (Eqs.~\ref{eq:Makarov_u} and~\ref{eq:Makarov_v}) with a slow variation in the control parameter given by an additional dynamical equation for $b$.  The presence of a slowly varying control parameter is a feature common to a number of MMO models including the SC model, and is necessary to reproduce some of the features observed in IF models. Additional comments about the model of Eqs.~\ref{eq:Makarov_u} and~\ref{eq:Makarov_v} relative to the augmented models are included in Subsec.~\ref{ssec:intra_LMFHN}.

The first augmented model we consider is an IF model with  a linear ramp for $b$ plus reset at a fixed threshold value for $u$: 
\begin{equation}
\dot{b}=\epsilon_2 .
\label{linb}
\end{equation}
Once the trajectory surpasses $u_{\rm th}$, $b$ is reset to $b_{\rm rs}$ and we choose to reset $u$ and $v$ such that the system is on the nullcline ($u_{\rm rs}=b_{\rm rs}$ and $v_{\rm rs}=b_{\rm rs}(b_{\rm rs}-a)(1-b_{\rm rs})$) or very close to it (see Subsec.~\ref{ssec:intra_LMFHN} for a discussion of the dynamics when reset values are off of the nullcline). This system, which we reference as the linearly augmented modified FN model (LMFN), generates only MMOs within the parameter values considered. The stochastic LMFN model exhibits features of both a SPHB and CR, depending on the value of $\epsilon_2$ as well as where the trajectory is reset. This allows for a comparison with simple IF models, highlighting the effect of noise  on the dynamics for different ramp speeds of the control parameter $b$ and reset values.

The second augmentation of the MFN model considered  is a nonlinear $u$-dependent equation for $b(u)$ with a functional form similar to that in the SC model:
\begin{equation}
\dot{b}= \left[ \frac{2}{1+\exp\left(\frac{u-0.2}{0.1}\right)}-c_b b\right] / \left[5\exp\left( -\frac{u-0.8}{0.15}\right)+1\right].
\label{eq:nlinb}
\end{equation}
This augmentation allows for a comparison of cases where the dependence of the control variable on $u$ results in qualitatively different stable behaviors for the deterministic system. Fig.~\ref{fig:bifurcation_Ma} shows the  three possible long time deterministic behaviors ($D=0$) depending on the parameter $c_b$,  keeping the other constants in Eq.~\ref{eq:nlinb} fixed. For $c_b > c_{b,\rm H} \approx 1.53$ there is a stable steady state, for intermediate values $1.04\lesssim c_b \lesssim 1.53$ there are sustained STOs,  and for $c_b \lesssim 1.04$ the attracting state is  MMOs.  The number of STO periods between spikes depends on $c_b$ and on the other constants in Eqs.~\ref{eq:Makarov_u}, \ref{eq:Makarov_v}, and~\ref{eq:nlinb}. We refer to this augmentation of the MFN model as nonlinearly augmented modified FN (NLMFN) and in the noisy version of this model similar patterns as in the SC model can be observed. For $c_b>c_{b,\rm H}$ noise can excite CR-type STOs and for  $c_b\lesssim 1.04$ the trajectory passes through the underlying HB and canard transition thereby displaying features of the SPHB.

\subsection{Measures}
\label{ssec:tools}

For the analysis of the MMOs and specifically the STOs in the ISIs, we use a number of measures: ISI density, amplitude of the STOs preceding a spike, power spectral distribution (PSD) of the STOs, and a coherence measure. The time series were generated by numerically integrating the coupled differential equations given in the preceeding subsections using a simple Euler-forward routine with constant time step. For the output, it was ensured that the data was sampled at a high enough rate $1/\Delta t$, such that the STOs could clearly be identified ($40\Delta t \lesssim T_{\rm STO}$).

The density of the interspike intervals (ISI) was obtained in the following way: In IF-type models (3DSC and LMFN), ISI duration is measured from the reset to the crossing of the threshold corresponding to initiation of the spike. The typically long refractory period without STOs in the 3DSC model was removed. 
In spiking models (7DSC, NLMFN), it is measured as the time between two consecutive crossings of a threshold (i.e., the time between two spikes including a spike). 

To compare the different models (with different time scales), we measure the ISI duration in $T_{\rm STO}$, the approximate mean STO period. For simplicity, we use only one $T_{\rm STO}$ for each of the model versions and obtain these approximately from one time series. We use the following approximate values for $T_{\rm STO}$: 3DSC: 107.5, NLMFN: 0.47, LMFN: 0.45, but note that the actual average $T_{\rm STO}$ vary by about $\pm 5\%$ among the classes introduced in the next section. Average ISI duration is used as a coarse measurement to calibrate FN-type models to the biophysical (SC) model, and the ISI density is used to compare models and parameter choices, including influence of the noise.

A second characteristic of the STOs used to calibrate FN-type models to the biophysical model is average STO amplitude trend before a spike. For this, the maxima of the low-pass filtered time series were extracted and averaged over many ISIs (for details see App.~\ref{app:measures}).

To obtain the power spectral distribution (PSD) of the STOs, the spikes were removed from the time series and it was ensured that the normalization of the PSD is consistent within each model. From the PSD, we compute a measure of coherence, namely~\cite{Gang1993}
\begin{equation}
\beta = h_p\frac{f_p}{\Delta f}
\label{eq:beta}
\end{equation}
where $f_p$ is the frequency of the peak in the PSD corresponding to the STOs, $h_p$ is its associated maximal power and $\Delta f$ is its full width at half maximum. Since the different models use different units, a comparison of the absolute value of $\beta$ across models was not performed. For details regarding the computation of the PSD and $\beta$, see App.~\ref{app:measures}.



\section{Analysis of MMO characteristics: Inter-model comparison}
\label{sec:analysis_inter}

In this section we present an analysis of the characteristics of MMOs with computational measures, focusing on the STOs, produced by the different models as introduced in Section~\ref{sec:models}. STO amplitude trend before a spike and average ISI provide a coarse classification of the MMOs in the subsections below. For each class of behavior the parameters of the 3DSC model are chosen to produce a type of STO observed in the 7DSC model, and the parameters for the FN-type models are chosen such that they reproduce the mean ISI and amplitude trend similar to the 3DSC. The results illustrate features of the MMOs that can be reproduced by both the biophysical and phenomenological models. The measures introduced in Subsec.~\ref{ssec:tools} provide a finer comparison of the three broad classes. We introduce different noise levels to highlight the mathematical elements of both the underlying deterministic model and stochasticity that influence the MMO characteristics.

\subsection{Class 1: Short ISI, increasing STO amplitude}
\label{ssec:analysis_inter_class1}

\begin{figure}
\center \includegraphics[width=.32\textwidth]{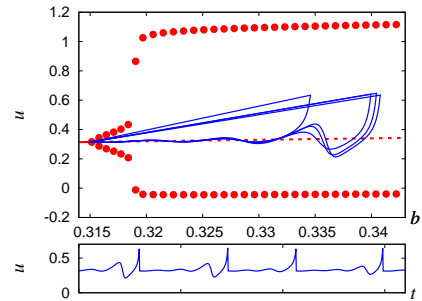}
\caption{Class 1 MMOs in the LMFN model: trajectory in the underlying 2D bifurcation diagram and time series. $\epsilon_2=0.0147$, $D=10^{-8}$, reset to $b_{\rm rs}=u=0.315, v=-0.12603$. The scale in the time series is 2.}
\label{fig:MMO_timeseries_1}
\end{figure}

\begin{figure}
\centering
\includegraphics[width=0.32\textwidth]{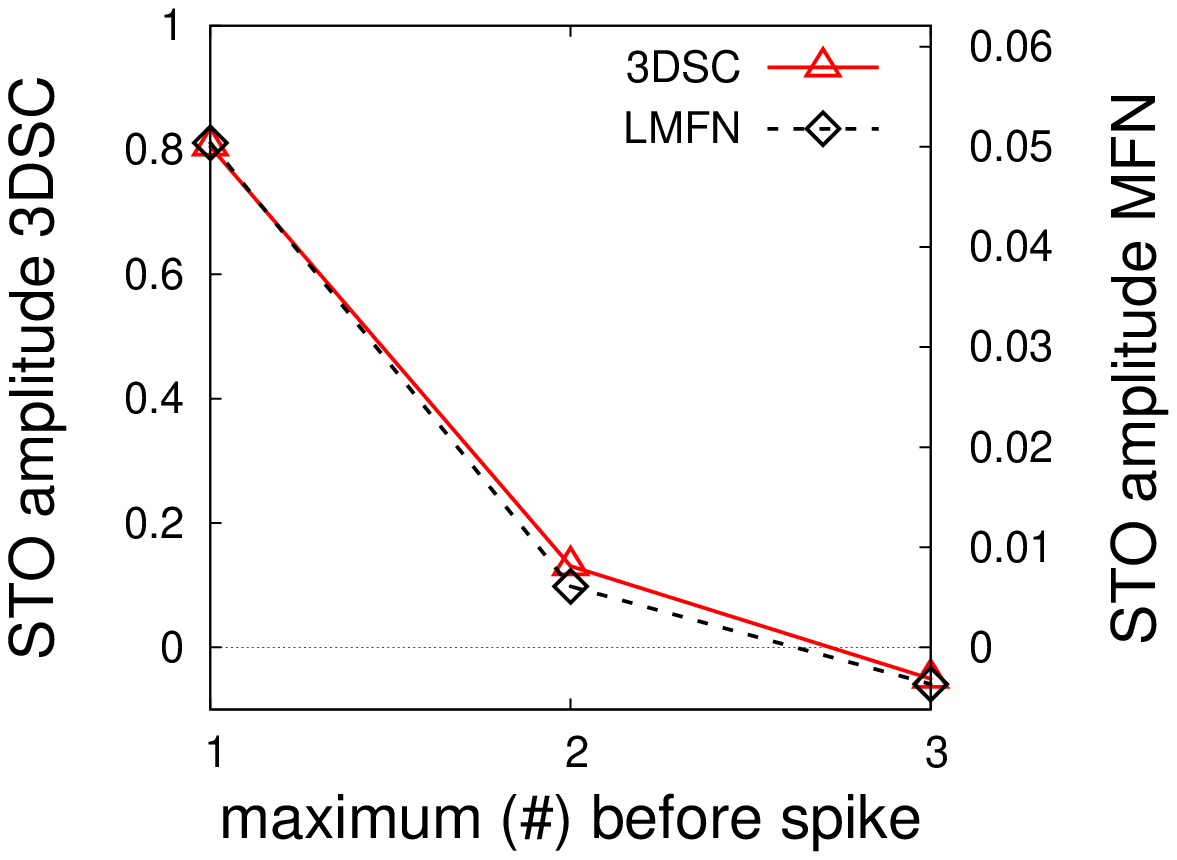}
\includegraphics[width=0.32\textwidth]{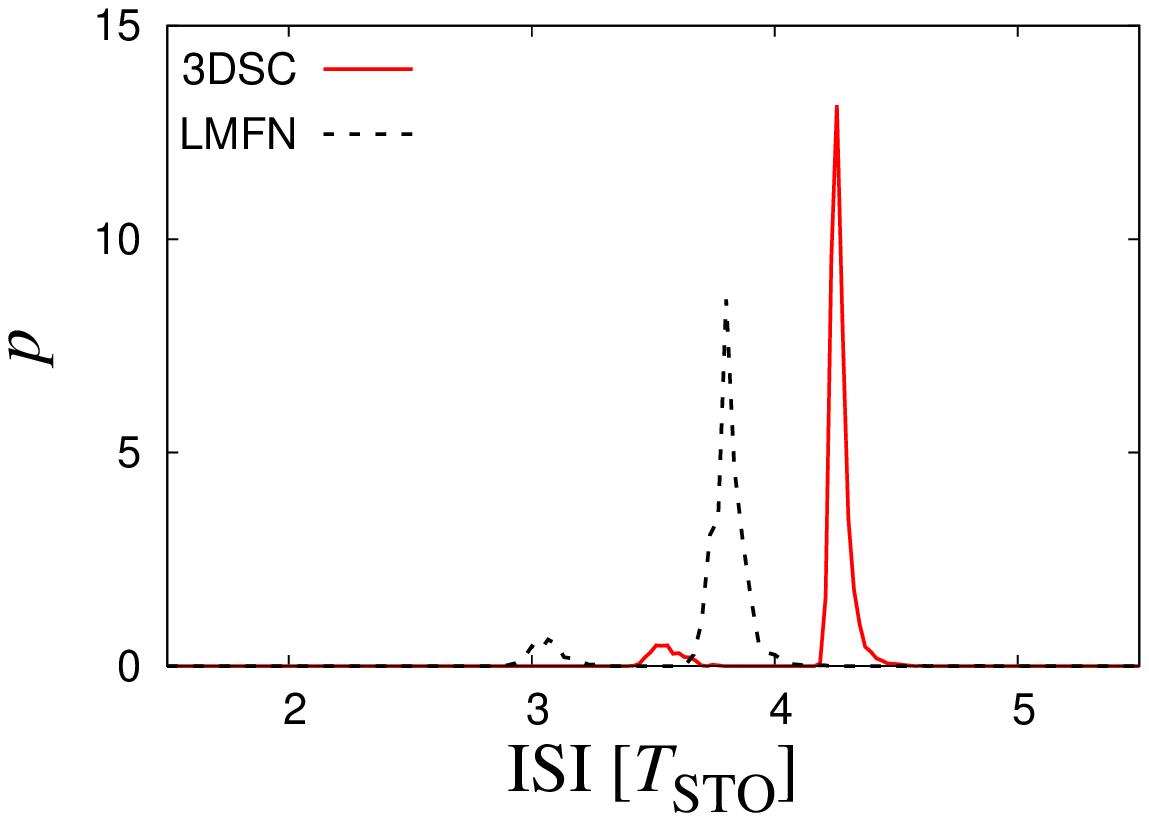}
\caption{{\bf Top:} Average amplitude before a spike (obtained by finding the maxima (see Subsec.~\ref{ssec:tools})) for time series of Class 1. $\triangle$ (solid red line): 3DSC (left ordinate); $\diamond$ (dotted black line): LMFN (right ordinate). {\bf Bottom:} Densities of ISI lengths for long ($\sim$ 5500 spikes) versions of the time series shown in Figs.~\ref{fig:bifurcation_sc} (top panel) and~\ref{fig:MMO_timeseries_1}. Solid line (red): 3DSC; dashed line (blue): LMFN. Parameters: $I_{\rm app}=-2.45$, $D=10^{-9}$ (3DSC); $\epsilon_2=0.0147$, $D=10^{-8}$, reset to $b_{\rm rs}=u=0.315, v=-0.12603$ (LMFN).}
\label{fig:class1_ISI}
\end{figure}

In Class 1, we analyze MMOs with short ISIs during which the amplitude of the STOs increases. Examples for this class of MMOs are shown in Fig.~\ref{fig:bifurcation_sc} (top panel) for the 3DSC model and Fig.~\ref{fig:MMO_timeseries_1} for the LMFN model. The comparison  of the amplitude dynamics of the STO and ISI density is shown in Fig.~\ref{fig:class1_ISI}, for 3DSC and LMFN, where the ISI density and mean agree up to a shift due to different reset conditions. This Class 1 behavior is not readily reproducible within the NLMFN model where the choices of $c_b$ needed for short ISIs lead to STOs with a larger amplitude throughout the ISI.

As shown in Sec.~\ref{sec:models}, the underlying dynamics for both the 3DSC and the LMFN model corresponds to a slow passage through a Hopf bifurcation. In the SC model, the increasing amplitude of the STOs are generated when the system spirals away from an unstable fixed point for $r_s$ values beyond the subcritical HB. In the LMFN model, the trajectory passes through the underlying supercritical HB at $b_{\rm H}$ followed by an increasing STO amplitude, with escape from STOs to a spike appearing for relatively large $b$ (around 0.34) beyond the canard point $b_c$. The significance of the speed of variation of $b$ in the LMFN model is discussed further in Subsec.~\ref{ssec:intra_LMFHN}.

The noise levels shown in Figs.~\ref{fig:bifurcation_sc} (top panel) and~\ref{fig:MMO_timeseries_1} are among the lowest at which stochastic effects are observed, with the STO dynamics dominated by the deterministic trend of increasing amplitude. Increasing the noise strengths has very similar effects in both models. 

\begin{figure}
\centering
\includegraphics[width=0.32\textwidth]{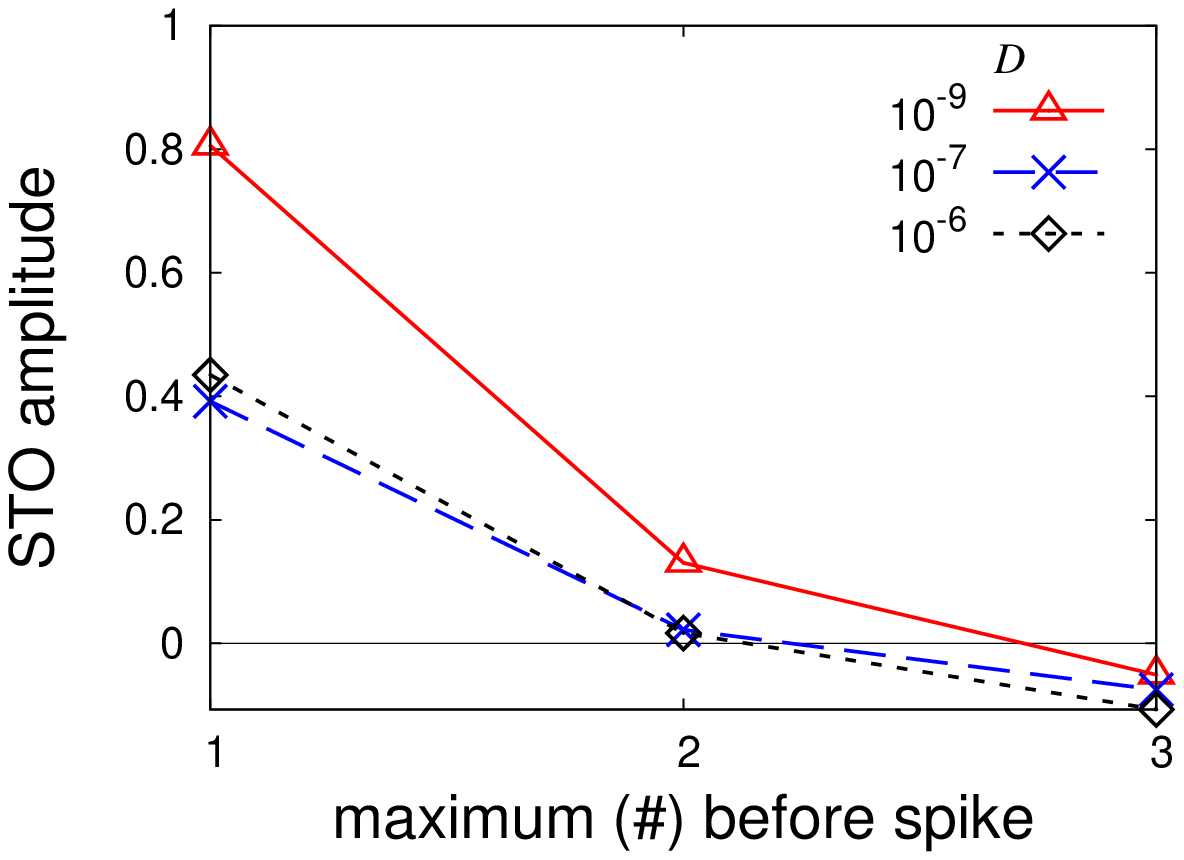}
\includegraphics[width=0.32\textwidth]{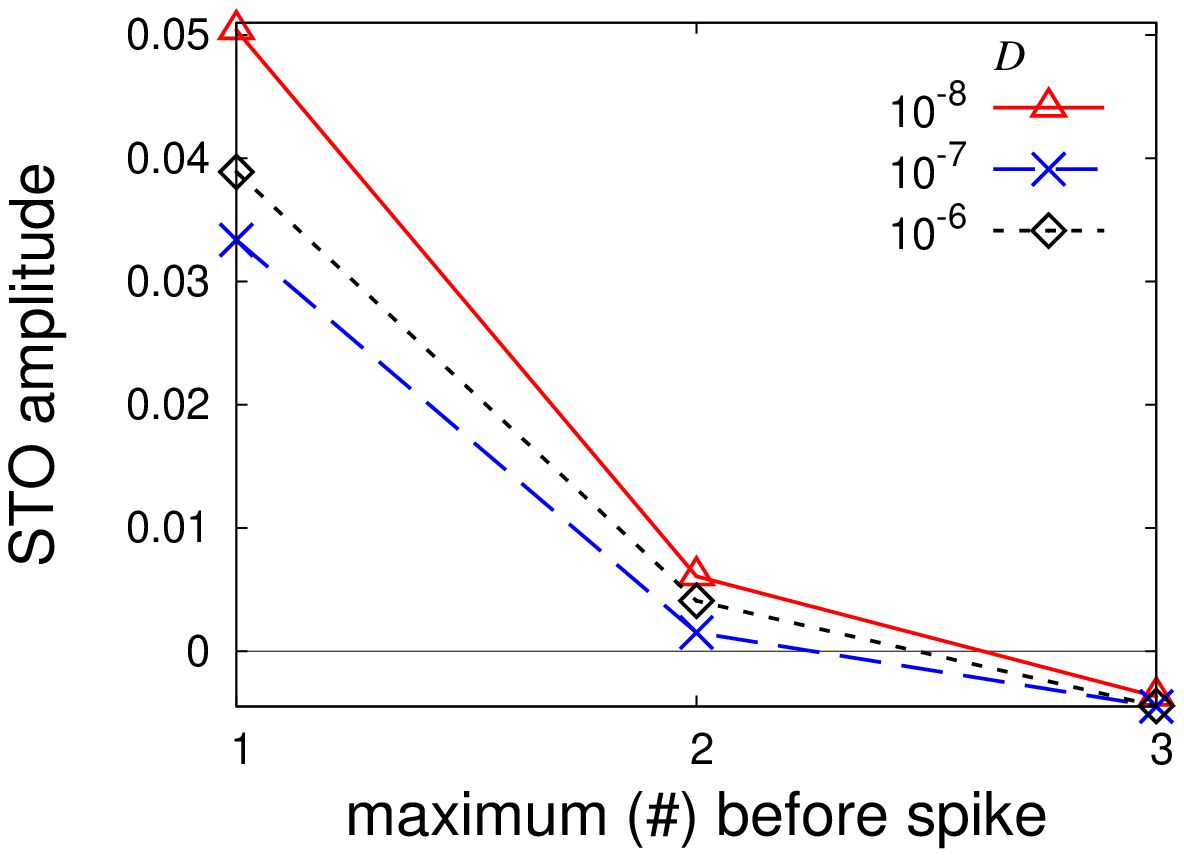}
\caption{Average amplitude before a spike for various values of $D$. {\bf Top:} 3DSC; {\bf bottom:} LMFN. The parameters are as of Fig.~\ref{fig:class1_ISI}.
}
\label{fig:class1_amp_noise}
\end{figure}

\begin{figure}
\centering
\includegraphics[width=0.32\textwidth]{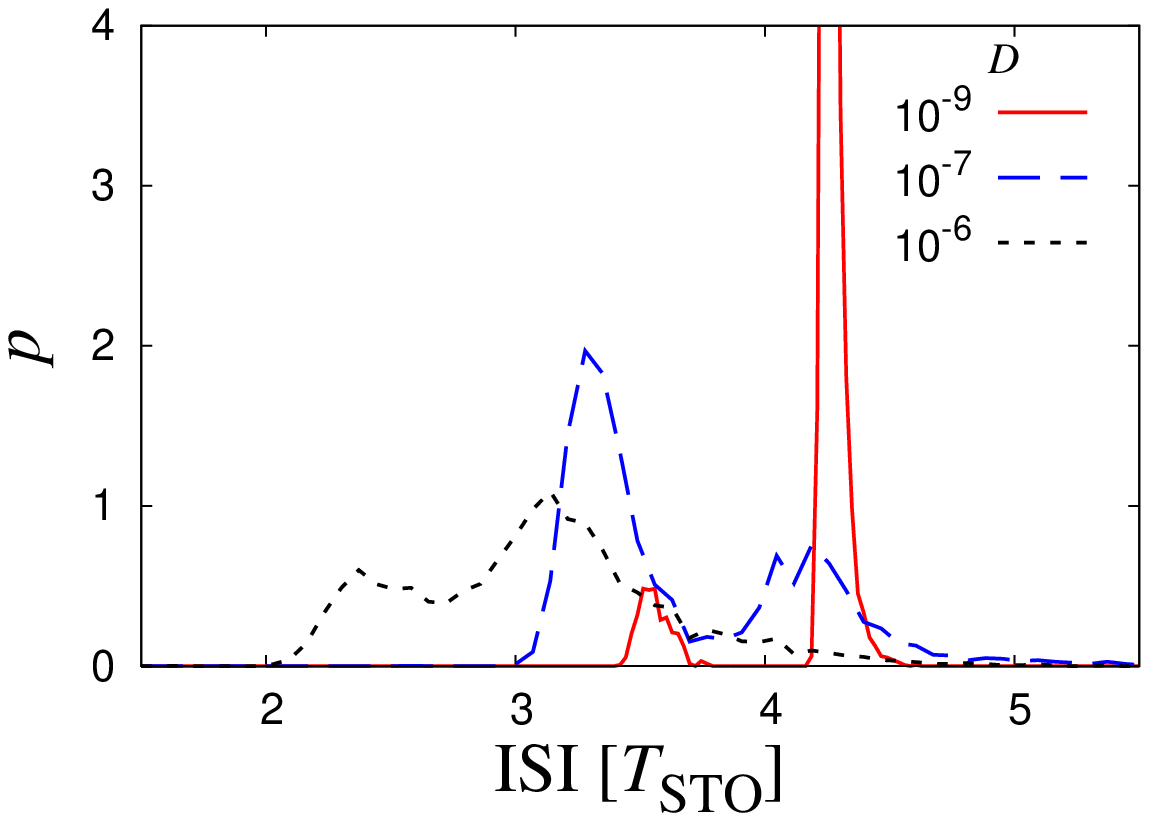}
\includegraphics[width=0.32\textwidth]{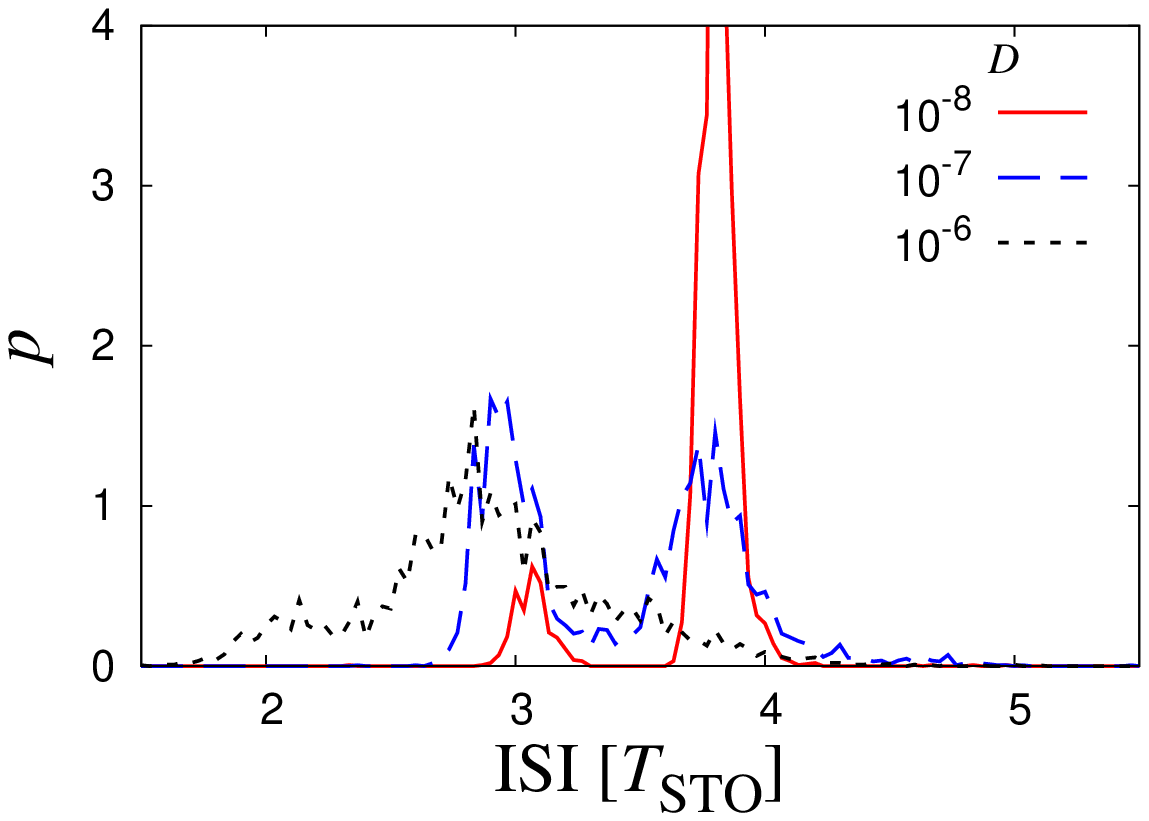}
\caption{Densities of ISI lengths for various values of the noise strength $D$. {\bf Top:} 3DSC; {\bf bottom:} LMFN. The parameters are as of Fig.~\ref{fig:class1_ISI}.}
\label{fig:class1_ISI_noise}
\end{figure}

$\bullet$
Typical for a SPHB, increased noise drives an earlier escape, indicated by shorter ISIs (Fig.~\ref{fig:class1_ISI_noise}) and lower average amplitude before escape (Fig.~\ref{fig:class1_amp_noise}). Differences due to the nature of the underlying Hopf bifurcation (sub-/supercritical) are evident only in the amount of reduction of the amplitude of the STOs. This reduction is more significant in the 3DSC, where the STOs are unstable for the underlying 2D system with $r_s>r_{s,\rm H}$, as compared to the LMFN model,  where the supercritical Hopf bifurcation yields stable STOs for $b>b_{\rm H}$.

\begin{figure}
\centering
\includegraphics[width=0.32\textwidth]{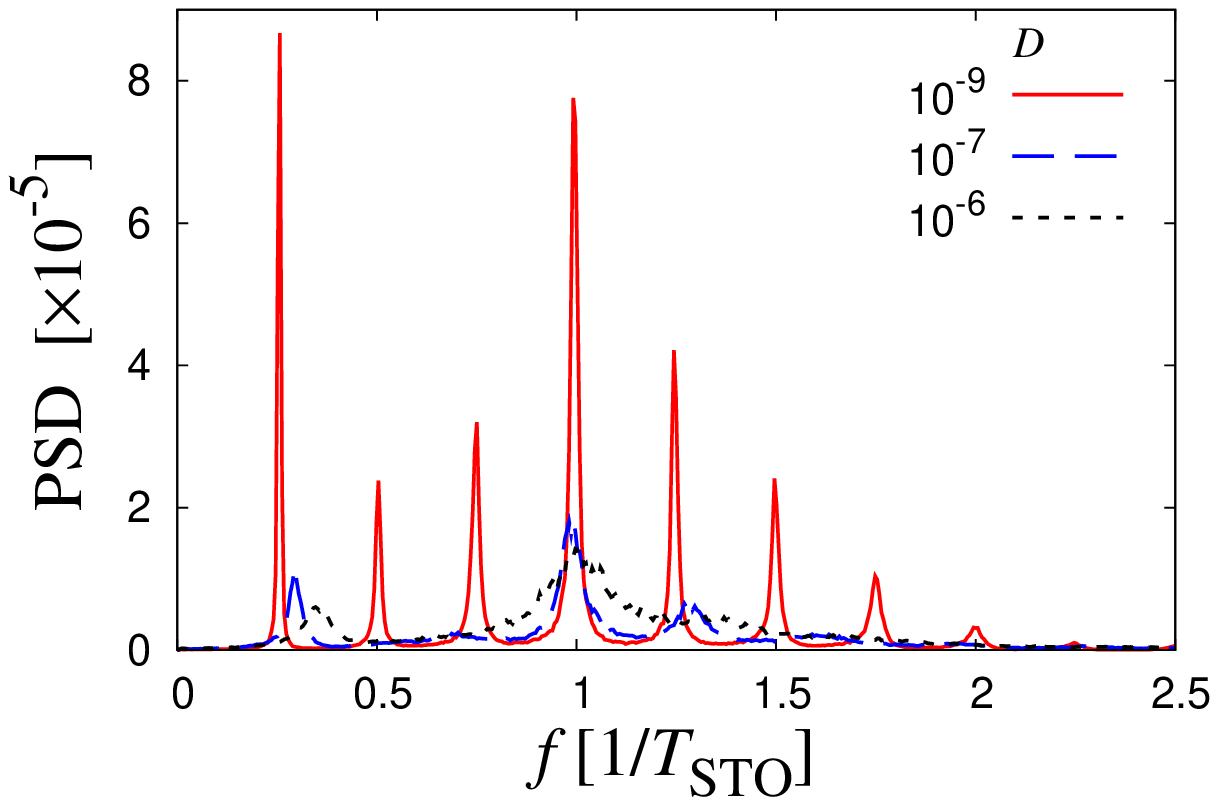}
\includegraphics[width=0.32\textwidth]{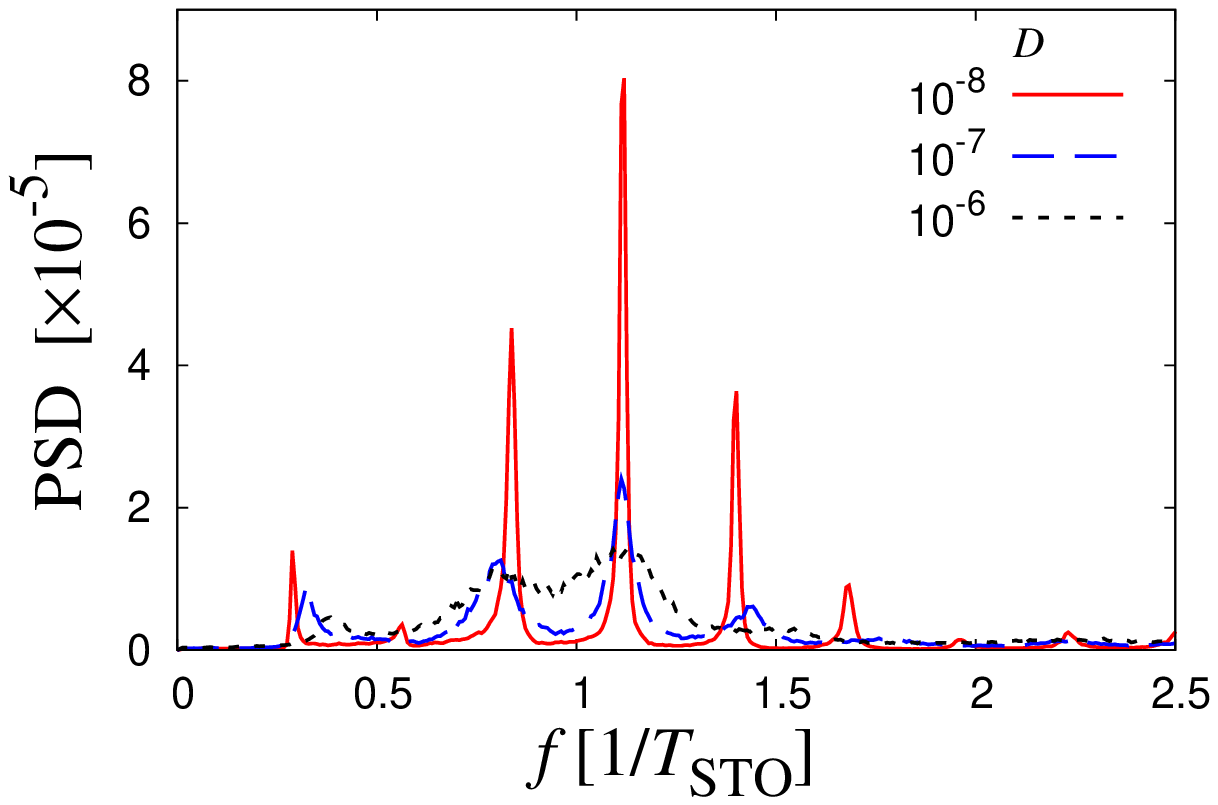}
\caption{Power spectral density (PSD) of the time series in Class 1 for various values of $D$. {\bf Top:} 3DSC ($I_{\rm app}=-2.45$); {\bf bottom:} LMFN ($\epsilon_2=0.0147$, reset to $b_{\rm rs}=u=0.315, v=-0.12603$). For details of how we computed and normalized the PSD see App.~\ref{app:measures}.}
\label{fig:class1_PSD_noise}
\end{figure}

$\bullet$ 
Increased noise shifts the ISI density to include peaks at shorter durations, consistent with earlier escape described above. Contributions at longer ISIs are due to a small fraction of trajectories for which the noise disrupts the transition and lengthens the ISI. Class 1 behavior survives only for low to moderate noise levels, as larger noise regularly eliminates the STOs. This is illustrated by the considerable spread of both the ISI density and the PSD (Fig.~\ref{fig:class1_PSD_noise}) for large noise levels.

$\bullet$
The coherence measure $\beta$ is not clearly defined for MMOs of Class 1, since the strong
 amplitude trend of the deterministic dynamics within the short ISI periods yields PSDs with many well-defined peaks (one of them at the STO frequency -- see Fig.~\ref{fig:class1_PSD_noise}).  The SC model shows more peaks due to larger amplitudes of the STOs across the ISI. The strong peak at low frequencies comes from the spiking (i.e., thresholding and concatenation).



\subsection{Class2: Long ISI, increasing STO amplitude}

Next we analyze STOs appearing in longer ISIs with a characteristic increasing amplitude before a spike. We consider two examples from the 3DSC model that have similar amplitude trends but different ISI behavior in the presence of noise. As discussed  in Sec.~\ref{sec:models} the choice of $I_{\rm app}$ for these two examples, $I_{\rm app} = -2.55$ and $I_{\rm app} = -2.58$, corresponds to different  underlying deterministic behavior, regular MMOs and stable steady state, respectively. We refer to these as Class 2A and Class 2B, respectively.

\begin{figure}
\centering
\includegraphics[width=.32\textwidth]{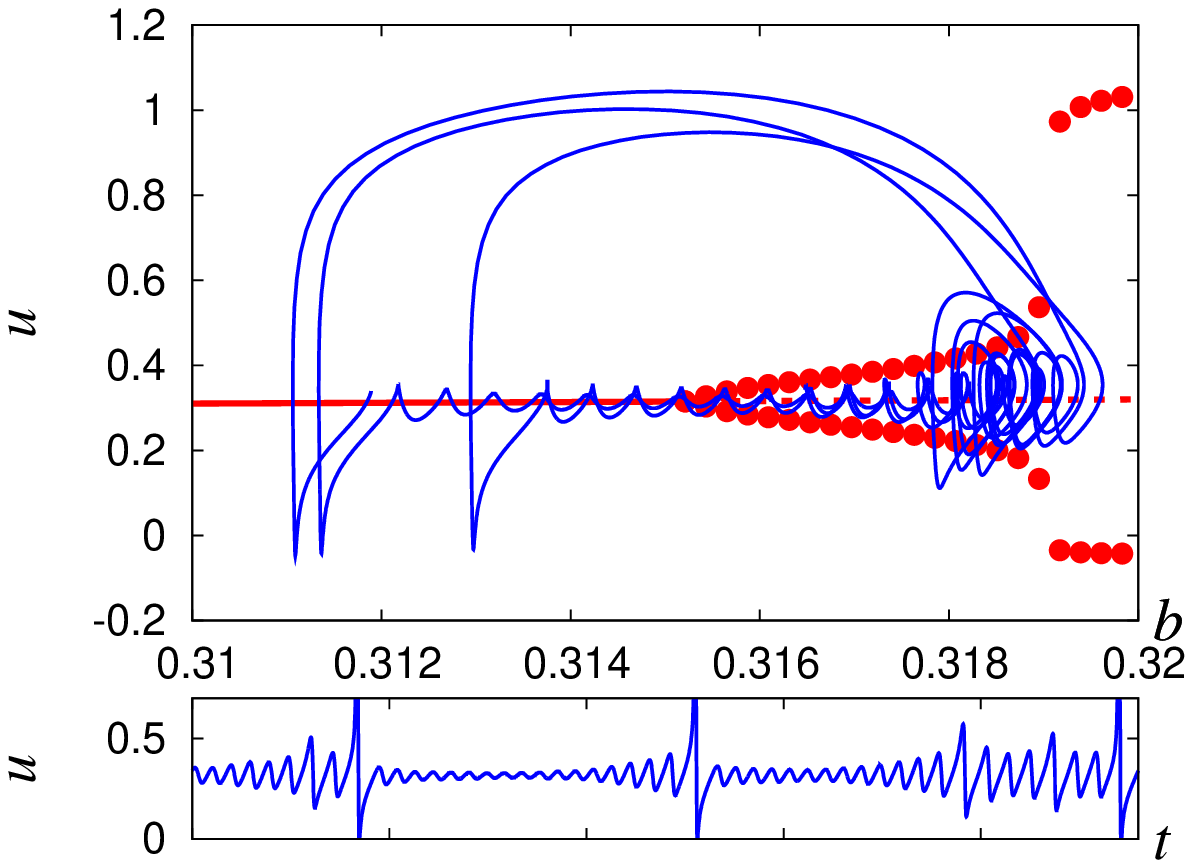}
\includegraphics[width=.32\textwidth]{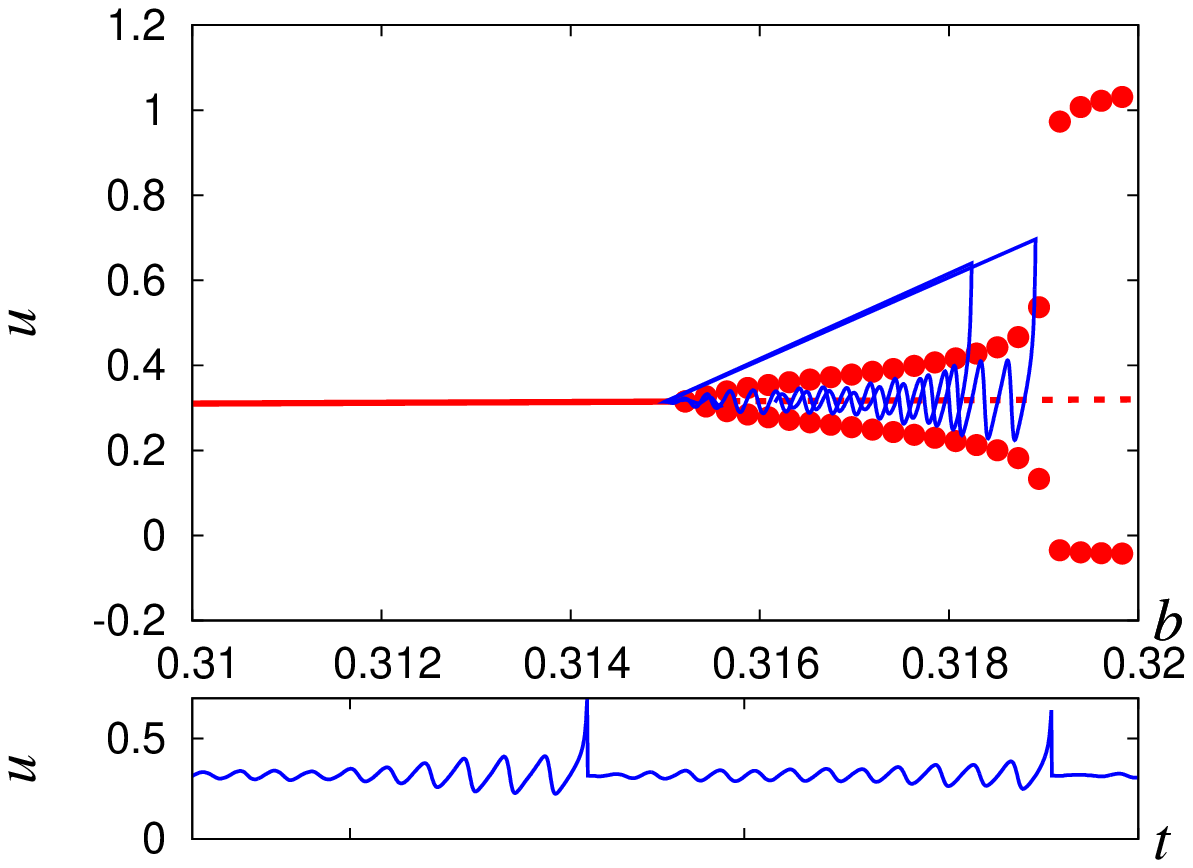}
\caption{Examples of the dynamics of the MFN model for Class 2. {\bf Top:} NLMFN with $c_b=1.1$, $D=5\times 10^{-8}$; {\bf bottom:} LMFN with $\epsilon_2=0.00055$, $D=3\times 10^{-7}$, reset to $b_{\rm rs}=0.315$. Trajectory in the underlying 2D bifurcation diagram and time series. The scale in the time series is 5.}
\label{fig:Ma_class2}
\end{figure}

\begin{figure}
\centering
\includegraphics[width=0.32\textwidth]{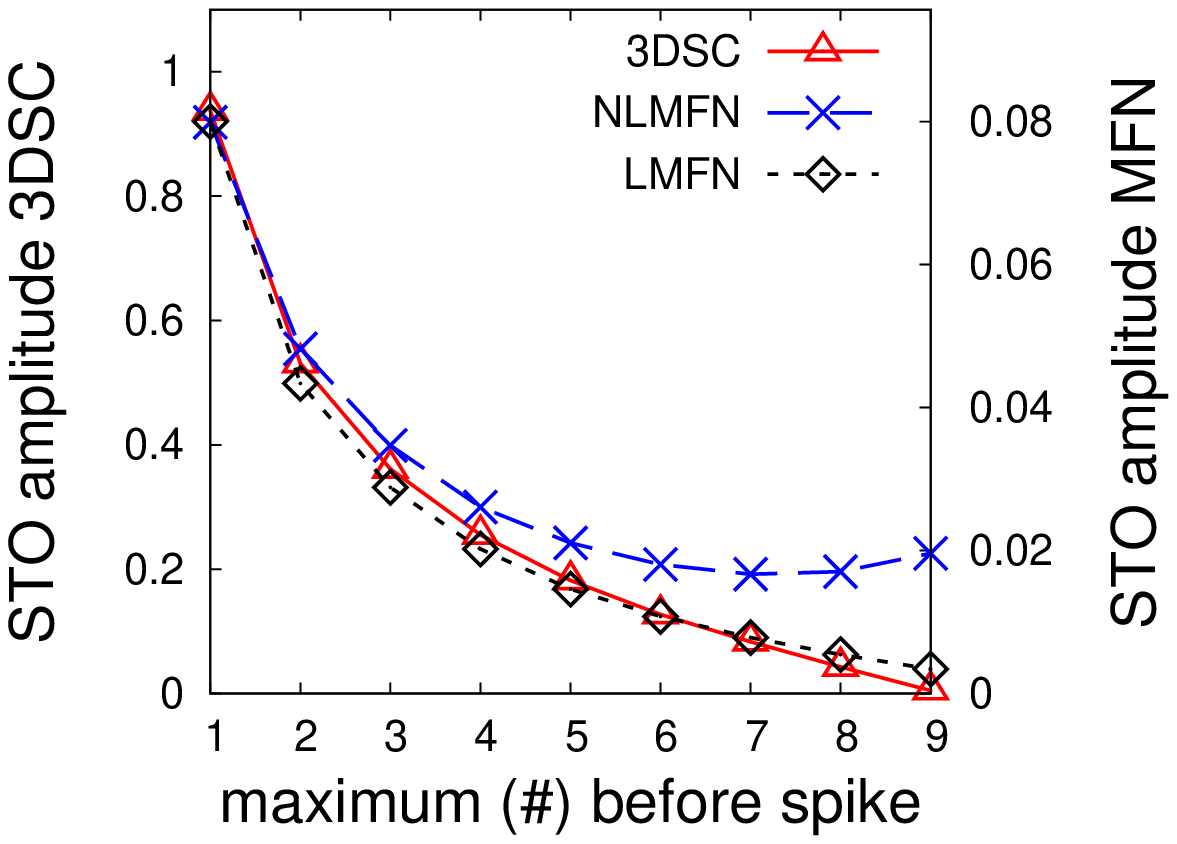}
\includegraphics[width=0.32\textwidth]{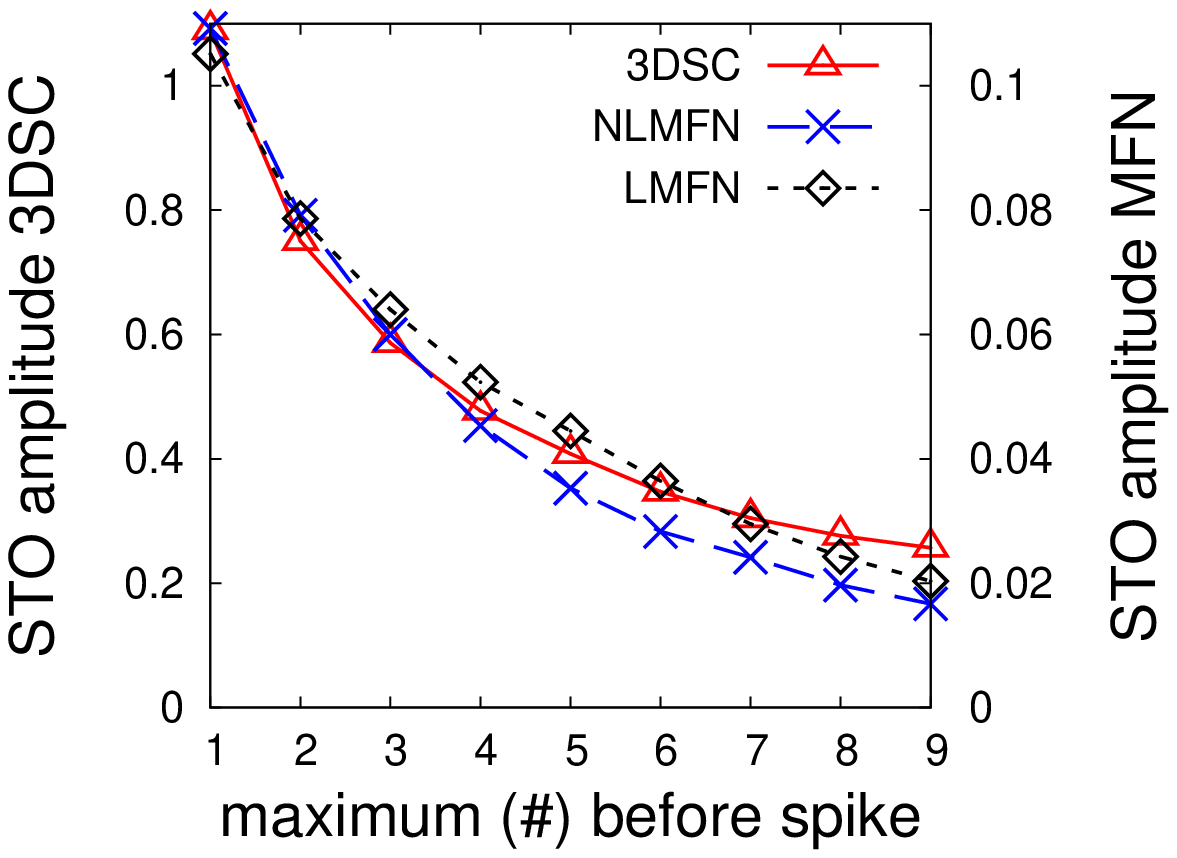}
\caption{Average amplitude before a spike for time series from Class 2 from three different models: $\triangle$ (solid red line): 3DSC ($\sim 1400$ spikes); $\times$ (dashed blue line): NLMFN ($\sim 2500$ spikes); $\diamond$ (dotted black line): LMFN ($\sim 1000$ spikes). {\bf Top:} Class 2A: 3DSC: $I_{\rm app}=-2.55$, $D=10^{-7}$; NLMFN: $c_b=0.95$, $D=10^{-8}$; LMFN: $\epsilon_2=0.001$, $D=10^{-7}$, $b_{\rm rs}=0.3155$; {\bf bottom:} Class 2B: $I_{\rm app}=-2.58$, $D=5\times 10^{-6}$ (Fig.~\ref{fig:bifurcation_sc}, bottom panel); NLMFN: $c_b=1.1$, $D=5\times 10^{-8}$ (Fig.~\ref{fig:Ma_class2}, top panel); LMFN: $\epsilon_2=0.00055$, $D=3\times 10^{-7}$, $b_{\rm rs}=0.315$ (Fig.~\ref{fig:Ma_class2}, bottom panel).
}
\label{fig:class2_amp}
\end{figure}

\begin{figure}
\centering
\includegraphics[width=0.32\textwidth]{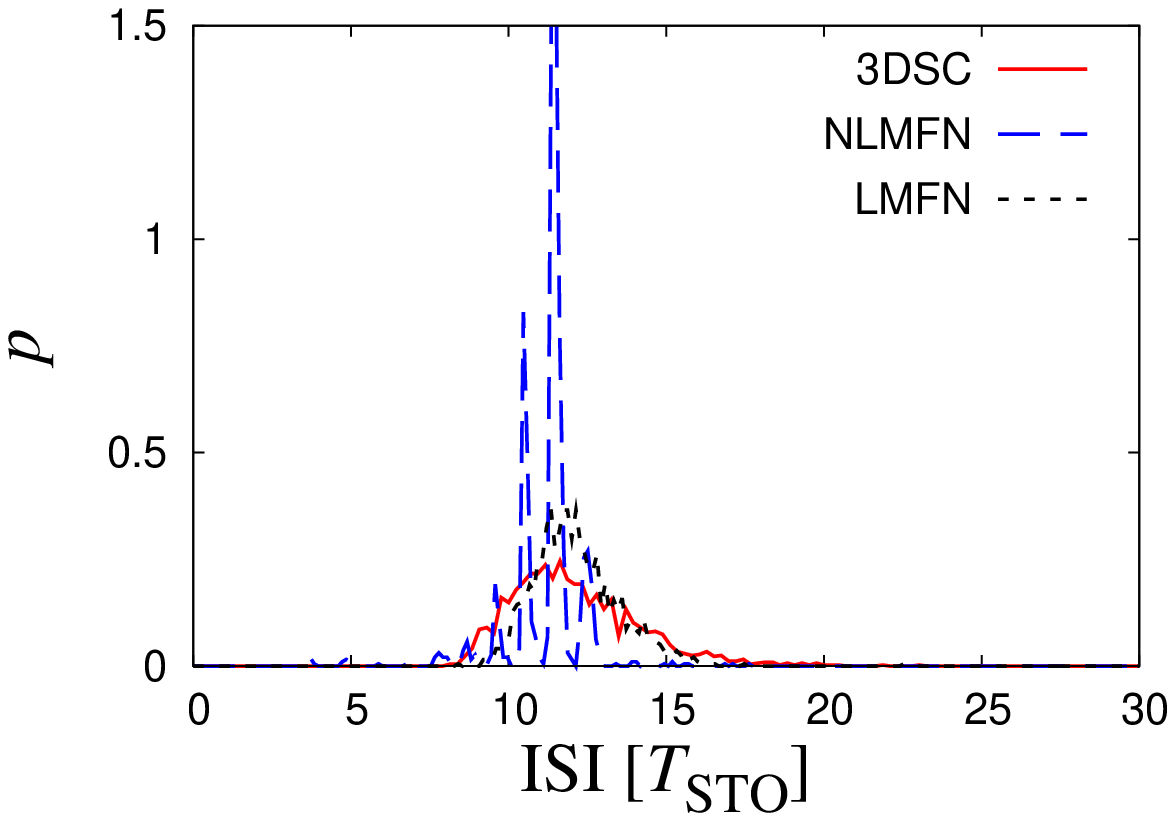}
\includegraphics[width=0.32\textwidth]{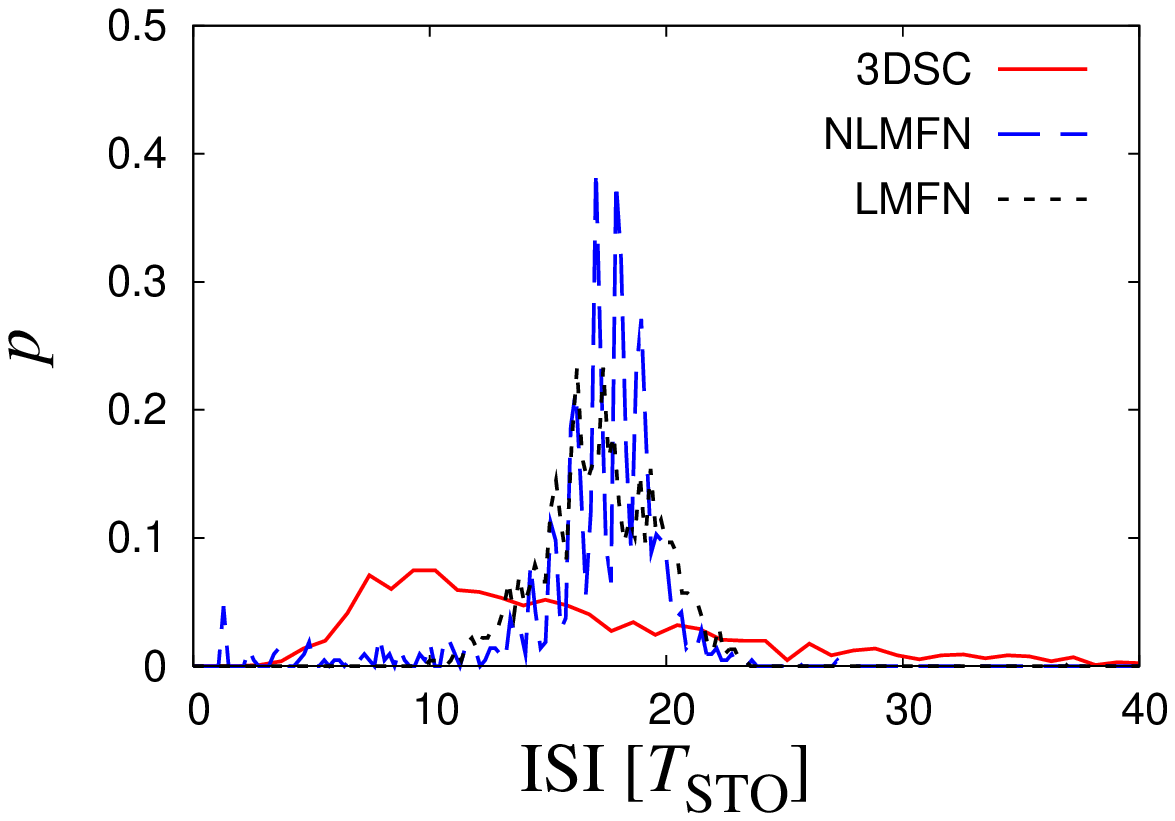}
\caption{Densities of ISI lengths for Class 2 MMOs. Solid line (red): 3DSC; dashed line (blue): NLMFN; dotted line (black): LMFN. {\bf Top:} Class 2A; {\bf bottom:} Class 2B. For parameters, see Fig.~\ref{fig:class2_amp}.}
\label{fig:class2_ISI}
\end{figure}

Sections of the time series as well as the trajectories within the underlying bifurcation are shown in Figs.~\ref{fig:bifurcation_sc} (bottom panel) and~\ref{fig:Ma_class2} for the 3DSC, NLMFN, and LMFN, respectively. These time series were calibrated for similar average amplitude trend and average ISI durations. Fig.~\ref{fig:class2_amp} shows that for low noise levels, it is possible to find parameter values for LMFN that compare well in amplitude behavior with 3DSC for both Class 2A and Class 2B. For NLMFN, the amplitude behavior is similar to 3DSC for Class 2B but differs in Class 2A for STOs well before the spike. In the refractory period following the spike in the NLMFN  the system relaxes near but not on the left null cline with $b<b_{\rm H}$,  yielding slowly damped STOs in the first part of the ISI. For the LMFN, the reset value for $u$ and $v$ is chosen very close to the fixed point, avoiding this part of STOs with decaying amplitude. Comparable average ISIs are easily obtained in the LMFN with $\epsilon_2$ chosen an order of magnitude smaller than in Class 1, and for  NLMFN with $c_b$ near unity for weak noise. Differences in the ISI density are apparent from Fig.~\ref{fig:class2_ISI} and are related to differences in the deterministic dynamics as discussed in Secs.~\ref{sec:analysis_intra} and~\ref{sec:intra_FHN}.

As we see below, by tuning parameters and noise levels, the FN-type model can capture  some but not all of the amplitude, ISI and coherence behavior for Class 2. Results for different noise levels point to differences in the underlying structure of the models.

$\bullet$ 
For noise levels increased by an order of magnitude,  both subclasses for 3DSC show roughly a 20-30\% decrease in the average amplitude of the largest STO immediately preceding a spike, while the FN-type models show a slightly greater decrease (Fig.~\ref{fig:class2_amp_noise}). For Class 2B, the entire average amplitude curve of 3DSC shifts to lower levels with increased noise, while both FN-type models show both shifts and changes in the average amplitude curve shape. For Class 2A, both 3DSC and LMFN show an increase of the average amplitude of STOs well before the spike for increased noise, indicating STOs driven both by CR and  by SPHB. For the NLMFN model, the return sets $b<b_{\rm H}$ following an earlier spike, so that the combination of damped STOs for $b<b_{\rm H}$ and CR effects yields robust STOs with less variation with noise level than for for the LMFN model.

$\bullet$ 
In Class 2B, the 3DSC model has longer tails in the ISI density for both small and intermediate noise levels. These longer tails are not  seen in the FN-type models due to differences in the underlying deterministic dynamics (see Sec.~\ref{sec:analysis_intra}). For weak noise, the multi-modal ISI density for the NLMFN is due to the stronger attraction to the underlying deterministic MMO or STO dynamics, with larger STOs initially in the ISI. In the LMFN model the reset onto the nullcline avoids damped STOs initially in the ISI, allowing a unimodal ISI density. For both Class 2A and 2B, 3DSC and LMFN show a shift to shorter ISIs and a similar shape of the ISI density for larger  noise, with a  larger  variance  for LMFN in Class 2A. For NLMFN there is no shift in the ISI density with increased noise, just an increased variance both for Class 2A and 2B, reflecting the dynamic return mechanism following the spike.

\clearpage

\begin{widetext}

\begin{figure}
\centering
\includegraphics[width=0.32\textwidth]{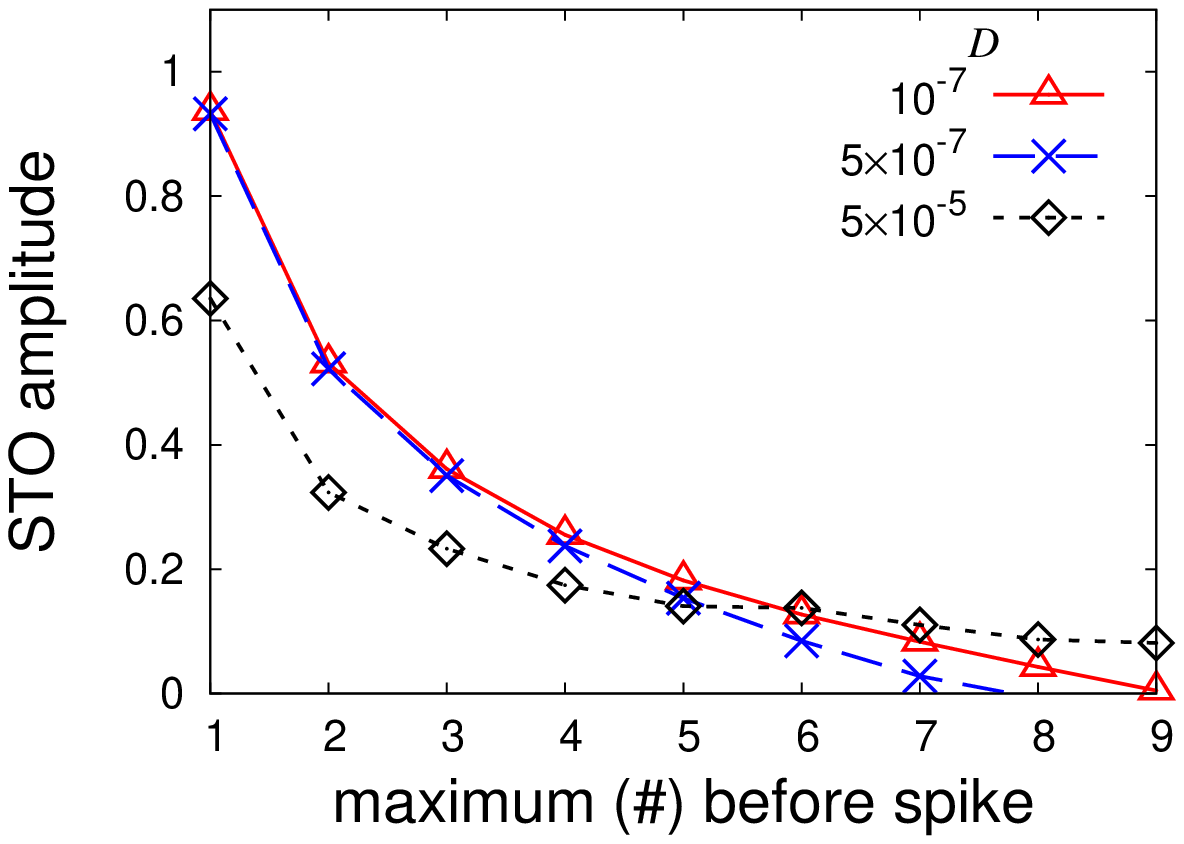}
\includegraphics[width=0.32\textwidth]{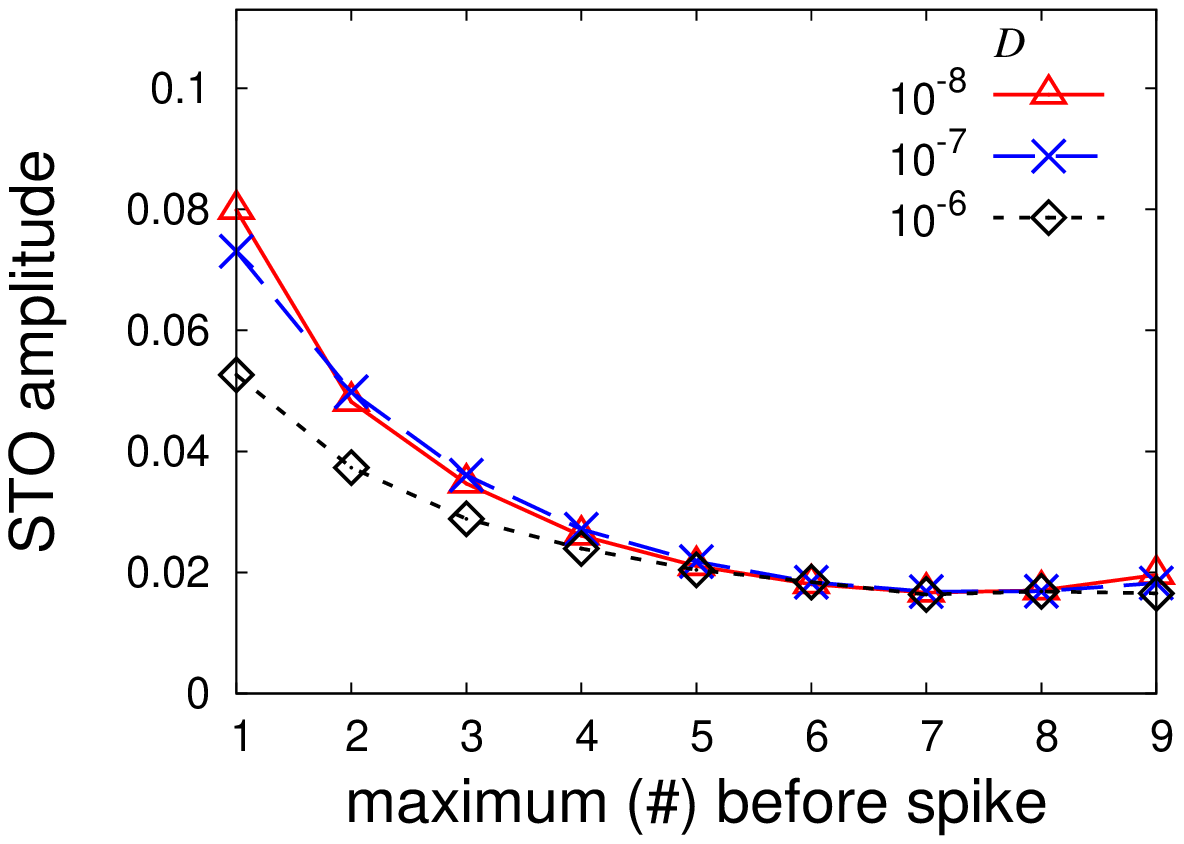}
\includegraphics[width=0.32\textwidth]{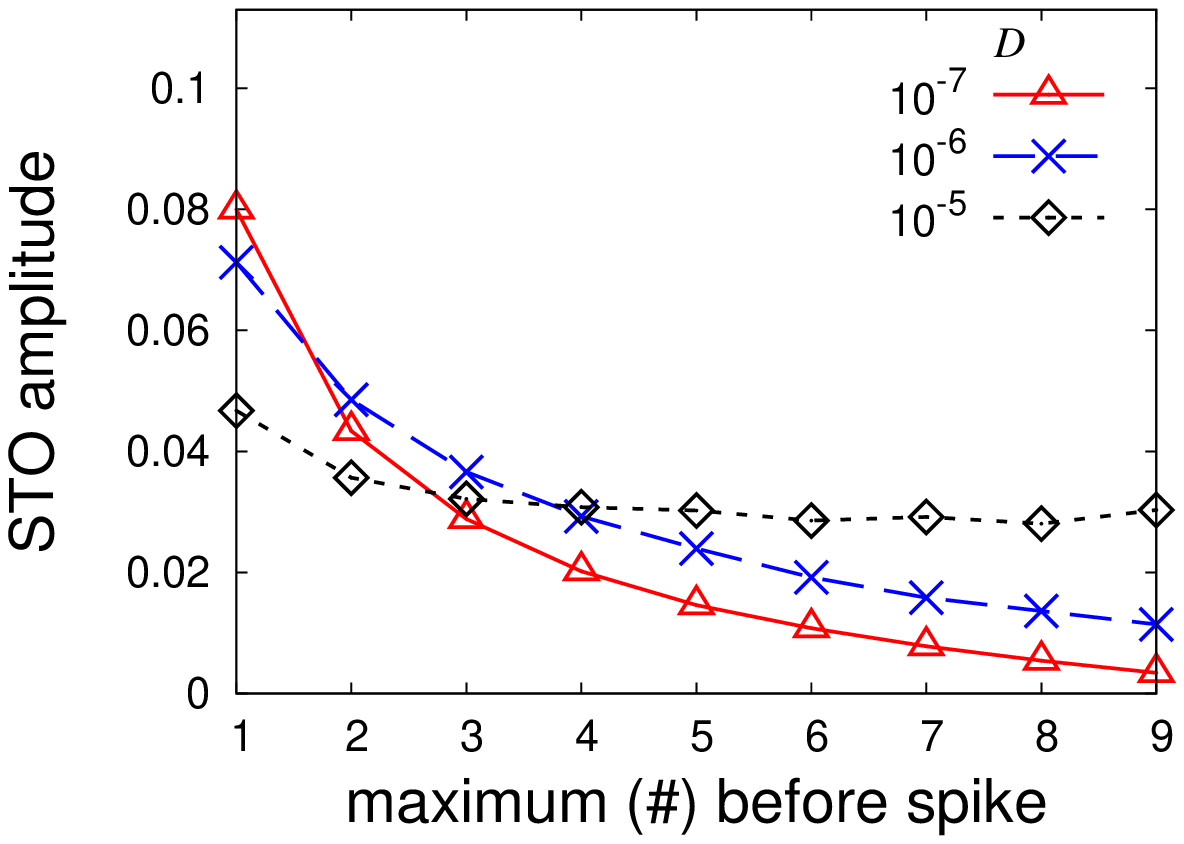}
\\
\includegraphics[width=0.32\textwidth]{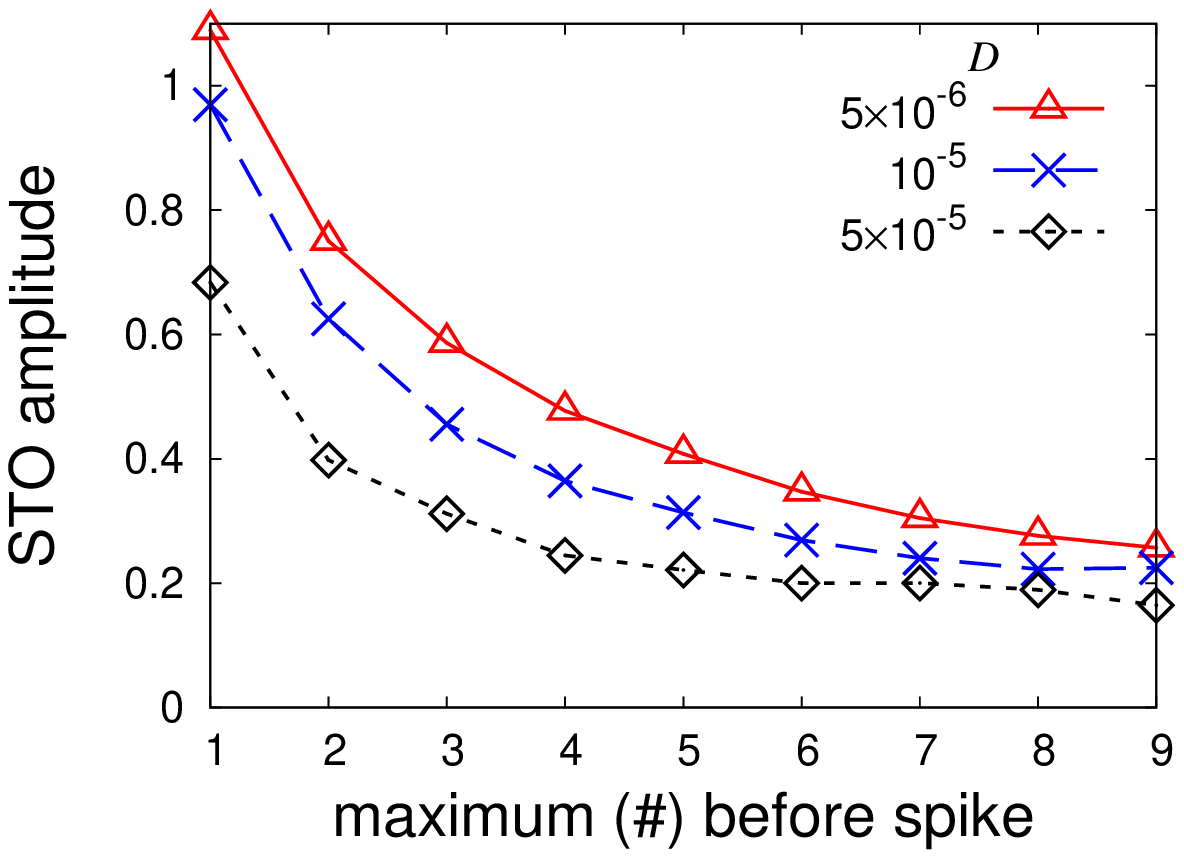}
\includegraphics[width=0.32\textwidth]{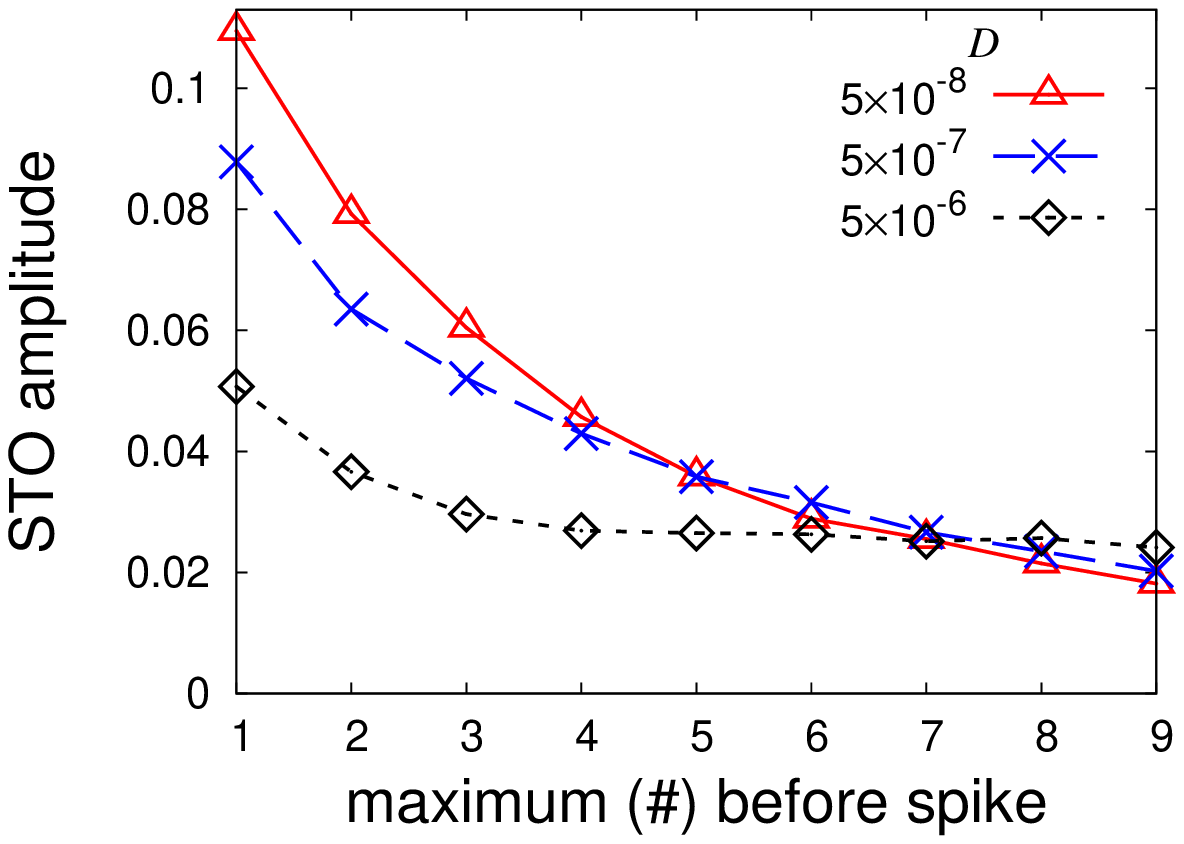}
\includegraphics[width=0.32\textwidth]{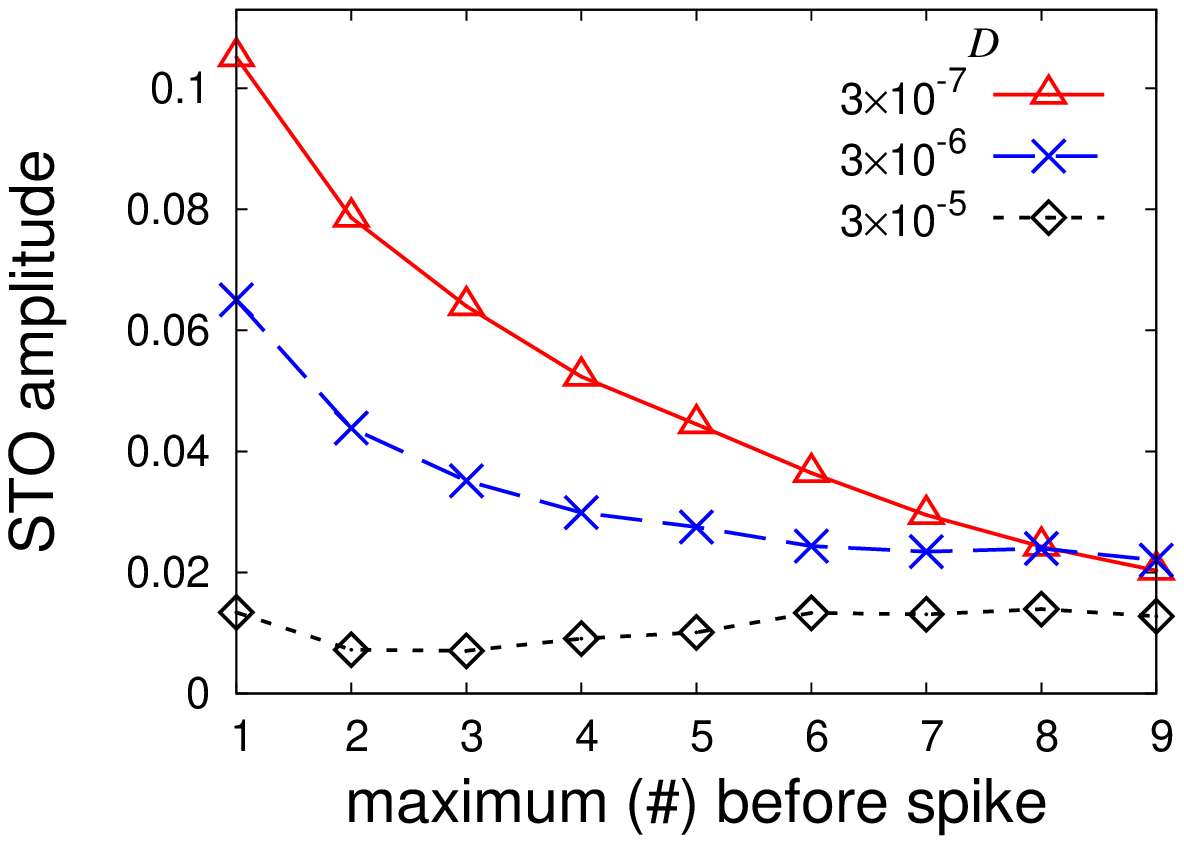}
\caption{Average amplitude before a spike in the two subclasses of Class 2 for various values of $D$. {\bf Left column:} 3DSC; {\bf middle column:} NLMFN; {\bf right column:} LMFN. {\bf Top row:} Class 2A; {\bf bottom row:} Class 2B. For parameters, see Fig.~\ref{fig:class2_amp}.}
\label{fig:class2_amp_noise}
\end{figure}
\begin{figure}
\centering
\includegraphics[width=0.32\textwidth]{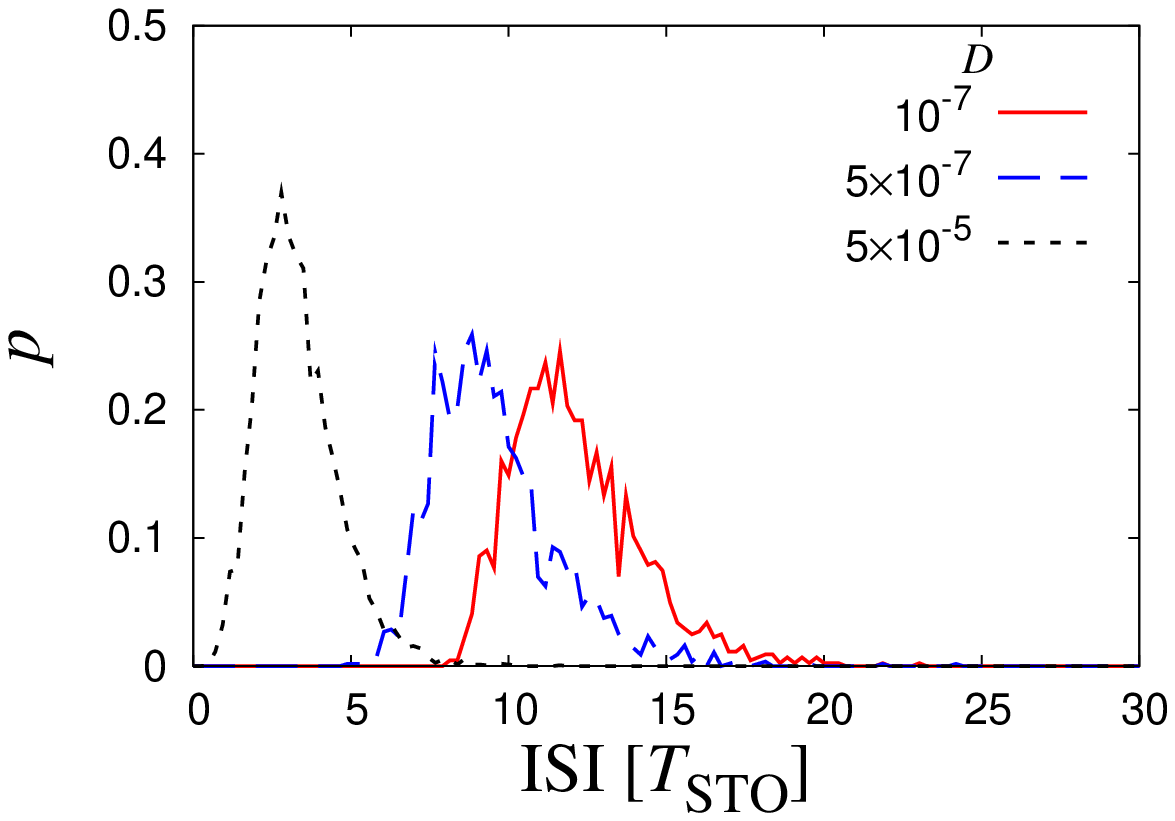}
\includegraphics[width=0.32\textwidth]{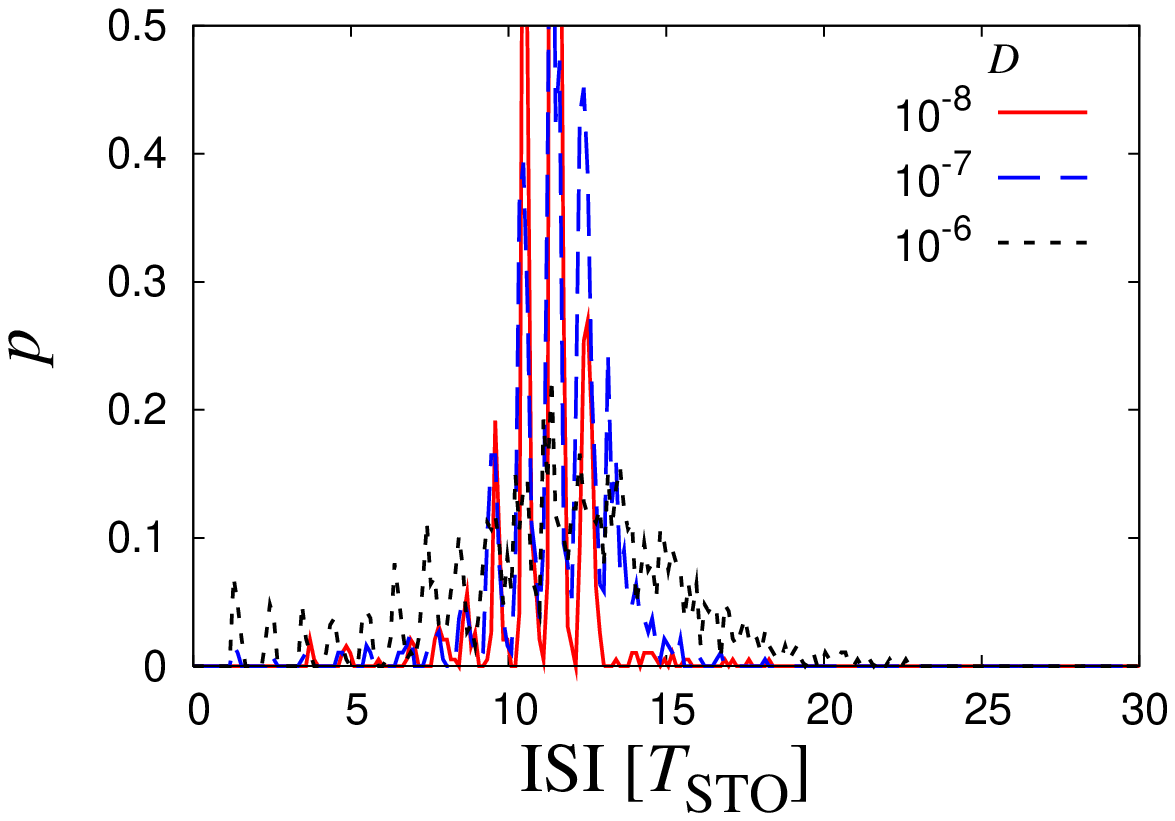}
\includegraphics[width=0.32\textwidth]{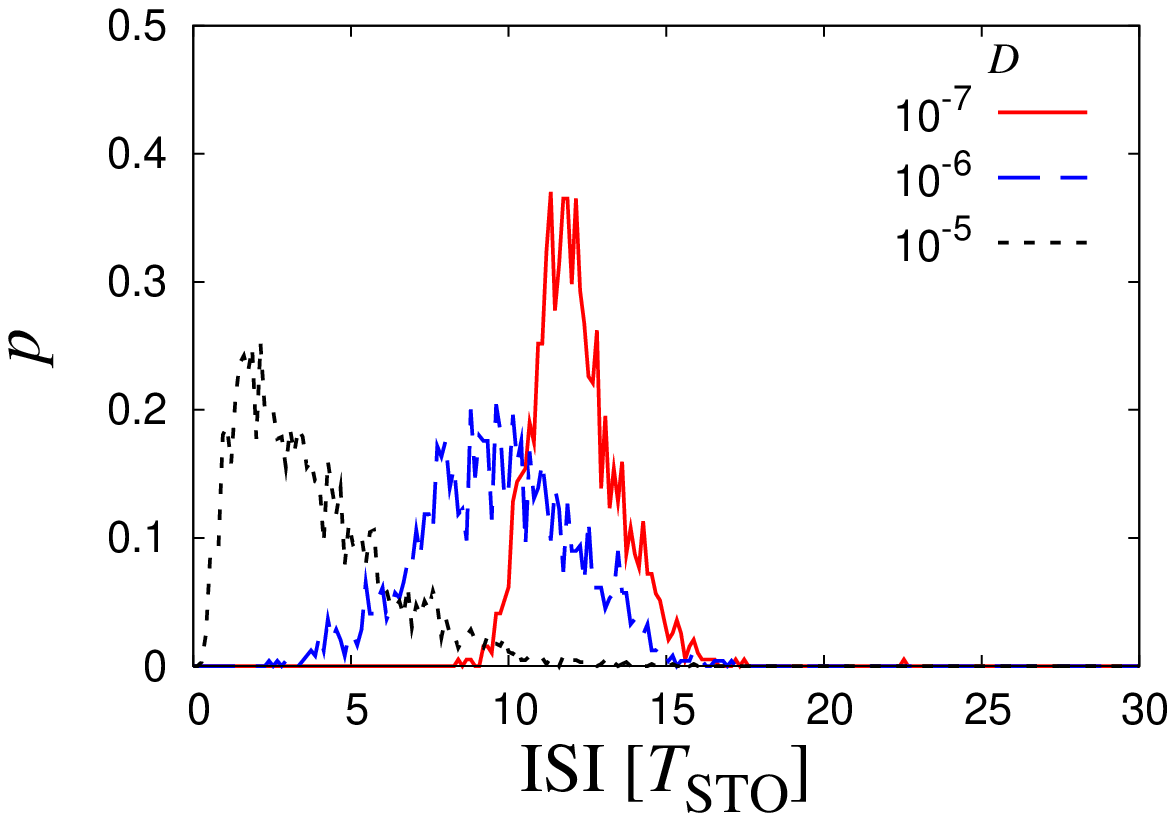}
\\
\includegraphics[width=0.32\textwidth]{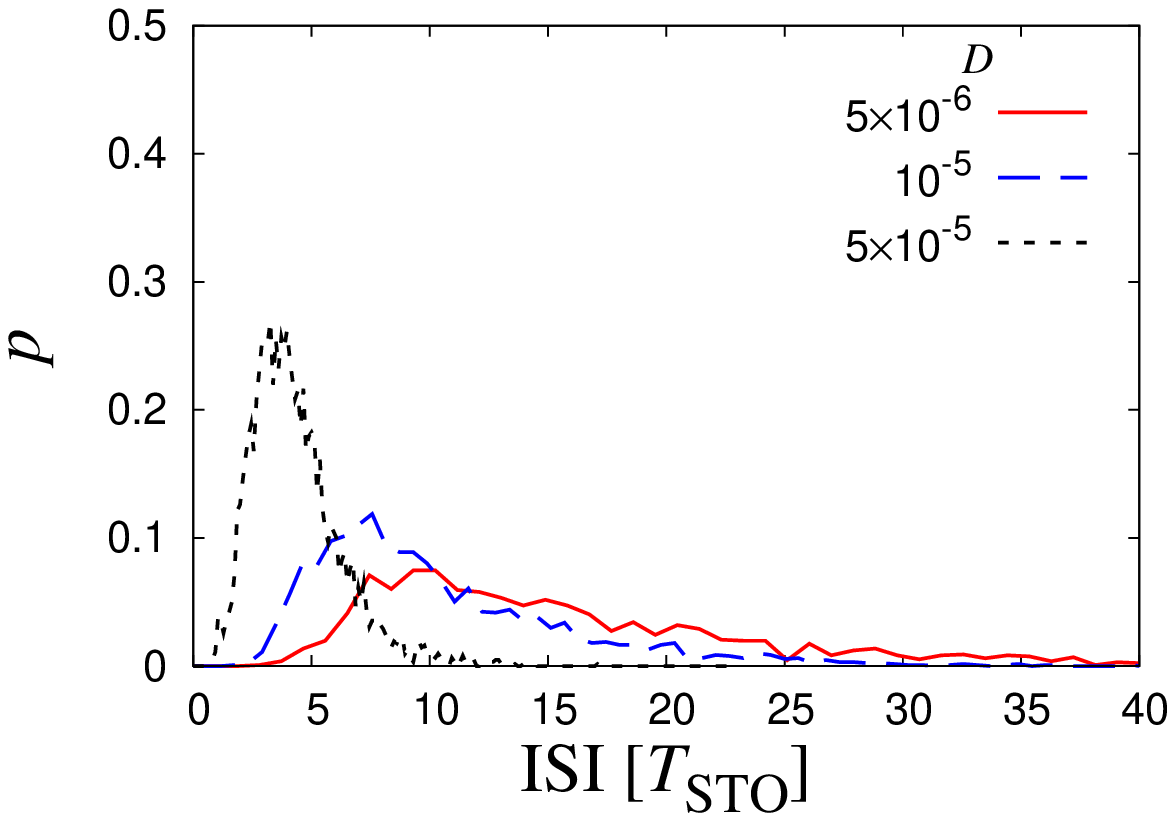}
\includegraphics[width=0.32\textwidth]{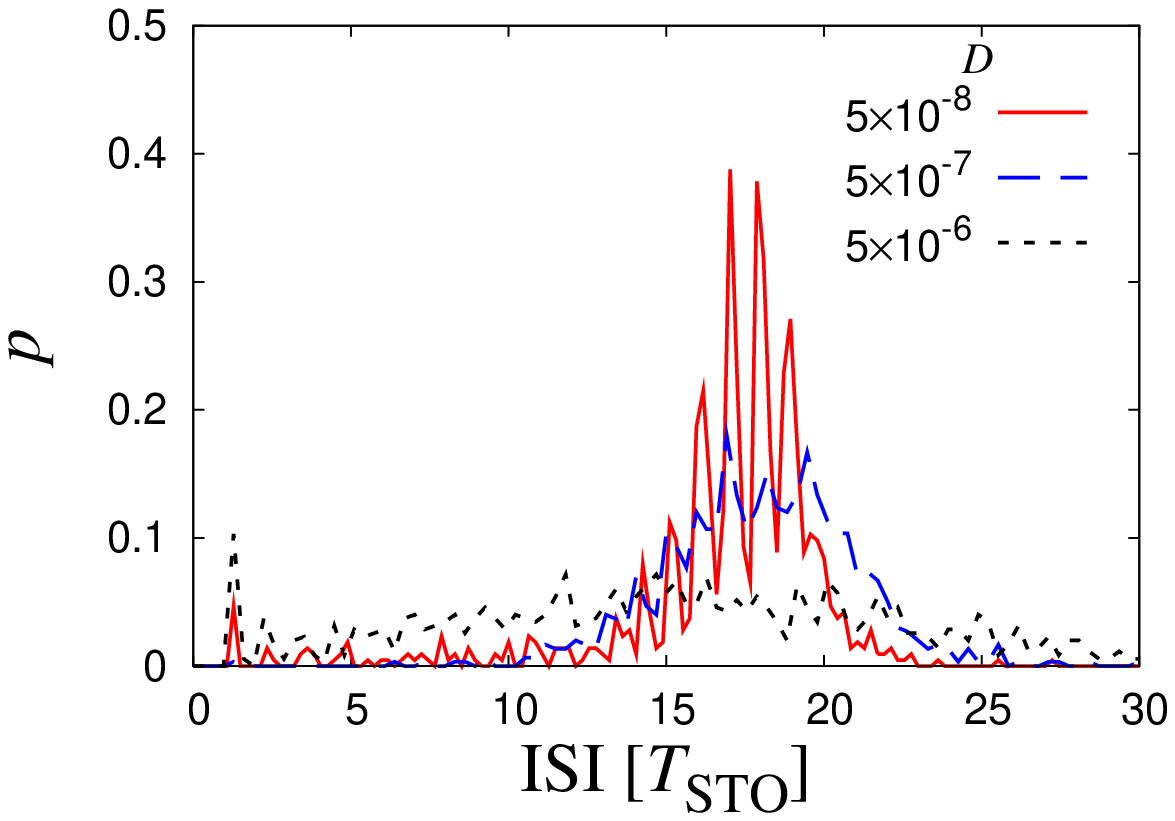}
\includegraphics[width=0.32\textwidth]{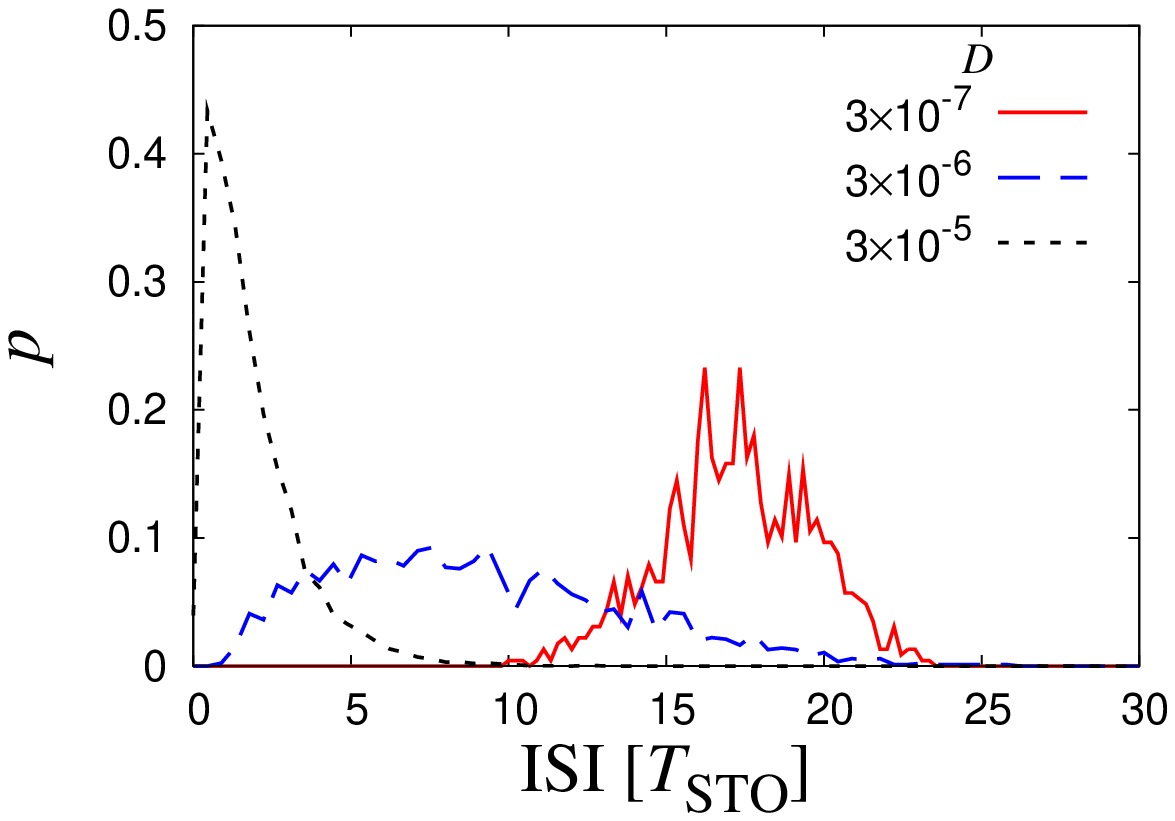}
\caption{Densities of ISI lengths in the two subclasses of Class 2 for various values of noise parameter $D$. {\bf Left column:} 3DSC; {\bf middle column:} NLMFN; {\bf right column:} LMFN. {\bf Top row:} Class 2A; {\bf bottom row:} Class 2B. For parameters, see Fig.~\ref{fig:class2_amp}. In the top row, middle panel, values for large $p$ are cut off.}
\label{fig:class2_ISI_noise}
\end{figure}
\end{widetext}

\begin{figure}
\centering
\includegraphics[width=0.32\textwidth]{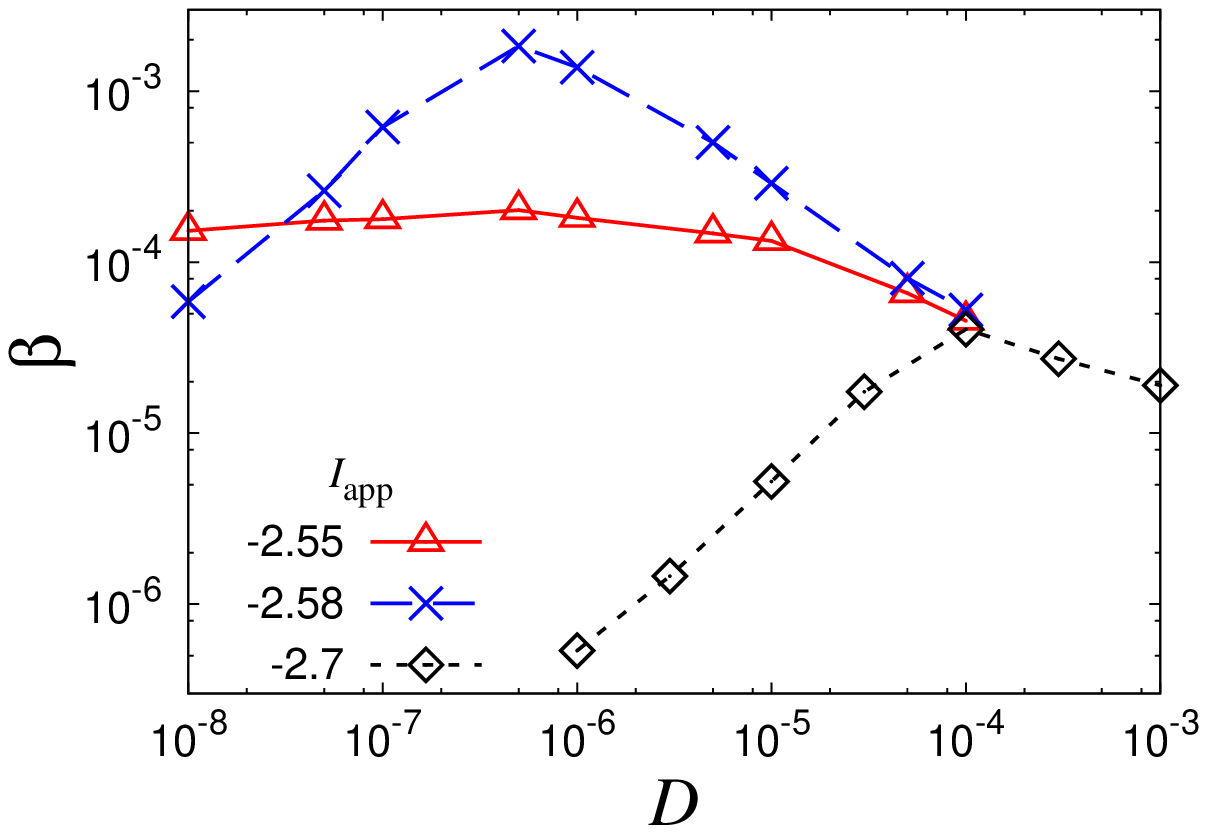}
\includegraphics[width=0.32\textwidth]{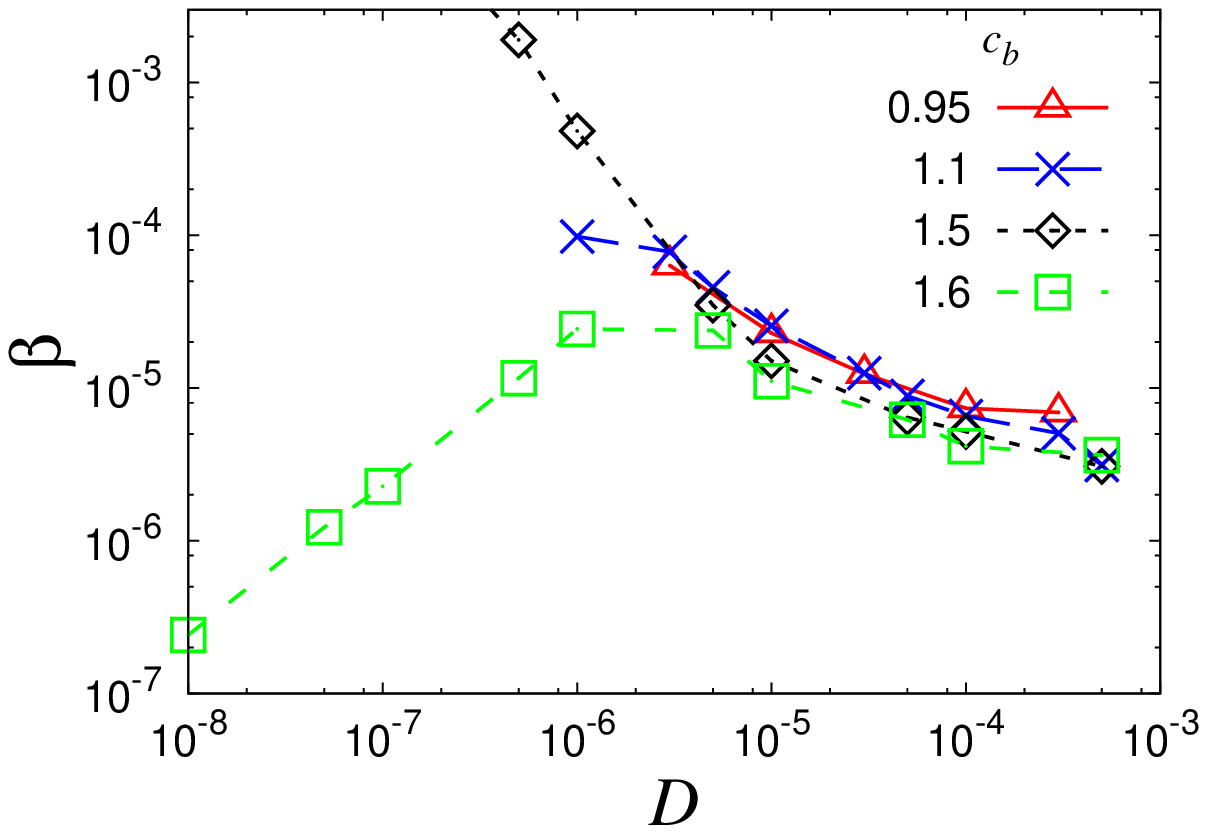}
\includegraphics[width=0.32\textwidth]{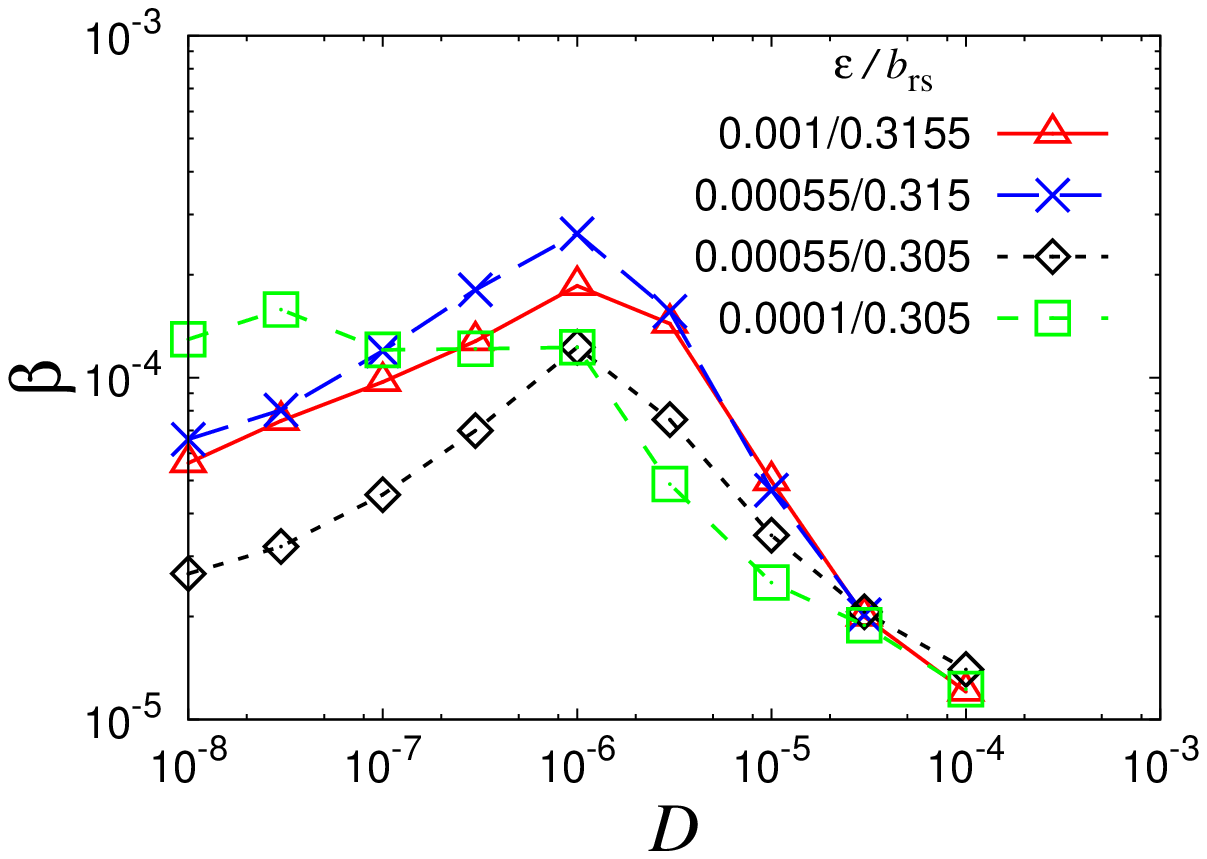}
\caption{Coherence measure $\beta$ (see Subsec.~\ref{ssec:tools}) for the 3DSC model ({\bf top}), NLMFN ({\bf middle}) and the LMFN model ({\bf bottom}) for various values of noise strength $D$. Class 2A is within the red solid lines, Class 2B within the blue dashed lines, and Class 3 within the black dotted lines (blue dashed for NLMFN).}
\label{fig:class2_beta}
\end{figure}

$\bullet$ 
Fig.~\ref{fig:class2_beta} shows the differences between the coherence measure $\beta$ (Subsec.~\ref{ssec:tools}) for the three different models for various noise levels. For 3DSC in Class 2B the STOs are driven by CR and there is a peak in $\beta$ (indicating CR), while for Class 2A noise reduces $\beta$ in general, since the STOs are driven by a SPHB. For low noise levels in NLMFN, the robust STOs are represented by a multi-peaked PSD (equivalent to the ISI density -- see above), for which $\beta$ is not defined.  For larger noise levels in Class 2, $\beta$ is defined, but noise only disrupts the underlying STOs, so $\beta$ decreases with noise. For LMFN, a moderate  noise level can enhance the STOs, particularly at the beginning of an ISI, yielding an optimal noise level for coherence in both types of Class 2. The influence of the underlying bifurcation structure on the coherence measure is discussed further in Sec.~\ref{sec:analysis_intra}.

\subsection{Class 3: Long ISI, constant average STO amplitude}

\begin{figure}
\centering
\includegraphics[width=.32\textwidth]{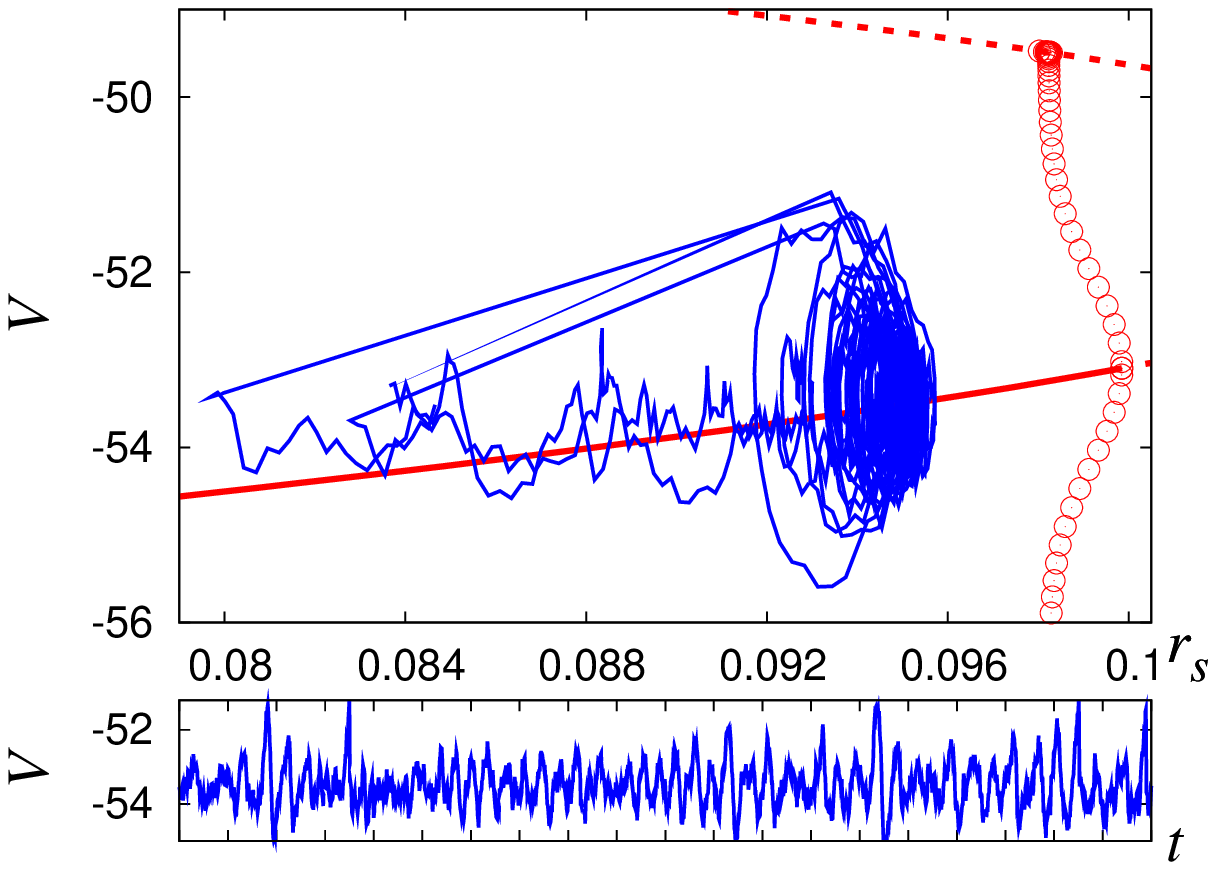}
\includegraphics[width=.32\textwidth]{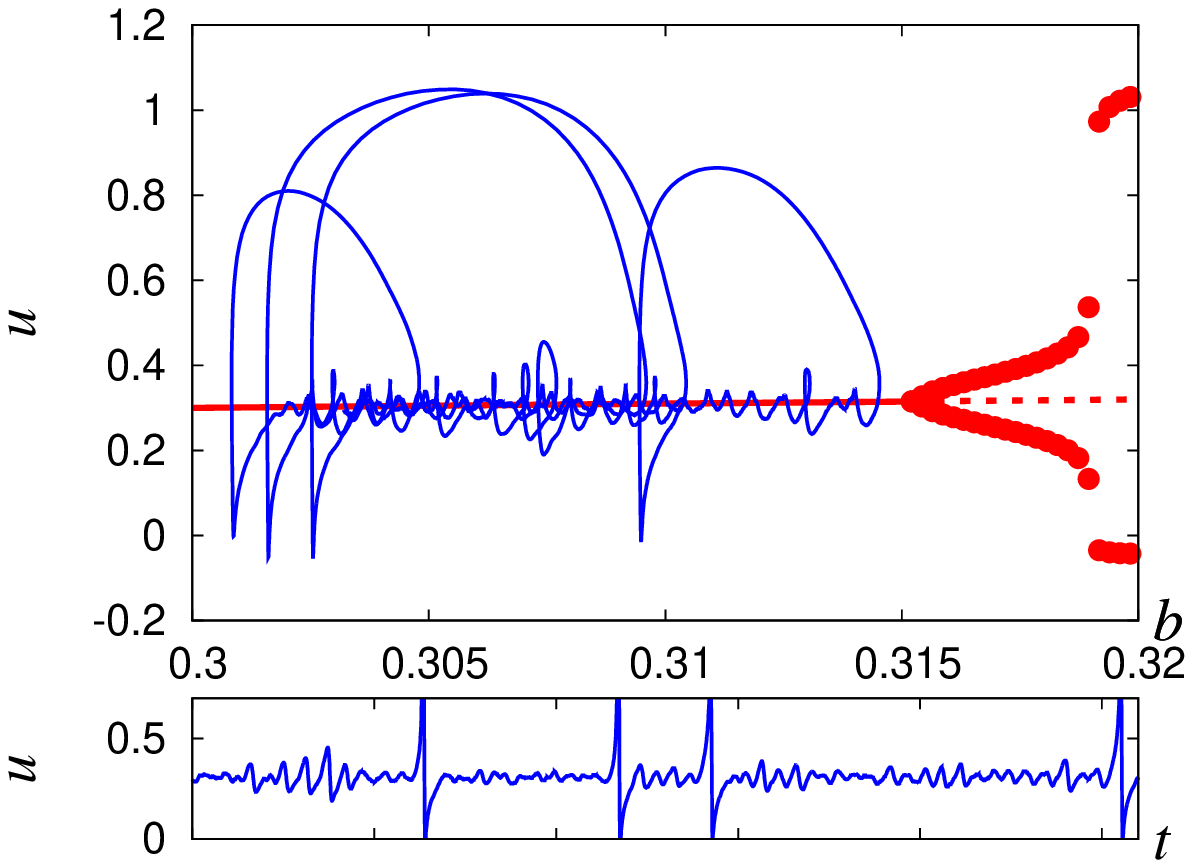}
\includegraphics[width=.32\textwidth]{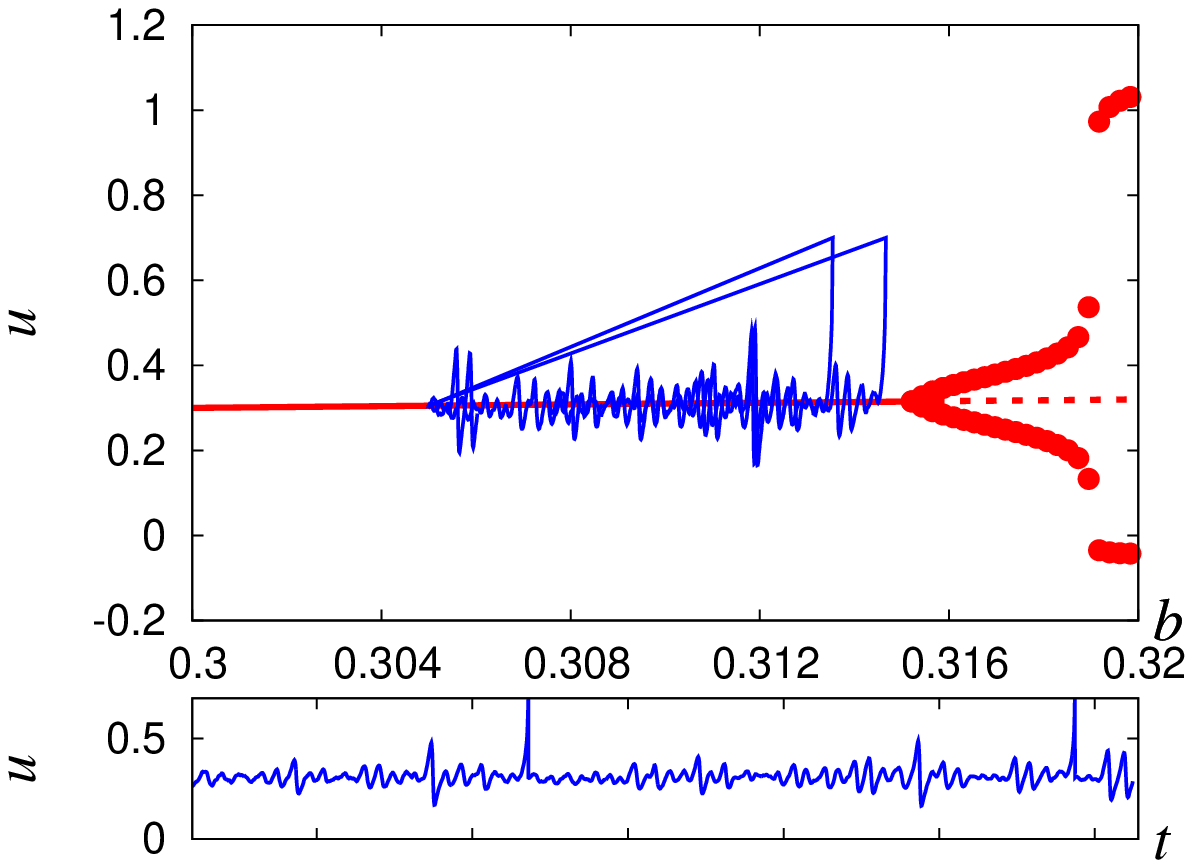}
\caption{Trajectories of Class 3 MMOs generated with the following models: {\bf top:} 3DSC ($I_{\rm app}=-2.7$, $D=10^{-4}$); {\bf middle:} NLMFN ($c_b=1.1$, $D= 10^{-5}$); {\bf bottom:} LMFN ($\epsilon_2=0.00055$, $D= 10^{-5}$, $b_{\rm rs}=0.305$). The scales in the time series are 250 (SC) and 5 (MFN).}
\label{fig:MMO_timeseries_3}
\end{figure}

Here we analyze STOs with an average ISI similar in length to Class 2, but with a constant average amplitude preceding a spike. Fig.~\ref{fig:MMO_timeseries_3} shows the trajectories from the 3DSC, NLMFN, and LMFN calibrated for similar average amplitude behavior and average ISI duration. Class 3 can  be observed in 3DSC only at higher noise levels compared with Classes 1 and 2.

\begin{figure}
\centering
\includegraphics[width=0.32\textwidth]{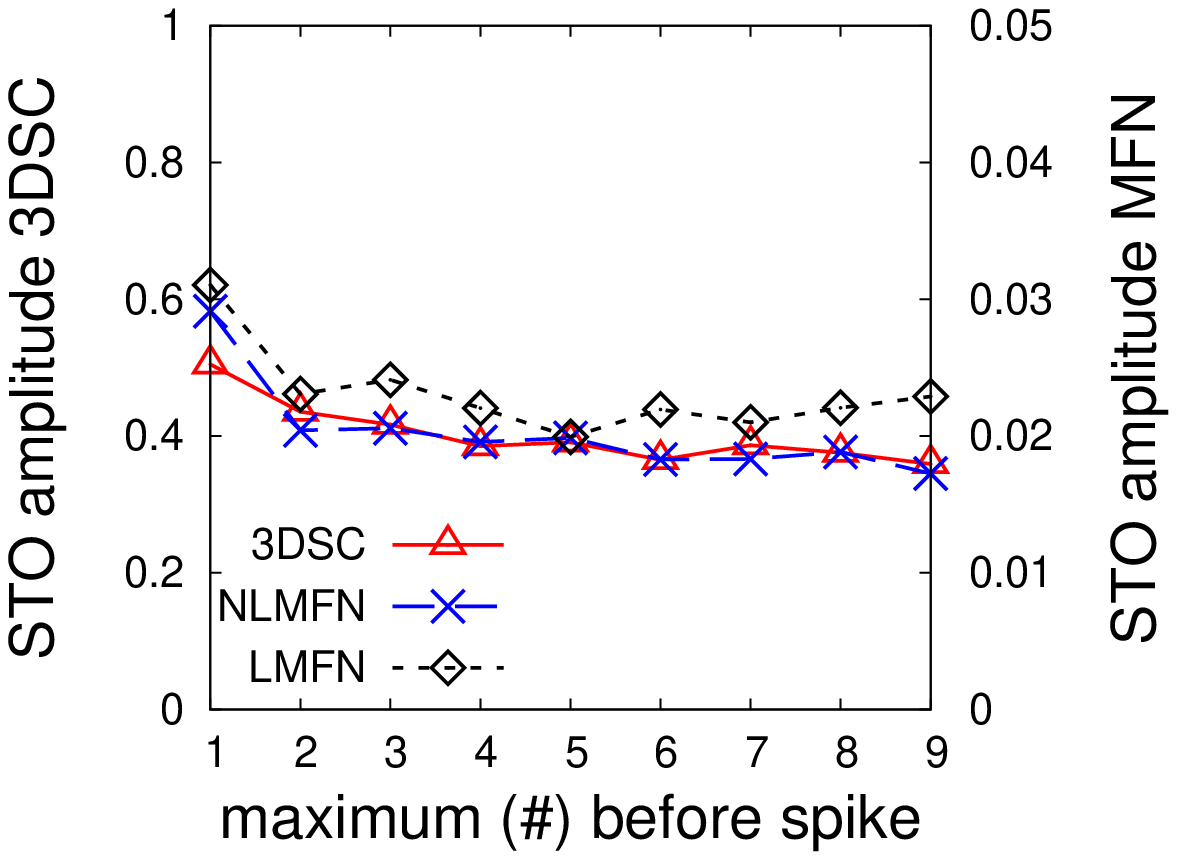}
\includegraphics[width=0.32\textwidth]{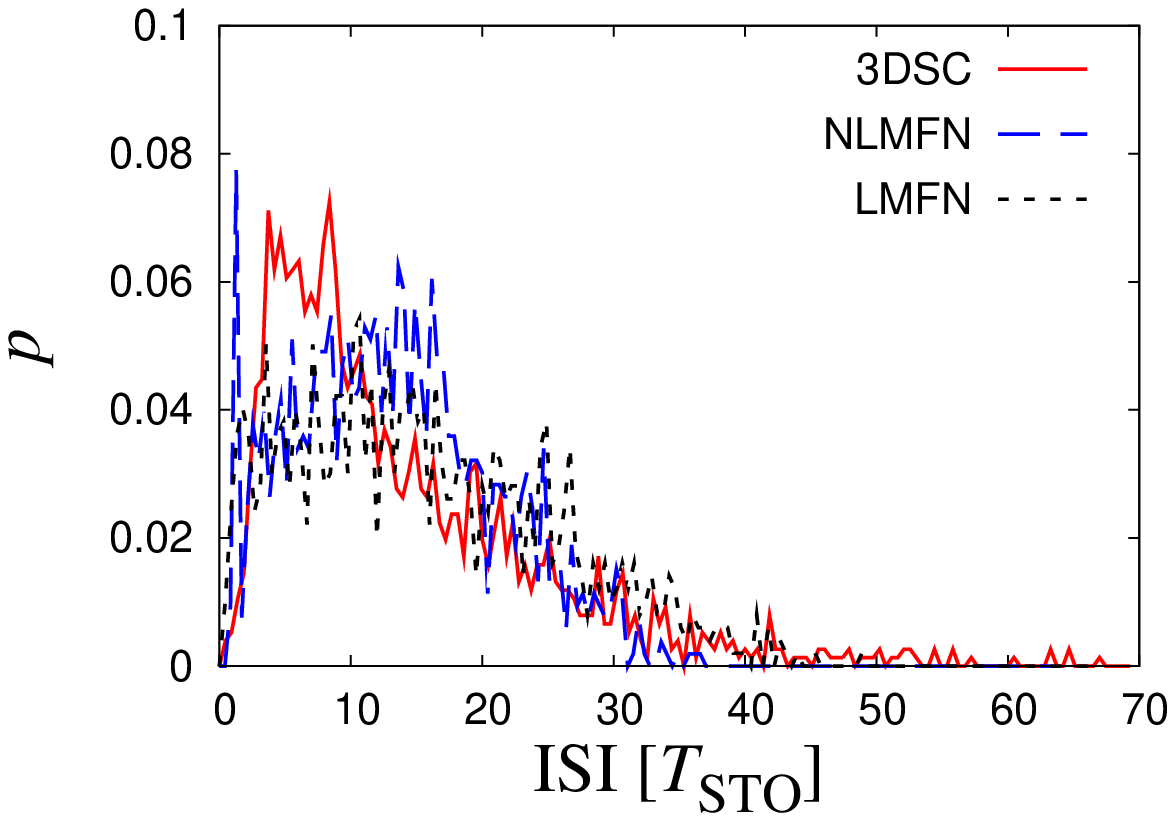}
\caption{{\bf Top:} Average amplitude before a spike for the time series from Class 3: $\triangle$ (red solid line): 3DSC ($\sim 1600$ spikes); $\times$ (blue dashed line): NLMFN ($\sim 1200$ spikes); $\diamond$ (black dotted line): LMFN ($\sim 1200$ spikes).
{\bf Bottom:} Densities of ISI lengths for Class 3 MMOs.
Solid line (red): 3DSC
; dashed line (blue): NLMFN
; dotted line (black): LMFN
. For parameters, see Fig.~\ref{fig:MMO_timeseries_3}.}
\label{fig:class3_amp}
\label{fig:class3_ISI}
\end{figure}

Fig.~\ref{fig:class3_amp} shows that the average STO amplitude for Class 3 is constant for at least the last ten periods of STOs with a slight amplitude increase in transition to the spike. This nearly constant average amplitude results from strong variation in amplitude with a constant ensemble average. A clear single peak in the PSD (data not shown) confirms well-defined STOs with variability in phase and amplitude.

The amplitude and average ISI characteristics of these MMOs are  readily reproduced with the FN-type models at stronger noise levels with the average amplitude lower by roughly a factor of two compared to Classes 1 and 2. Class 3 behavior can also be obtained  for the MFN model (Eqs.~\ref{eq:Makarov_u} and~\ref{eq:Makarov_v}) with constant control parameter $b$ as in Ref.~\onlinecite{Makarov2001}. The ISI densities shown in Fig.~\ref{fig:class3_ISI} are similar to each other, with a slightly more significant tail in the ISIs  for 3DSC.

\begin{figure}
\centering
\includegraphics[width=0.32\textwidth]{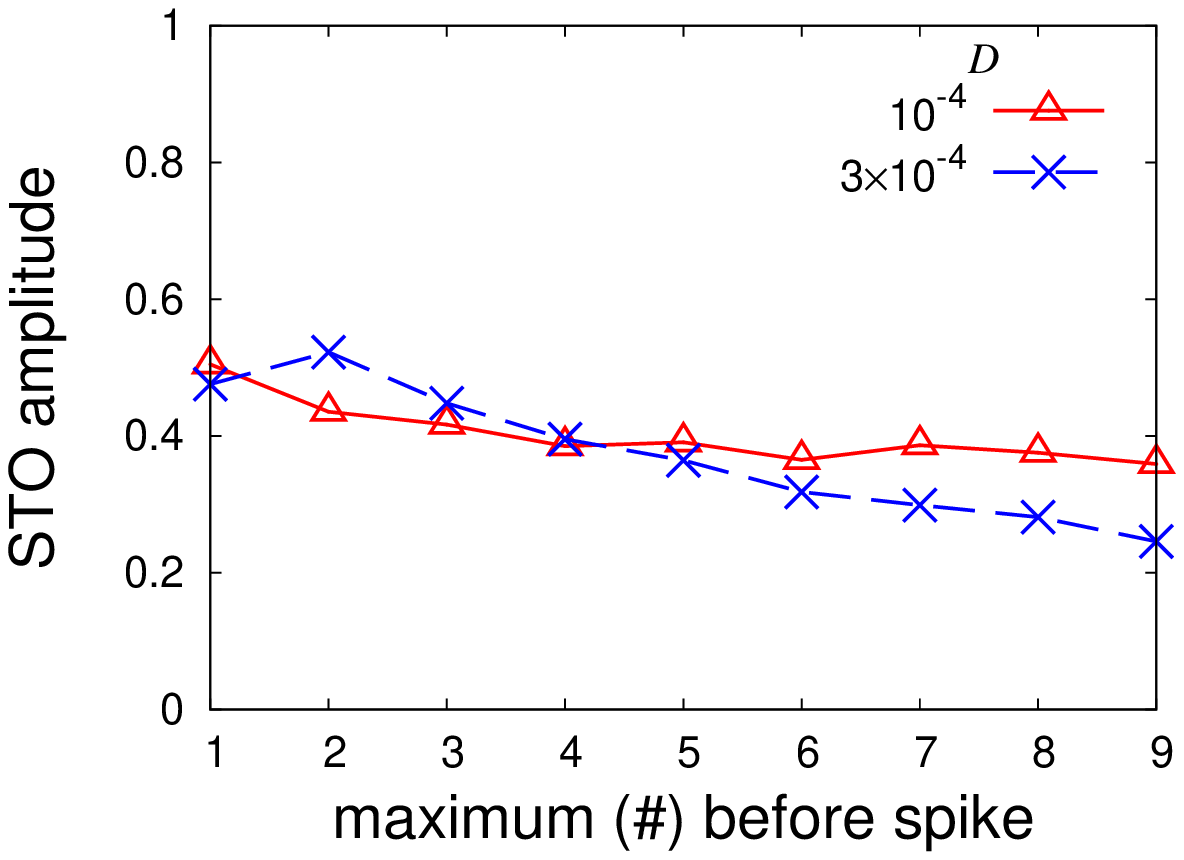}
\includegraphics[width=0.32\textwidth]{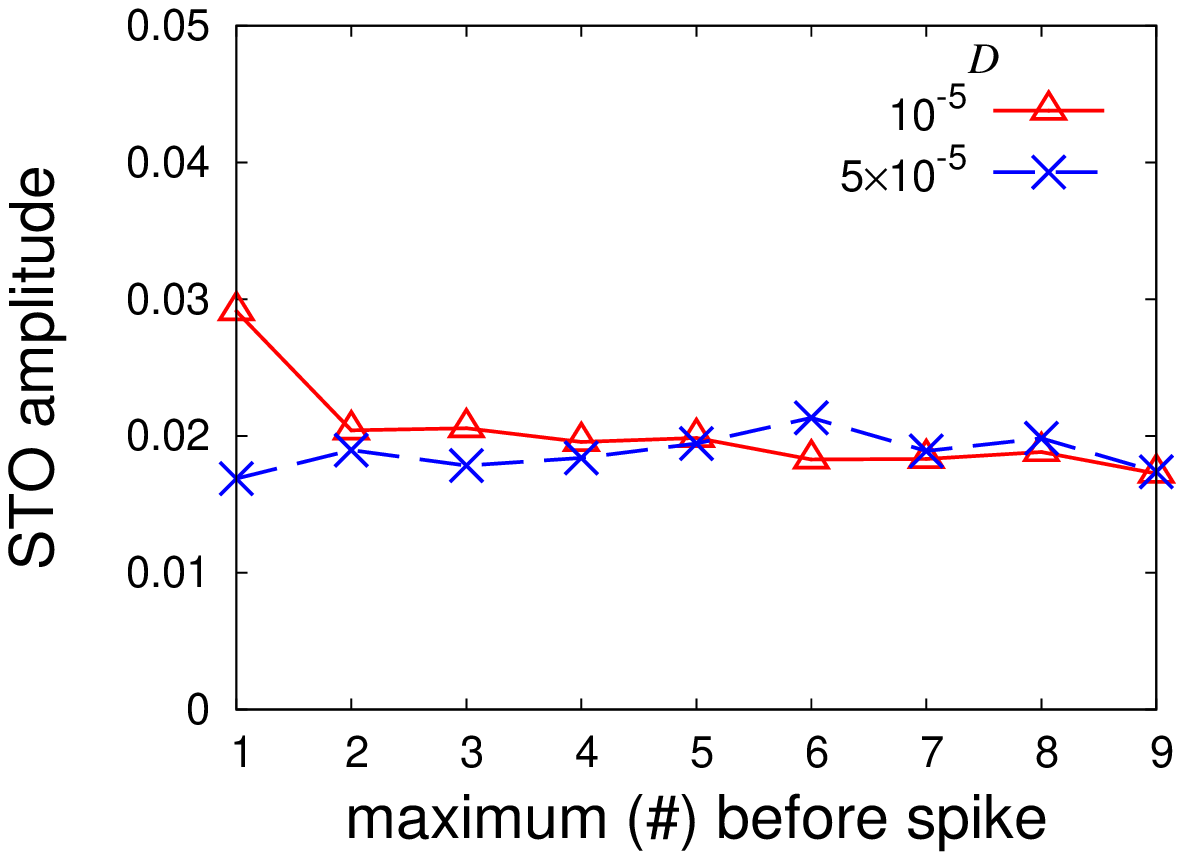}
\includegraphics[width=0.32\textwidth]{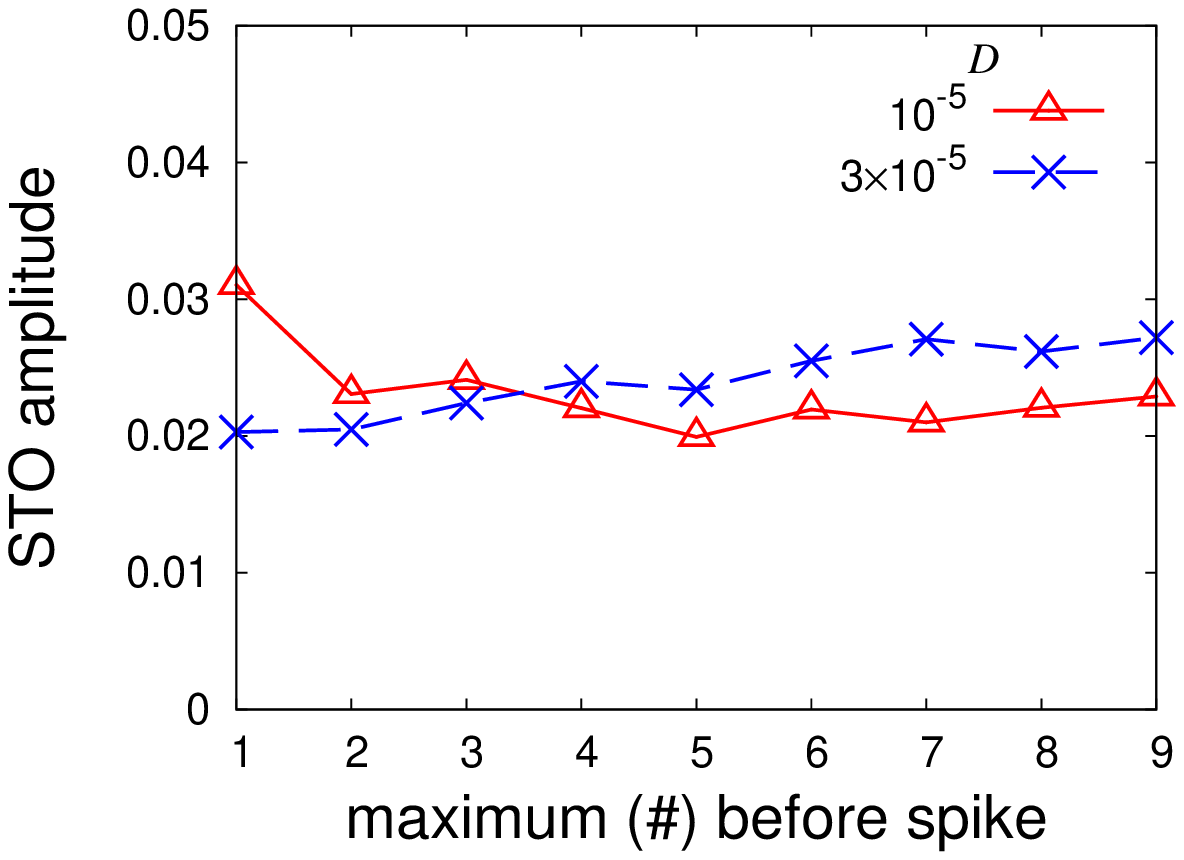}
\caption{Average amplitude before a spike in Class 3 for various values of $D$. {\bf Top:} 3DSC ($I_{\rm app}=-2.7$); {\bf middle:} NLMFN ($c_b=1.1$); {\bf bottom:} LMFN ($\epsilon_2=0.00055$, $b_{\rm rs}=0.305$).}
\label{fig:class3_amp_noise}
\end{figure}

\begin{figure}
\centering
\includegraphics[width=0.32\textwidth]{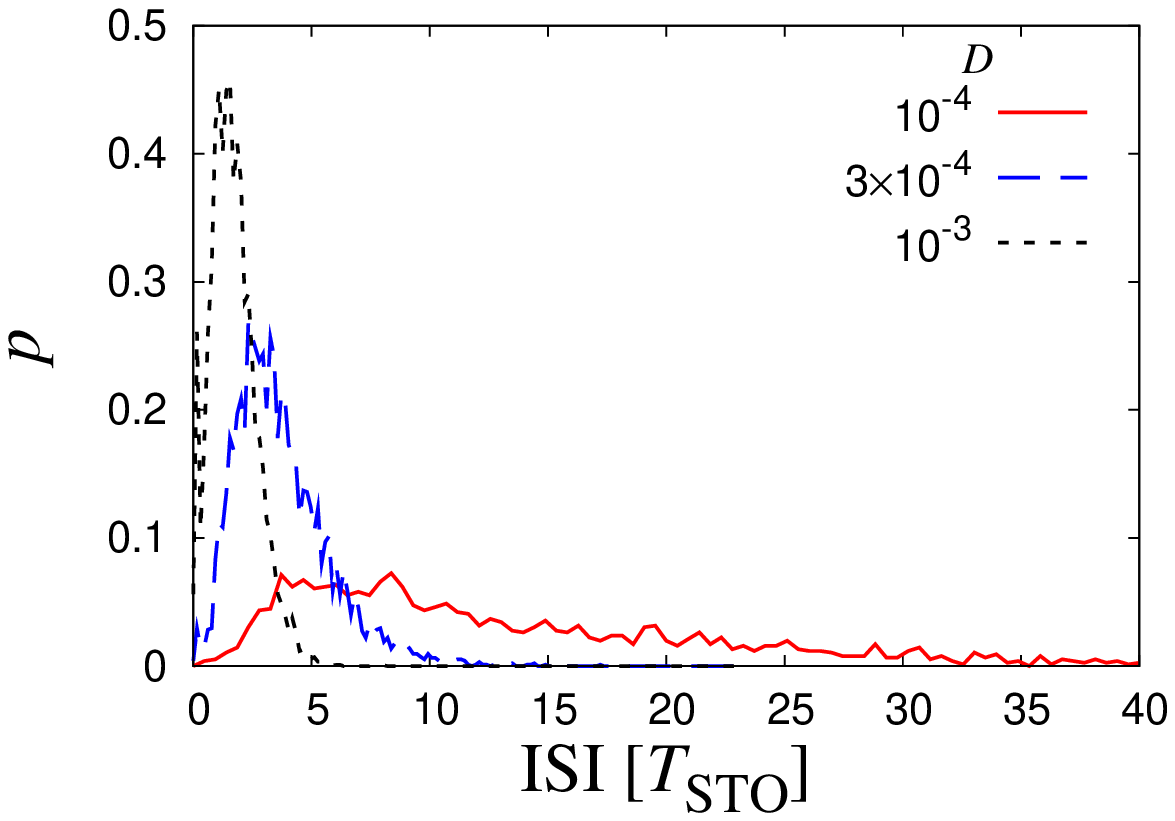}
\includegraphics[width=0.32\textwidth]{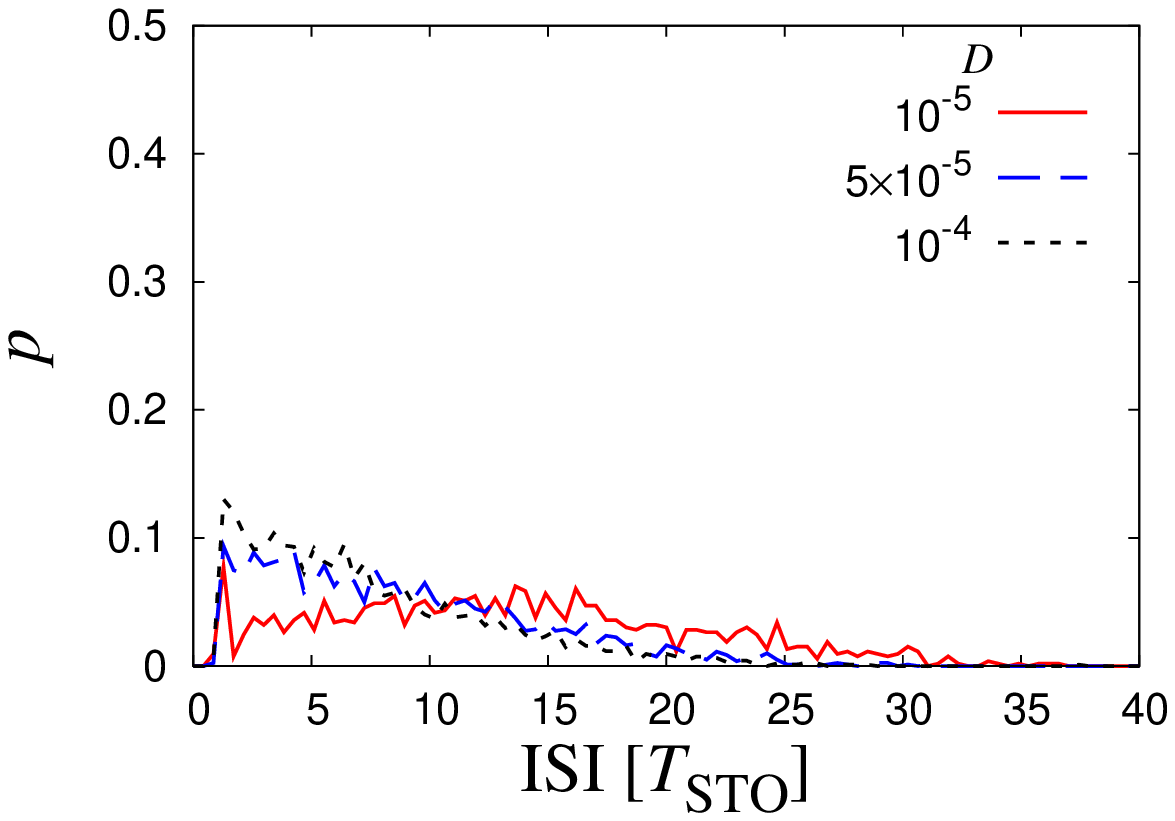}
\includegraphics[width=0.32\textwidth]{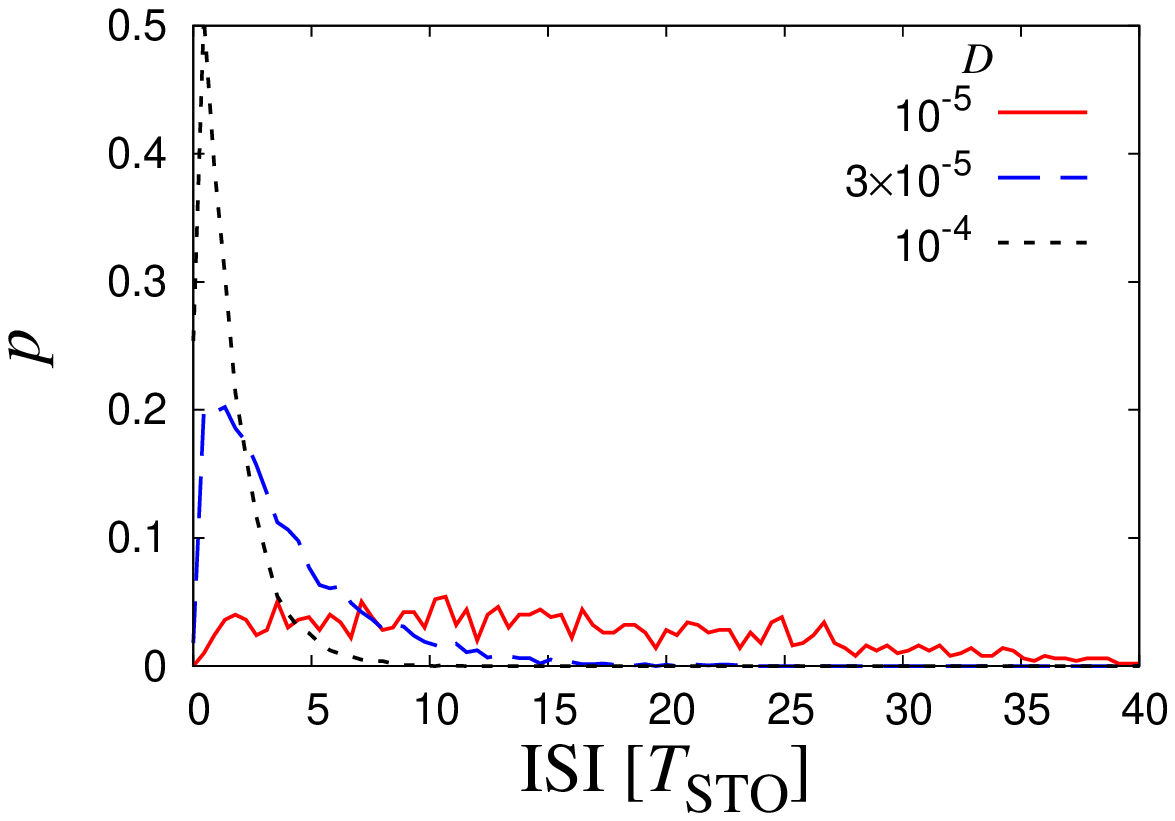}
\caption{Densities of ISI lengths in Class 3 for various values of the noise strength $D$. {\bf Top:} 3DSC; {\bf middle:} NLMFN; {\bf bottom:} LMFN. For parameters, see Fig.~\ref{fig:class3_amp_noise}.}
\label{fig:class3_ISI_noise}
\end{figure}

Figs.~\ref{fig:class3_amp_noise} and~\ref{fig:class3_ISI_noise} show the trend of the STO amplitude and the density of ISI when noise levels increase.

$\bullet$ 
For stronger noise the increase in amplitude directly before the spike in the FN-type models is eliminated, with stronger noise levels rather than deterministic dynamics dominating the transition to spikes. At higher noise levels, distinguishing stochastic fluctuations from regular STOs
is less dependable, so that the amplitude dynamics for the largest noise levels are not shown in Fig.~\ref{fig:class3_amp_noise}. In the 3DSC model, there is a slight increase in average amplitude during the ISI which is due to our algorithm (Subsec.~\ref{ssec:tools}) not fully eliminating the trend of increasing $V$ in steady state with increasing $r_s$ for $r_s<r_{s,\rm H}$. 

$\bullet$ 
The ISI densities for 3DSC and LMFN are more concentrated at shorter ISI durations, while the ISI density for NLMFN remains spread over a large range of ISIs. This is due to the underlying stable STOs for NLMFN, together with the nonlinear return mechanism that typically returns $b$ to values well below $b_{\rm H}$ following a spike, thus allowing for a range of ISI durations. With the ISI density concentrated at low values, clustered spikes are more frequent in 3DSC and LMFN with nearly tonic spiking at the higher noise level (cf. Ref.~\onlinecite{Kuske2009}).

$\bullet$ 
For Class 3, $\beta$ for the 3DSC model shows a similar behavior as for Class 2B (Fig.~\ref{fig:class2_beta}), with a clear maximum. The values of $\beta$ in Class 3 are lower than in Class 2 for 3DSC, mainly because of smaller amplitudes of the STOs (being inversely proportional to the distance from $r_{s,\rm H}$ for CR~\cite{Yu2006} -- also see Subsec.~\ref{ssec:tools}). Similarly, the trend for $\beta$ in Class 2B and 3 is the same in LMFN. For the NLMFN model, the value of $c_b$ is the same for Class 2B and Class 3 and the behavior of $\beta$ is shown in Fig.~\ref{fig:class2_beta}.

\section{Intra-model comparisons for stellate cells}
\label{sec:analysis_intra}

The comparisons in the previous section  point to the role of the underlying bifurcation structure on the characteristics of the STOs and MMOs.  We summarize the results for the 3DSC model from Sec.~\ref{sec:analysis_inter}, comparing within that model. We also discuss alternative values of $I_{\rm app}$ for the 7DSC model in order to compare with previous studies of the full model.

\subsection{Comparison for different ranges of $I_{\rm app}$}

Within the 3DSC model (and similarly the 7DSC model) differences in the stochastic behavior can be related to the differences in the deterministic behavior for different ranges of $I_{\rm app}$. For $I_{\rm app}< I_{\rm app,H} \approx -2.575$ in 3DSC (see Subsec.~\ref{ssec:scmodel}), the stable steady state has  a value of  $r_s =r_{s,0}$ well below $r_{s,\rm H}$. For  $I_{\rm app}> I_{\rm app,H}$ there is no steady state for the deterministic system, and the STOs are driven by SPHB with  $r_s$ as the slowly varying control parameter. In the case $I_{\rm app}\approx I_{\rm app,H}$, characteristics of both CR and SPHB are observed.

$\bullet$
Amplitude: For $I_{\rm app}$ well below $I_{\rm app,H}$, $r_{s,0}$ is well below $r_{s,\rm H}$, so only high noise levels can drive MMOs, and STOs are  purely of the CR-type without any trend in the average amplitude (as in Class 3). In contrast in Class 1, with $I_{\rm app}$ clearly above $I_{\rm app,H}$ the STOs due to the SPHB have an increasing trend in amplitude before the spike.  Well-defined periods of STOs survive only for lower noise levels, as the slow passage is sensitive to very weak noise. For values of $I_{\rm app}$ closer to $I_{\rm app,H}$, as in Class 2, both CR and SPHB can affect the dynamics, so that periods of STOs with increasing amplitude trend can survive over a large range of noise levels.


$\bullet$
ISI density: Significant differences are seen at lower noise levels, where longer tails in the ISI density correspond to STOs of the CR-type, not observed in STOs driven by SPHB. 

$\bullet$
Coherence measure $\beta$: CR-driven STOs in Class 2B and Class 3 are  responsible for the more pronounced peak in the coherence measure $\beta$ in Fig.~\ref{fig:class2_beta}. 

Intermediate values of  $I_{\rm app} = -2.5, -2.65$ were also analyzed (not shown), and as expected intermediate behavior between the three classes is observed. For $I_{\rm app} = -2.5$, STO and ISI behavior was closer to that of Class 2B than Class 1, without contributions to longer ISIs, and for intermediate to larger noise levels very weak coherence is observed with a mild peak in $\beta$ around $D\approx 10^{-5}$.

\subsection{Families of MMOs for $I_{\rm app}>I_{\rm app,H}$ at weak noise}
\label{ssec:intra_sc_weak_noise}

Here we consider the effect of noise on MMO families in the full 7DSC model in the range $I_{\rm app}>I_{\rm app,H}^{\rm 7D}$ (i.e., within the regime of deterministic MMOs), as illustrated through the ISI density. The 3DSC model shows a similar behavior, with a shift in the values of $I_{\rm app}$. Recall that for any fixed $I_{\rm app}>I_{\rm app,H}^{\rm 7D}\approx -2.702$, the underlying deterministic dynamics is a slow passage of $r_s$ through $r_{s,\rm H}$. We restrict our study to very low noise levels and consider trajectories that fit into Classes 1 and 2A (Sec.~\ref{sec:analysis_inter}) when stronger noise is added. For the sake of comparison with the analysis in Ref.~\onlinecite{Wechselberger2009}, we use an alternative equation for the dynamics of $r_s$. Instead of Eq.~\ref{eq:sc_rs}, we use Eq.~\ref{eq:sc_rs_58} given in App.~\ref{app:sc_eq}, which alters the numerical values of the model but not the qualitative features. 

\begin{figure}
\centering
\includegraphics[width=.32\textwidth]{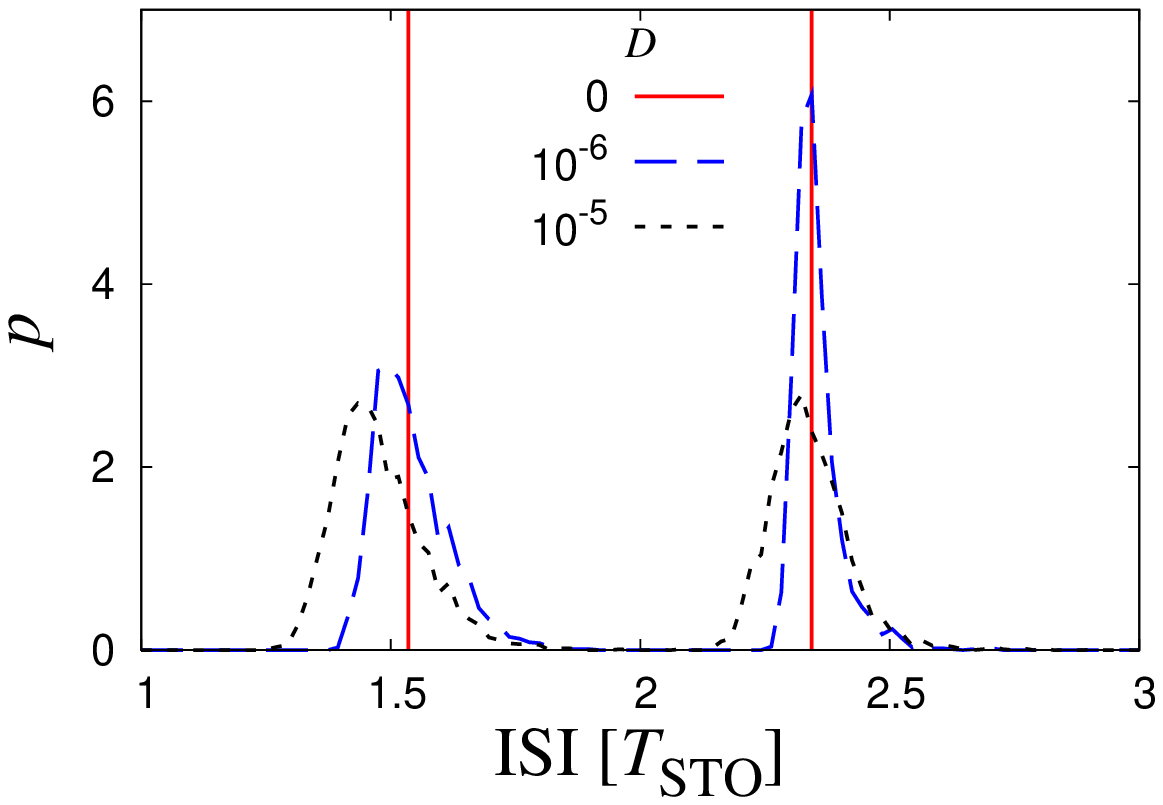}
\includegraphics[width=.32\textwidth]{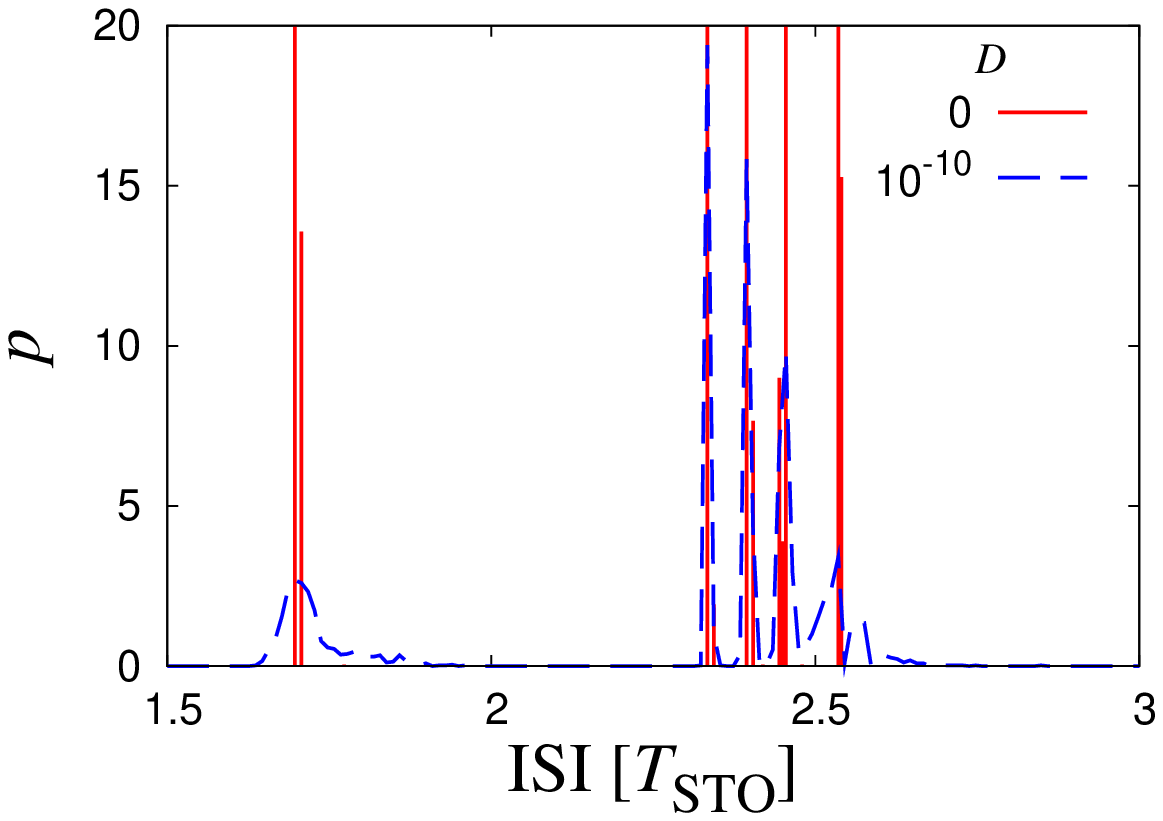}
\caption{ISI densities from time series generated with the 7D version of the SC model (using the alternative equation (Eq.~\ref{eq:sc_rs_58})) with various values of $I_{\rm app}$ ($-2.435$ in the top panel corresponding to family $J_0(J_1)^1$; $-2.4416$ in the bottom panel corresponding to family $J_0(J_1)^4$). The deterministic case ($D = 0$) is shown as a histogram (solid red lines) whereas the data for simulations with noise ($D\ne 0$) is shown as line plots (dashed blue/dotted black lines). We adopt the notation for the MMOs from Ref.~\onlinecite{Wechselberger2009}. $T_{\rm STO}\approx 99$.}
\label{fig:sc_7D_ISI}
\end{figure}

Fig.~\ref{fig:sc_7D_ISI} shows examples of the ISI density for different families of MMOs. The $1^s$ families (for notation, see Ref.~\onlinecite{Wechselberger2009}) have MMOs with $s$ STOs in each ISI, yielding ISI densities with a single point mass. The MMOs for a second, more complex family, $J_i(J_j)^l$, have an ISI with $i$ STOs followed by $l$ ISIs each with a different type of $j$ STOs that increase in amplitude and period for each of the subsequent $l$ ISIs. The  $J_i(J_j)^l$ families then  have $1+l$ point masses in the ISI density. We consider only those MMO families that are stable in the deterministic model. In the range $-2.55<I_{\rm app}< -2.4$ stable $1^s$ and $J_i(J_j)^l$ families are found for alternating ranges of $I_{\rm app}$. For $I_{\rm app}\lesssim -2.55$, the stable solutions are only $1^s$ families densely filling the space in $I_{\rm app}$~\cite{Wechselberger2009}. The time series analyzed earlier as Class 1 in this notation would be a mixture of $1^2$ and $1^3$.

\begin{figure}
\centering
\includegraphics[width=.32\textwidth]{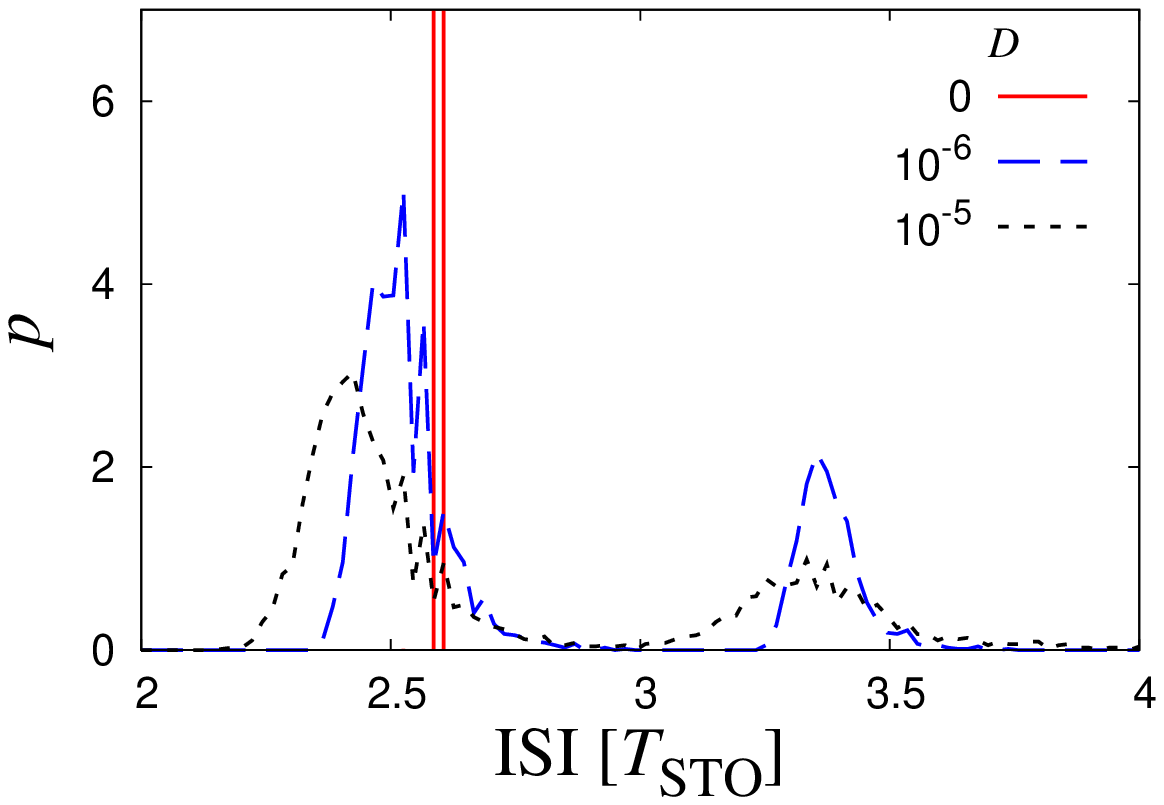}
\includegraphics[width=.32\textwidth]{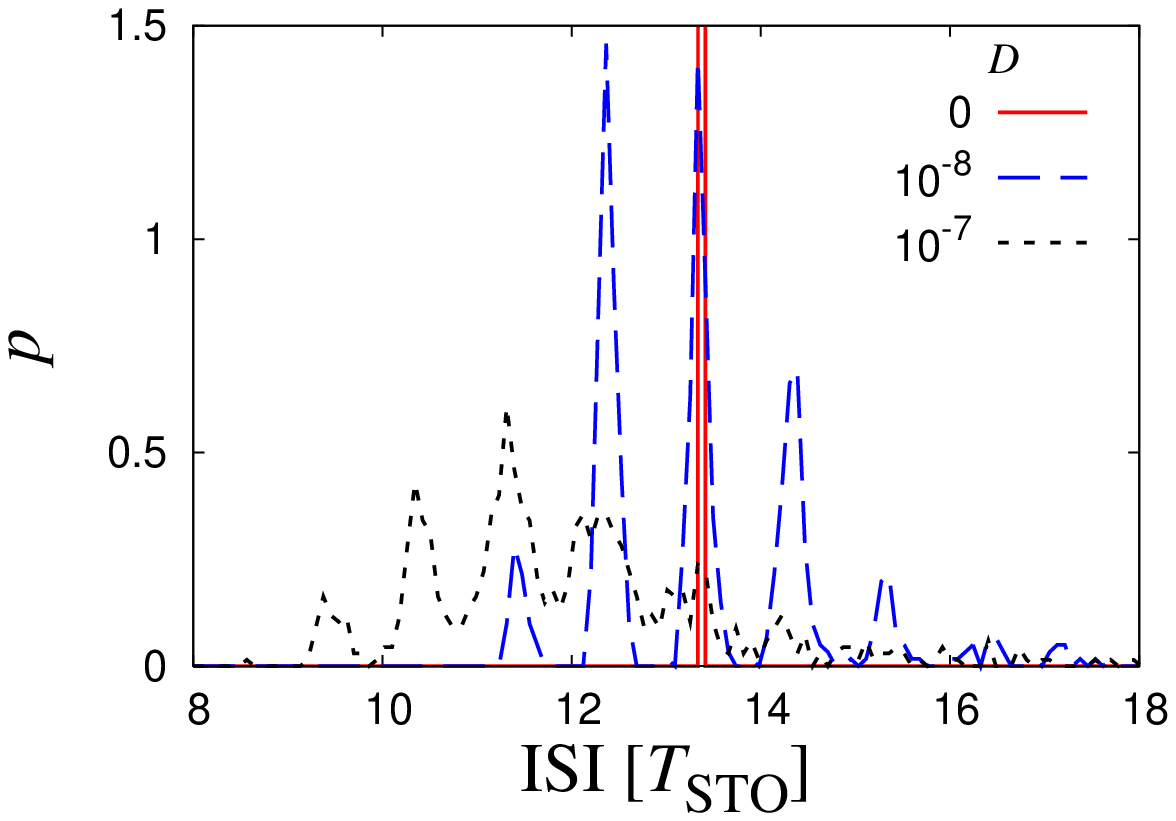}
\caption{As of Fig.~\ref{fig:sc_7D_ISI} with $I_{\rm app}=-2.5$ (deterministic family $1^1$ -- {\bf top}) and $I_{\rm app}=-2.66$ (deterministic family $1^{12}$ -- {\bf bottom}).}
\label{fig:sc_7D_ISI_2}
\end{figure}

For values of $I_{\rm app}$ near the center of the stability range for a particular type of MMO, well separated from other families, weaker noise  broadens  the ISI density and stronger noise shifts the density towards shorter ISIs when noise drives an early escape typical in a slow passage through a HB, as shown in Fig.~\ref{fig:sc_7D_ISI} for $J_0(J_1)^1$. The bottom panel of Fig.~\ref{fig:sc_7D_ISI} shows that for values of $I_{\rm app}$ near the edge of a stability range, very weak noise can lead to a mixture of close families of MMOs.  As noise drives a faster escape to spiking the peak corresponding to the longest ISI and largest STO of $J_0(J_1)^4$ is reduced as the nearby MMOs of $J_0(J_1)^3$ occur, appearing as stronger nearby peaks in the ISI density. There is also considerable sensitivity in the ISI density when stability ranges are small and densely packed. Noise can drive a sampling of a mixture of families close in terms of the bifurcation parameter, yielding clearly defined peaks that are not present in the deterministic case, as shown in Fig.~\ref{fig:sc_7D_ISI_2}. In the top panel,  the deterministic MMO is  $1^1$, and noise excites a second peak for a family $1^2$ or a mixture of families between $1^1$ and $1^2$.  At $I_{\rm app}=-2.66$ the stable solution is $1^{12}$ and weak noise drives a sampling of the six nearby MMO solutions $1^{s}$ with $10\le s \le 15$.  At stronger noise the longer ISIs are lost and the ISI density spreads out and shifts towards shorter ISIs.

In conclusion, even very weak noise can alter the dynamics of a MMO-generating system significantly, particularly if the deterministic solutions are close  in parameter space. Weak noise drives a sampling between a few or many of these solutions. Then it can be difficult to distinguish between complex deterministic solutions like the $J_i(J_j)^l$ families and a noisy trajectory sampling a few different MMO families.



\section{Intra-model comparisons for the modified FN model}
\label{sec:intra_FHN}

As in the SC model, the underlying bifurcation structure and deterministic behavior influence the source and characteristics of the STOs. For the FN-type models, the underlying bifurcation is a supercritical Hopf, with stable STOs in the 2D reduced system for $u$-$v$ with constant $b_{\rm H}<b<b_c$.

\subsection{Linearly augmented modified FN model}
\label{ssec:intra_LMFHN}

In the LMFN model with appropriate reset the slowly varying control variable sweeps through an underlying HB and canard transition, which always yields MMOs in the deterministic setting. We consider  two aspects of the control parameter that can be varied to change the characteristics of the STOs, the rate of change in $b$, $\dot{b}=\epsilon_2$, and its reset $b_{\rm rs}$, following a spike/threshold crossing.

In the different classes studied in Sec.~\ref{sec:analysis_inter}, we analyzed four different combinations of $\epsilon_2$ and $b_{\rm rs}$: slower variation in $b$ given by smaller $\epsilon_2=0.00055$, with reset $b_{\rm rs}<b_{\rm H}$ (Class 3) or $b_{\rm rs}\approx b_{\rm H}$ (Class 2B), and larger values of $\epsilon_2= 0.001$ (Class 2A) and $\epsilon_2=0.0147$ (Class 1) both with $b_{\rm rs}\approx b_{\rm H}$. 

\subsubsection{Characteristics for $\epsilon_2$, $b_{\rm rs}$ in Section~\ref{sec:analysis_inter}}

The main underlying feature that influences  amplitude dynamics for these different parameter combinations, is time spent in the parameter range $b>b_{\rm H}$.  
$\bullet$
Amplitude increase is more pronounced for low noise levels and smaller values of $\epsilon_2$. For low noise levels,   the system  follows the underlying STO dynamics, less likely to be driven to spiking by noise.   For smaller values of $\epsilon_2$, the control parameter behaves almost as a constant relative to the oscillation frequency. This allows attraction to larger STOs for slowly increasing $b_{\rm H}<b<b_c$, with the possibility of reaching values of $b>b_c$ before spiking, so that larger amplitudes are typical before escape. For larger values of $\epsilon_2$, as in Class 1 or Class 2A, $b$ increases through the STO region less slowly. Then the trend of increasing amplitude consistent for $b_{\rm H}<b<b_c$  survives but increased variation in $b$ limits the attraction to the stable STOs as $b$ approaches and exceed $b_c$.  The behavior is closer to a series of transients that is more susceptible to noise-induced escape to spiking. For larger noise levels, the escape to spiking is noise-driven in or near the stable STO region. An increase in STO amplitude farther from the spike, for $b\approx b_{\rm H}$ is observed due to the role of CR together with the stable STOs. An earlier escape limits the opportunity for amplitude increase closer to the spike.

$\bullet$ 
The relation between $\epsilon_2$ and $b_{\rm rs}$ and ISI density is as expected: smaller values of $\epsilon_2$ or $b_{\rm rs}$ lead to greater likelihood of longer ISIs.

$\bullet$ 
The coherence measure $\beta$ shown in Fig.~\ref{fig:class2_beta} increases for long stretches of large amplitude oscillations, and decreases with fluctuating amplitude. Trajectories with the reset close to the HB $b_{\rm rs}\approx b_{\rm H}$ show the largest value of $\beta$ for moderate noise levels. Slower variation in $b$ in these cases also reduces fluctuation in amplitude, thus increasing coherence. For reset $b_{\rm rs}<b_{\rm H}$ there is fluctuation in amplitude due to oscillations driven by both CR and STOs, so $\beta$ is reduced somewhat overall. For larger values of noise  coherence is destroyed in all cases.

\subsubsection{Characteristics for slower increase in $b$ with $b_{\rm rs}<b_{\rm H}$}

\begin{figure}
\centering
\includegraphics[width=0.32\textwidth]{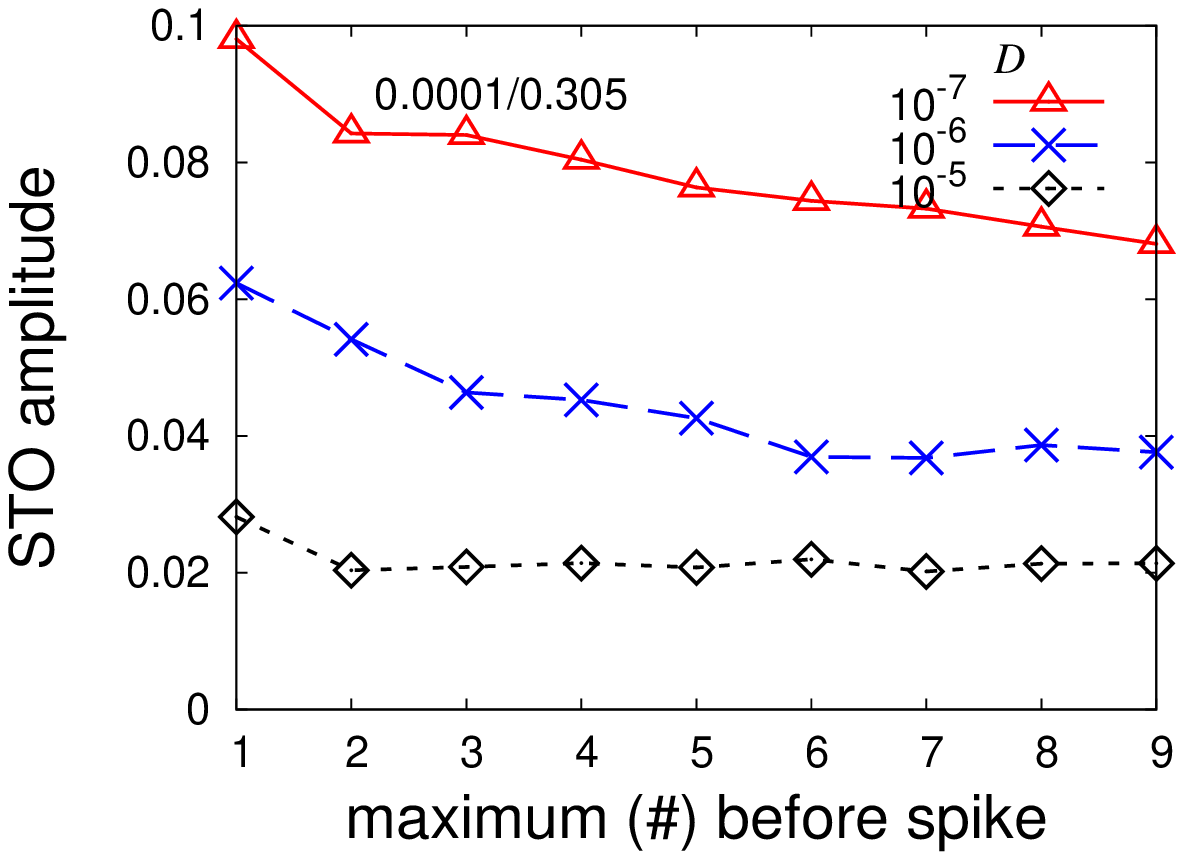}
\includegraphics[width=0.32\textwidth]{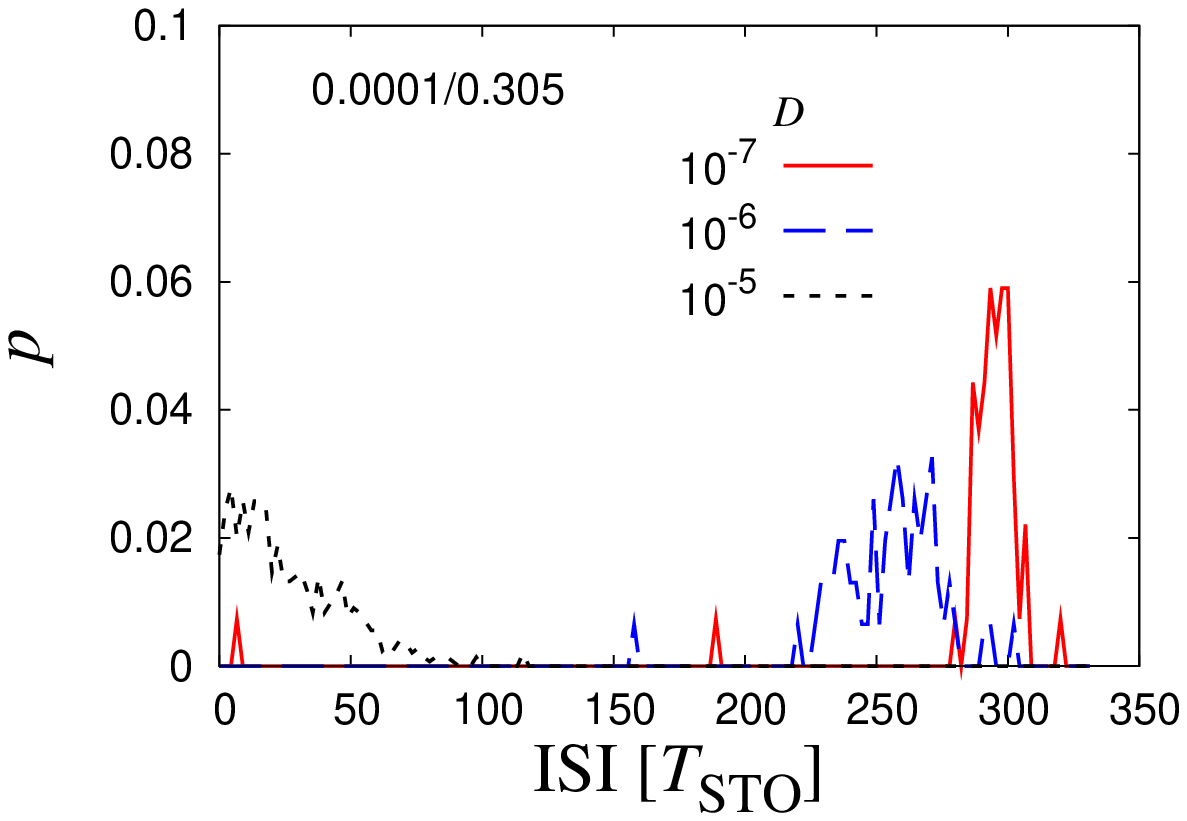}
\includegraphics[width=0.32\textwidth]{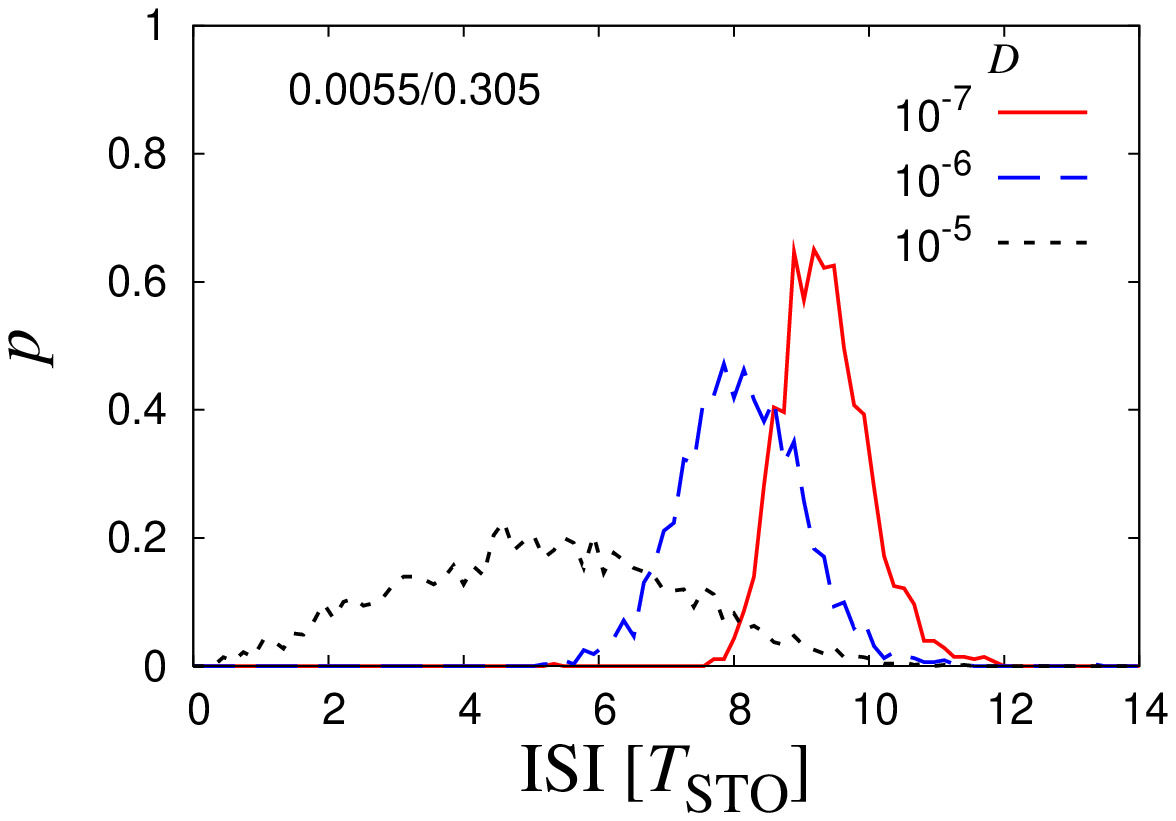}
\caption{Average amplitude ({\bf top}) and ISI density before a spike ({\bf middle}) for the LMFN model for a very slow sweep through the HB ($\epsilon_2=0.0001$) and reset before the HB ($b_{\rm rs}=0.305$). ISI density ({\bf bottom}) for an intermediate case ($\epsilon_2=0.0055$, $b_{\rm rs}=0.305$).}
\label{fig:intra_LMFHN}
\end{figure}

Fig.~\ref{fig:intra_LMFHN} shows the analysis of an additional parameter combination, a very slow sweep through the HB and reset before the HB ($\epsilon_2=0.0001$, $b_{\rm rs}=0.305$), whose coherence measure $\beta$ was included in Fig.~\ref{fig:class2_beta}.

$\bullet$ 
The difference in amplitude dynamics for lower and higher noise levels is as described above.  For lower noise levels and very slow dynamics of $b$, the system is attracted to larger amplitude STOs before spiking, so that larger amplitudes are typical for more  STOs  before escape.

$\bullet$ 
The ISIs are very long, and only for stronger noise  are they comparable to Class 3 (cf. Fig.~\ref{fig:class3_ISI}). In contrast to Class 2, only for stronger noise levels is the ISI density concentrated at shorter ISIs, since the trajectory takes longer to reach $b_{\rm H}$.

$\bullet$ 
The shape of the curve $\beta$ vs. $D$  is qualitatively similar to the classes studied in Sec.~\ref{sec:analysis_inter}. However, at very low noise values, $\beta$ is larger than other cases, caused by CR-driven STOs and slower variation in $b$ as described above. 

The last panel in Fig.~\ref{fig:intra_LMFHN} shows the ISIs for an intermediate case, with larger $\epsilon_2$ and $b_{\rm rs}< b_{\rm H}$. The ISI density then shows characteristics between Class 2 and Class 3, with a narrow ISI density for smaller noise levels, and longer tails for stronger noise, as oscillations of the CR-type appear for $b<b_{\rm H}$.

We add one additional note about resetting off of the nullcline. The choice of reset used in Sec.~\ref{sec:analysis_inter} (reset  very close to the nullcline), can be relaxed to a reset near the nullcline without qualitative changes in the measures used above. A choice of reset away from the nullcline modifies the amplitude dynamics of the STO at the beginning of the ISI.  For reset with $b_{\rm rs}$ well below $b_{\rm H}$, STOs with considerable amplitude follow the reset and  relax as $b$ approaches $b_{\rm H}$, similar to NLMFN. For Classes 1 and 2 this increased amplitude supports an earlier escape to spiking and reduced ISIs, particularly in Class 1. For low noise levels in Class 2 the ISIs are then narrower, since the larger STOs are less susceptible to small noise. For Class 3 the stronger noise dominates, so that variation in the reset has limited effect.

As noted in Subsec.~\ref{ssec:fhn}, the model of Eqs.~\ref{eq:Makarov_u} and~\ref{eq:Makarov_v} has a larger distance between $b_{\rm H}$ and $b_c$, as compared with vdP, yielding a larger region in parameter space  where well-defined periods of STOs are observed in the presence of noise. This behavior is  confirmed by comparisons of PSDs for MFN and for a rescaled version of vdP to effectively increase the canard value. This comparison illustrates that the distance between $b_{\rm H}$ and $b_c$ plays a critical role in the robustness of STOs and MMOs, in addition to the reset and ramp speed.

\subsection{Nonlinearly augmented modified FN model}

As described in Subsec.~\ref{ssec:fhn}, the long time behavior of the NLMFN deterministic model  includes different states depending on $c_b$: $c_b\gtrsim 1.53$: steady state; $1.04\lesssim c_b\lesssim 1.53$: stable STOs; $c_b\lesssim 1.04$: MMOs. In  Sec.~\ref{sec:analysis_inter} we have  considered only those values of $c_b$ that support phenomenological behavior observed in the SC model. Here we also discuss cases that differ qualitatively from those studied in Sec.~\ref{sec:analysis_inter}, and compare the amplitude, ISI density, and coherence measure within this model.

\subsubsection{Values of $c_b\approx 1$}

Values of $c_b =0.95$ and $c_b= 1.1$ were used in Sec.~\ref{sec:analysis_inter}  to approximate Class 2 behavior. For $c_b = 0.95$ the underlying deterministic dynamics is MMOs, with $b$ regularly  crossing $b_c$.  Then the  observed measures are consistent with a slow passage through a Hopf bifurcation and canard point as observed in the other models:
 
$\bullet$ 
Increasing amplitude before the spike over a range of noise levels.

$\bullet$
Multi-peaked ISI density for lower noise levels and concentrated ISI density for larger noise.

$\bullet$
Decreasing trend in $\beta$, defined only for larger values of the noise.

For values of $c_b<0.95$ the return value of $b$ following a spike takes values between $b_{\rm H}$ and $b_c$, thus shortening the ISI significantly.  As discussed above, we do not classify this behavior as Class 1, since the return value yields larger amplitude STOs and limited increase for low noise levels, similar to those observed for a related model in Subsec.~\ref{ssec:self-coupled_MFHN}.

In contrast, for $c_b= 1.1$, the deterministic system exhibits stable STOs where $b$ oscillates near $b_c$. The stochastic behavior is similar to that of the case with $c_b=0.95$, but differences due to the underlying deterministic dynamics are observed for lower noise levels and $c_b=1.1$:

$\bullet$ 
Increased noise drives an earlier escape to spiking. This results in a reset of $b<b_{\rm H}$, yielding both CR-type and SPHB-driven oscillations, and a limited opportunity for amplitude increase just before the spike.

$\bullet$ 
The ISI density is broader for $c_b=1.1$ (compared to $c_b=0.95$).

\subsubsection{Intermediate values:  $c_b= 1.3$}

The attracting deterministic behavior for $b$ consists of small oscillations midway between $b_{\rm H}$ and $b_c$, so characteristics include some elements from Class 2 for smaller noise, and similarities to Class 3 and $c_b= 1.5$ (see below) for larger noise:

$\bullet$
The amplitude dynamics is similar to Class 2 for lower noise, with a weaker increase.

$\bullet$
The ISI behavior has characteristics closer to Class 2.

$\bullet$
The coherence measure $\beta$ vs. $D$ behaves similarly to $c_b= 1.5$, with larger values of $\beta$ for small noise.

\subsubsection{Large values of $c_b= 1.5,1.6$}

\begin{figure}
\centering
\includegraphics[width=0.32\textwidth]{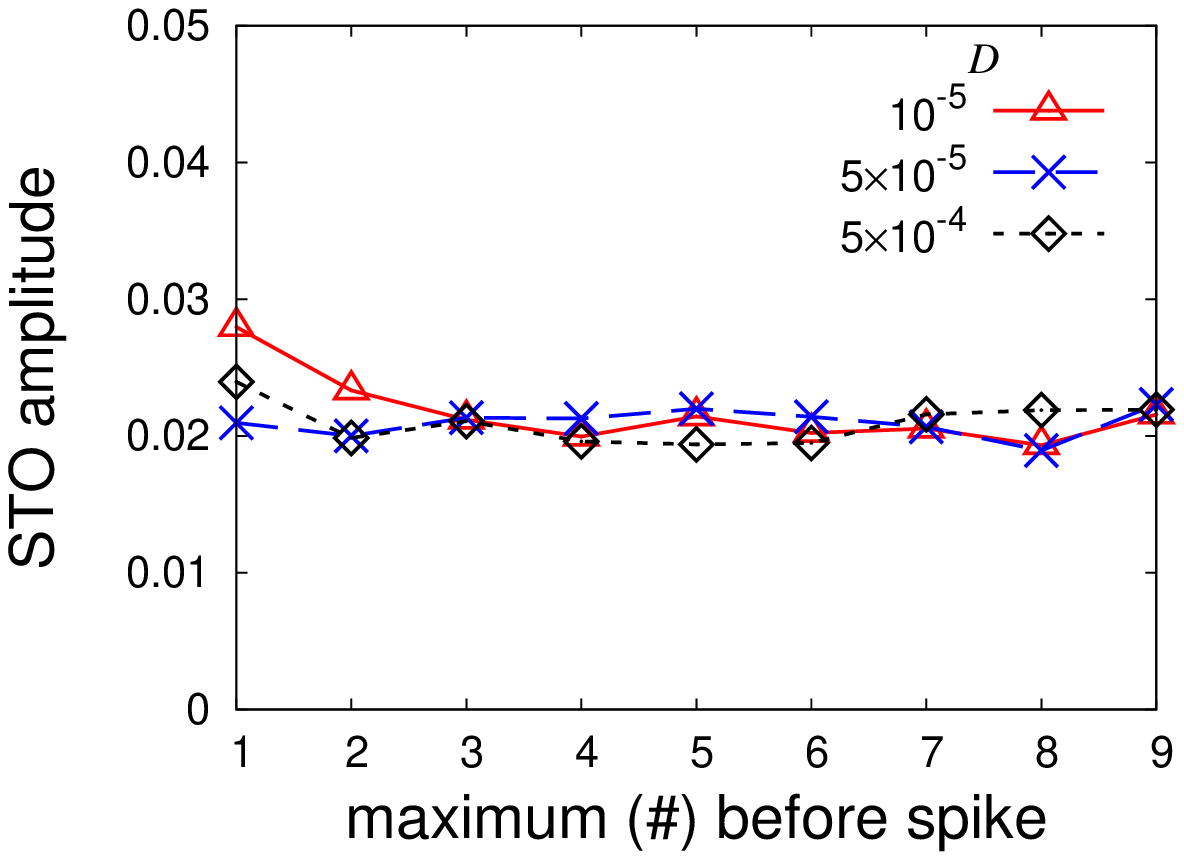}
\includegraphics[width=0.32\textwidth]{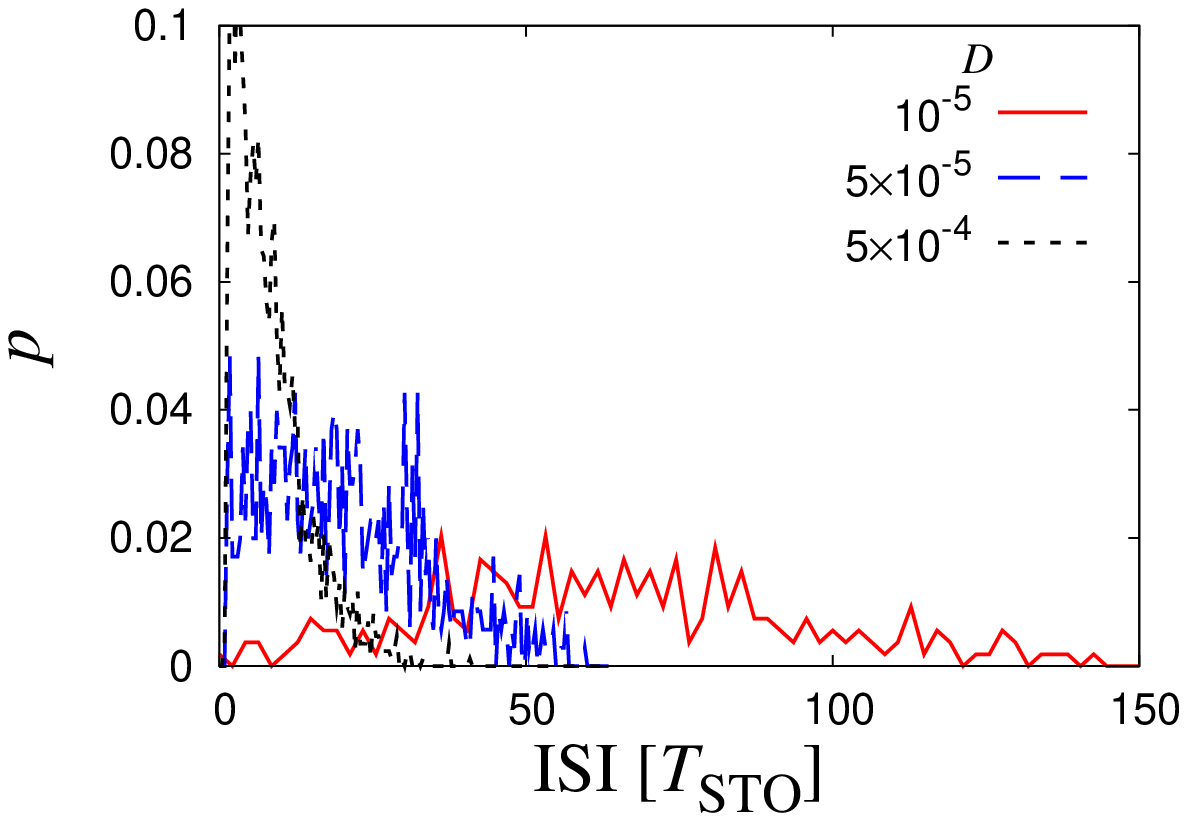}
\caption{Average amplitude before a spike ({\bf top}) and ISI density ({\bf bottom}) for the NLMFN model with $c_b=1.6$ at which the deterministic system is quiescent.}
\label{fig:intra_NLMFHN}
\end{figure}

For larger values of $c_b$, the attracting state of the deterministic system takes values of $b$ near $b_{\rm H}$. For $c_b= 1.6$ the long term deterministic behavior is quiescent with the steady state corresponding to $b$ just below $b_{\rm H}$. For $c_b= 1.5$ the deterministic system has stable STOs, with  small oscillations in $b$ just above $b_{\rm H}$. To drive MMOs in either of these cases, noise must be strong enough to cause excursions beyond the canard transition at $b_c$, certainly stronger than in the cases for $c_b\approx 1$ above. ISI and amplitude measures can therefore only be derived for these stronger values of noise, leading only to Class 3 MMOs with the following characteristics (Fig.~\ref{fig:intra_NLMFHN}):

$\bullet$
Primarily constant trend in average amplitude before a spike.

$\bullet$
Broad ISI density with long tails.

$\bullet$
A clear difference between $c_b = 1.5$ and  $1.6$ due to the underlying deterministic dynamics is seen in the coherence measure $\beta$ (Fig.~\ref{fig:class2_beta}).  For $c_b=1.5$ all  noise levels  disrupt the coherence of the STOs produced deterministically,  so that $\beta$ strictly decreases with noise.  For $c_b=1.6$, $\beta$ has a maximum indicating an  optimal noise level typical of CR-driven oscillations~\cite{Pikovsky1997,Yu2006}.

\section{Application to other models}
\label{sec:other_models}

\subsection{The 3DSC model with noise in $I_{\rm app}$}

The stochastic nature  of the SC system results from noise in ion channels and  noisy external signals. The model we consider above follows Ref.~\onlinecite{White1998}, including the primary noise source  in the gating variable of the persistent sodium channel. Here, we compare our results with those obtained with a different noise source in the 3DSC model, namely, in the applied current $I_{\rm app}$. The consideration of this additive noise is motivated by several factors. First, systems with multiple time scales often can be divided into regions of time or space where noise plays a more or less significant role.  This raises questions about where or when noise plays a critical role, and whether the type of noise can make a significant difference. Second, since $I_{\rm app}$ is the only variable in the system that can be easily experimentally controlled, this raises the question of whether varying $I_{\rm app}$ can be used as a controllable input to probe the dynamics. Providing noisy input is a technique used in experimental neuroscience (e.g., Ref.~\onlinecite{mainen1995}) to investigate the structure of the complex dynamics.  Finally, understanding whether different types of noise sources have similar or different effects contributes both to the  understanding of the mechanism for the dynamics and to model identification.

To test whether fluctuations in $I_{\rm app}$ can reproduce  dynamics similar to those generated by noise in the persistent sodium current, we consider the voltage equation in the interval near the end of the ISI, when the dynamics of $V$ is slow. This is the stage at which STOs are generated for parameters near the Hopf bifurcation of the reduced system. In that part of the cycle, $V$ can be approximated by a constant $V_0$ in Eq.~\ref{eq:sc_V}. This approximation suggests that even though the noise is parametric, at the end of the ISI when the STOs are generated, the noise behaves as additive noise to leading order with a coefficient $D'\approx (0.15 G_p(V_0-E_{\rm Na}))^2D$ and Eq.~\ref{eq:sc_V} can be approximated by
\begin{align}
\dot{V} &  \sim \frac{1}{C}\Big[ I_{\rm{app}} - G_L(V_0-E_L) - G_p \left( \frac{1}{1+\exp\left(-\frac{V_0+38}{6.5}\right)} \right)  \nonumber \\
& \quad \times (V_0-E_{\rm Na}) - G_h(0.65 r_f + 0.35 r_s)(V_0-E_h)\Big] \nonumber \\
& \quad +  \frac{1}{C}\sqrt{2D'}\eta(t).
\label{eq:sc_V_noiseI}
\end{align}
Eq.~\ref{eq:sc_V_noiseI} can be interpreted as having a noisy applied current with $I_{\rm app}+\sqrt{2D'}\eta(t)$ and $D'\approx 68D$ (for $V_0\approx -55$) instead of the original noisy gating variable. Then one would expect that the effects of different noise levels on the STOs as described in Sec.~\ref{sec:analysis_inter} could  be reproduced by varying the noise level in $I_{\rm app}$, as long as the results are scaled in a manner consistent with Eq.~\ref{eq:sc_V_noiseI}. Indeed for the ranges of noise levels considered in this paper, the results for stochastic $I_{\rm app}$  are essentially the same as described in the classification of Sec.~\ref{sec:analysis_inter}. With the above scaling, the plots for the coherence measure $\beta$ (as in the top panel of Fig.~\ref{fig:class2_beta}) overlap almost perfectly. We also reproduced similar behaviors of the ISI density and amplitude dynamics as analyzed for the 3DSC model in Sec.~\ref{sec:analysis_inter} (data not shown).

\subsection{A self-coupled FN-type model}
\label{ssec:self-coupled_MFHN}

Here, we analyze another MMO-generating model, a self-coupled modified FN model derived and used for the study of coupled Hodgkin-Huxley neurons~\cite{Wechselberger2005,Desroches2008}. Similar to the MFN model presented earlier, $v$ represents a voltage variable, $h$ is gating variable, and $s$ is a dynamic coupling between neurons. Compared to Refs.~\onlinecite{Desroches2008} and~\onlinecite{Wechselberger2005}, we add a noise term in the first differential equation:
\begin{align}
\dot{v} = & -0.5(v^3-v+1)+h-\gamma s v + \sqrt{2D} \xi(t) \label{eq:De_v}\\
\dot{h} = & \epsilon_{D}(2h+2.6v) \\
\dot{s} = & \beta H(v)(1-s)-\epsilon_{D}\delta s .
\label{eq:De_s}
\end{align}
$H(v)$ is the Heaviside function. This system of differential equations has multiple time scales similar to the models analyzed so far: the dynamic variable $s$ plays the role of a slowly varying control parameter, decreasing and passing through a HB at $s_{\rm H}$ for the underlying $v$-$h$ system. Also, the spikes are included in the model (not just a threshold and reset) but this is done using a sharp switch ($H(v)$) rather than a smooth nonlinear function for the control variable as in our NLMFN. The parameters used in Ref.~\onlinecite{Desroches2008} are as follows: coupling strength $\gamma =0.5$, activation rate $\beta=0.035$, slow time scale of the relaxation oscillations $\epsilon_{D} = 0.015$, and $\delta = 0.565$ regulating the speed of the dynamic coupling. Ref.~\onlinecite{Desroches2008} analyzes the deterministic ($D=0$) dynamics of the system in detail. By varying the parameters in Eqs.~\ref{eq:De_v}--~\ref{eq:De_s}, in particular $\delta$ and $D$, we find MMOs according to the classification scheme of Sec.~\ref{sec:analysis_inter}. We choose $\epsilon_D = 0.01$ and $\gamma = 0.9$ to obtain a sharper canard transition, similar to the MFN models analyzed earlier.  There are a number of differences between Eqs.~\ref{eq:De_v}--\ref{eq:De_s} and the MFN model. The slow variation of $s$ is nonlinear, slowing down significantly in the SPHB, a behavior somewhat similar to $b$ in the  NLMFN case with  $c_b=1.5$. Another noticeable difference between Eqs.~\ref{eq:De_v}--\ref{eq:De_s} and the MFN models is that the dynamics of $s$ yields a return value well below the Hopf point $s_{\rm H}$ following a spike.  After the return there is a significant interval during which $s$ approaches $s_{\rm H}$, which has implications for both the amplitude and ISI behavior.

\begin{figure}
\centering
\includegraphics[width=0.32\textwidth]{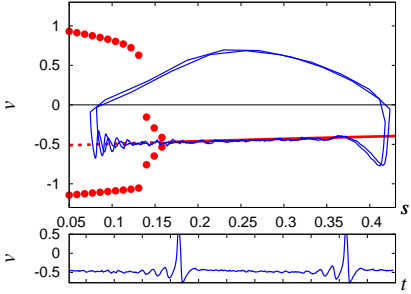}
\caption{The underlying (fixed $s$) bifurcation diagram of the self-coupled FN-type model with $\epsilon_D=0.01$ and $\gamma=0.9$ together with the trajectory and time series for $\delta=0.3$ and $D=10^{-5}$. At lower noise values, the trajectory crosses $v=0$ at lower values of $s$ (around $0.024$ for $D=0$).
Lines and symbols as in Fig.~\ref{fig:bifurcation_Ma}. The time scale in the time series is 100.}
\end{figure}

Class 1 time series are  obtained with $\delta=2.5$ and $D=10^{-7}$, yielding both the amplitude dynamics and the ISI density similar to those shown in Fig.~\ref{fig:class1_ISI}. One difference for the model of Eqs.~\ref{eq:De_v}--\ref{eq:De_s} is that increasing noise strength typically increases the amplitude of the STOs before a spike. This is consistent both with the slowing of $s$ near $s_{\rm H}$ and with the opportunity for CR-type STOs in the interval where $s>s_{\rm H}$. As in Class 1, we see the appearance of new well-defined peaks in the ISI density with low noise levels. For this particular choice of parameters, the first of these new peaks appears at longer ISIs, similar to the behavior in the upper panel of Fig.~\ref{fig:sc_7D_ISI_2} for the 7DSC model.

\begin{figure}
\centering
\includegraphics[width=.32\textwidth]{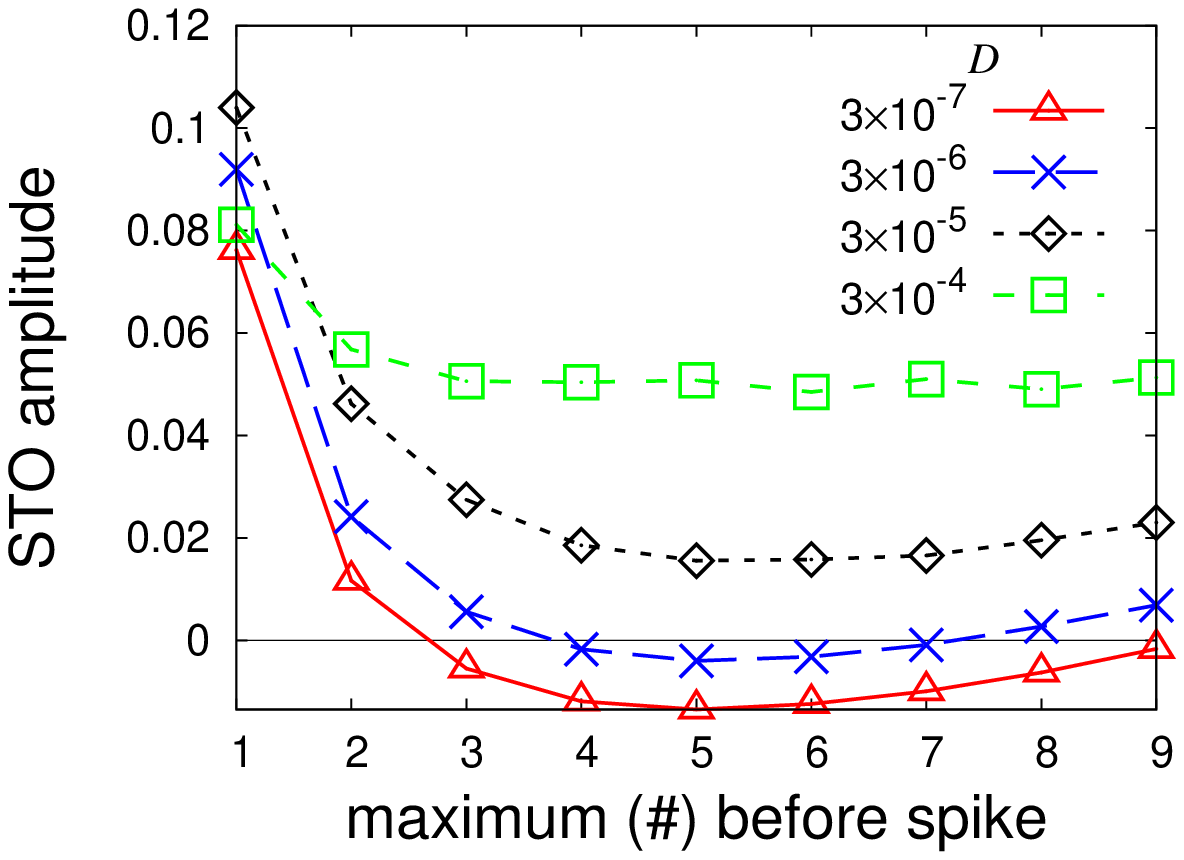}
\includegraphics[width=.32\textwidth]{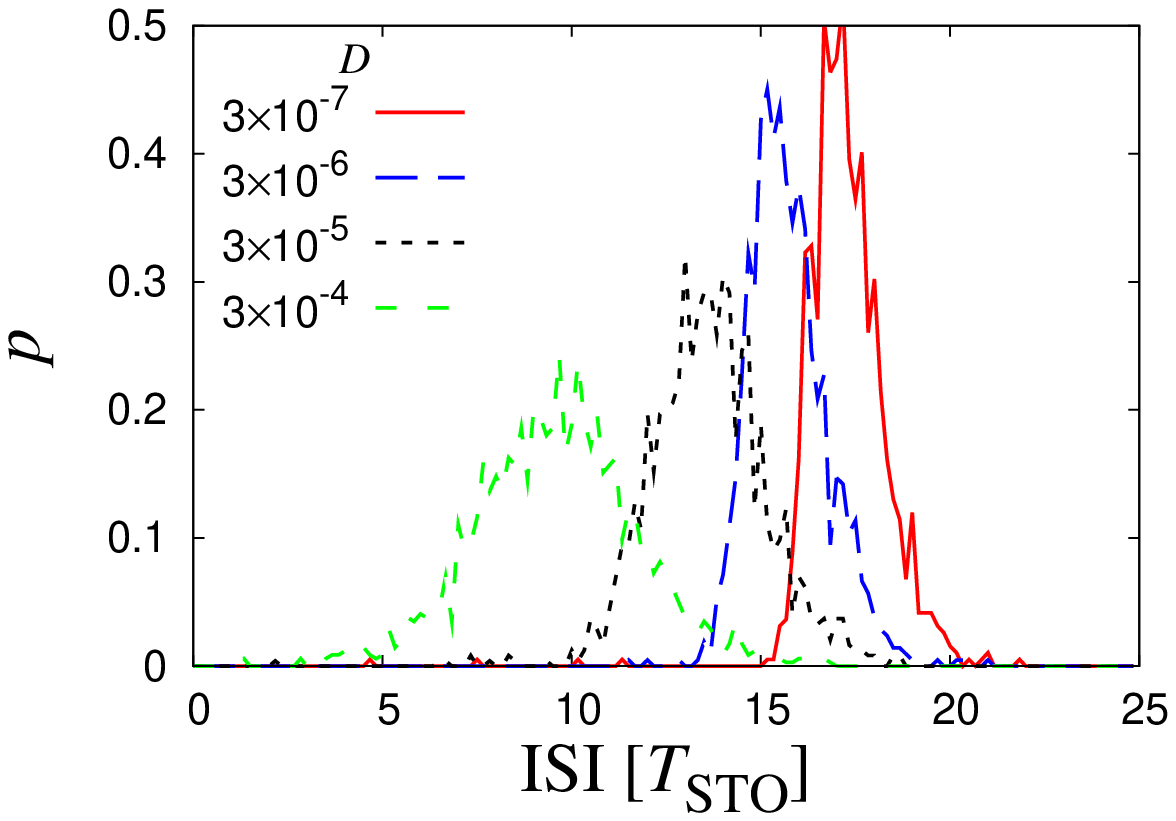}
\includegraphics[width=.32\textwidth]{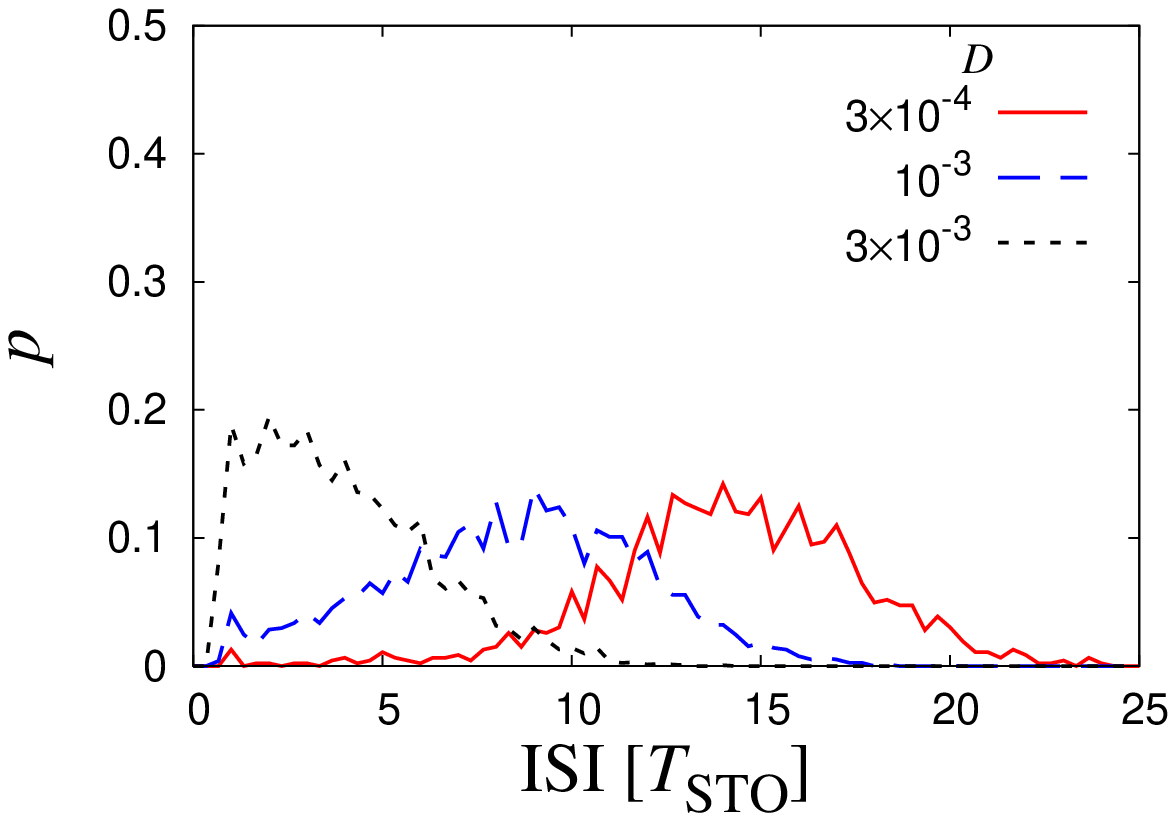}
\caption{Average amplitude before a spike ({\bf top}) and ISI density ({\bf middle}) for time series similar to Class 2 from the self-coupled FN-type ($\delta=0.3$) and various noise strengths. {\bf Bottom:} ISI density for Class 3-like time series from the self-coupled MFN ($\delta=0.17$). $T_{\rm STO} \approx 40$.}
\label{fig:De_amp_ISI}
\end{figure}

For time series with the characteristics of Class 2, we find $\delta=0.3$ and $D=3\times 10^{-6}$ in Eqs.~\ref{eq:De_v}--\ref{eq:De_s} yielding dynamics comparable to Sec.~\ref{sec:analysis_inter} in terms of average ISI and amplitude trend (Fig.~\ref{fig:De_amp_ISI}). For the STOs well before the spike, an increase of the amplitude results from the algorithm that subtracts the average $v$ as $v$ has a decreasing trend. The amplitude before the spike increases with noise level, once again due to the combination of CR-driven STOs before $s$ crosses $s_{\rm H}$, and a slowing of $s$.  Larger values of $D$ shift the ISI density towards smaller values and broadens it, yet a certain minimal ISI is maintained due to the return of $s$ well beyond $s_{\rm H}$. The effect of such a return mechanism was also observed in the ISI density in the NLMFN.

Reduced speed of $s$ ($\delta=0.17$) combined with stronger noise ($D=3\times 10^{-4}$) leads to time series that fit into our Class 3. The amplitude dynamics is very similar, but the ISI density is different from those shown in Fig.~\ref{fig:class3_ISI}. At this noise level, the ISI density is still centered around 15 $T_{\rm STO}$, again due to the return of $s$ well above $s_{\rm H}$.  Only for the highest noise levels considered ($D=3\times 10^{-3}$) the noise drives an early escape, reflected in the ISI density concentrated at shorter ISI lengths.

\begin{figure}
\centering
\includegraphics[width=.32\textwidth]{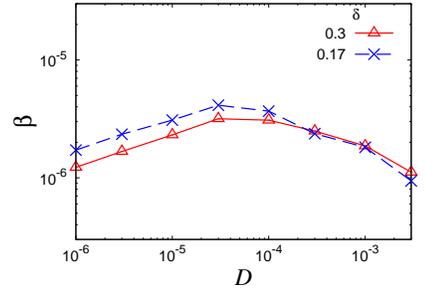}
\caption{The coherence measure $\beta$ (see Subsec.~\ref{ssec:tools}) for the self-coupled FN model at two values of $\delta$.}
\label{fig:beta_Desroches}
\end{figure}

In Fig.~\ref{fig:beta_Desroches} we show the coherence measure $\beta$ for the two parameter sets used for generating Class 2 and Class 3 time series. As with the earlier analyses, obtaining $\beta$ for Class 1 time series is not possible due to a multi-peaked PSD. For Classes 2 and 3, we neglect a strong  peak in the PSD resulting from the concatenation and compute $\beta$ as in previous examples. There is a weak peak in $\beta$ at intermediate noise levels, consistent with previous examples that exhibit MMOs related to both CR and SPHB. In Eqs.~\ref{eq:De_v}--\ref{eq:De_s} this combination of STO mechanisms results from the return of $s$ and its slower variation near the HB.

\section{Conclusion}

Attention towards oscillatory neural dynamics has expanded to include coherent subthreshold oscillatory activity in the voltage of some neuron classes as well as spikes~\cite{Erchova2008}. Combined, the sequence of subthreshold oscillations (STOs) followed by a spike comprises a class of  mixed-mode oscillations MMOs. As these MMOs have been recognized more frequently in both experimental and modeling settings there is an increasing challenge to distinguish between different MMO-generating mechanisms. For stochastic systems there are additional routes for MMOs that include  noise-induced oscillations and transients that are sustained with noise.  

In this article, we analyzed MMOs of this type both in a biophysical model for the dynamics of stellate cells and in augmented versions of phenomenological FN-type models. Our analysis focuses on the STO part of the signal where the slow dynamics are prominent and noise has the greatest impact on the STOs. In both model types we observe signatures of two distinct oscillation-generating mechanisms: slow passage through a Hopf bifurcation (SPHB) and noise-induced coherent  oscillations with frequencies close to that of the Hopf bifurcation. The latter of these appears in certain types of coherence resonance (CR).

For noise levels that are in the range of what is observed experimentally, features of the underlying deterministic models can be hidden or transformed in the stochastic time series.   A further complication in identifying the MMO mechanism in stochastic models is that substantially distinct models can easily be tuned to produce very similar time series.   Then model calibration based on time series alone must search through a larger parameter space or range of stochastic    models needed for real world data.

 We use a suite of measures to be able to distinguish between different MMO- and STO-generating mechanisms in the presence of noise, differentiating between the types of underlying models as well as  analyzing the influence of certain general model parameters. Identifying possible MMO mechanisms in this way  limits the broad range  of parameter or model  space for finding appropriate MMO mechanisms or calibrating a biophysical model. Furthermore  this type of comparison can also  identify appropriate classes of reduced models, with appropriate stochastic behavior over a range of parameters, to be used to approximate a full biophysical model. Such reduced models are used both for simplicity and for  computational speed within  larger modeling frameworks, as, e.g., in Ref.~\onlinecite{Izhikevich2008}.

The suite of these measures was chosen to focus on the STOs in the ISI, where noise has the greatest impact on the character of the MMOs. We show that the measures can also be used to exploit the noise to identify the route to the MMOs. Given the variety of routes to MMOs, more than one such measure is needed, and we use three distinct measures (interspike interval, trend of the STO amplitude and noise-dependent coherence). While the focus of these measures is on bifurcation parameters, noise levels, and slowly varying control parameters,  we found that these measures also reveal information about the refractory behavior or reset in the models. Furthermore, we have focused on using an approach that can easily be applied to time series, so that it can be used in both experimental and simulation settings.

We summarize the main characteristics obtained from these measures for different MMO generating mechanisms.

STOs of the CR-type have distinct features in these measures in comparison with those dominated by SPHB:

1. MMOs driven primarily by the deterministic behavior of SPHB display a strong trend in increasing STO amplitude, while those driven by CR have a weaker trend or no trend at all, for average amplitude.

2. For small noise ISI densities are highly concentrated for families driven by SPHB, in contrast to the STOs of the CR-type that have ISI densities with long tails. For larger noise values, these ISI densities are more similar and may depend on the return mechanism (see items 7--9 below).

3. The coherence measure has a clear optimal noise level for CR-driven STOs. PSDs are typically multipeaked for  SPHB for smaller noise, so that a coherence measure is not well defined for small noise.  For larger noise values, the coherence measure for SPHB is strictly decreasing.

While each model has a HB for the underlying 2D subsystem, differences in the criticality of these bifurcations can be observed depending on the underlying attracting deterministic dynamics.
 
4.  For STOs dominated by SPHB, an increase in noise level  typically drives a greater reduction in STO amplitude before the spike for a subcritical HB than a supercritical HB, since the latter has attracting STOs.

5. The underlying deterministic dynamics influence different behaviors of $\beta$ as a function of noise level. Systems with a  supercritical HB may have  underlying attracting behavior of small oscillations   that can support increased coherence. This type of deterministic behavior is typically not  seen for subcritical HB.

6. For STOs dominated by SPHB in the subcritical case, even very low noise levels can drive a sampling from a variety of MMO families, together with a shift and spread of ISI values, making it difficult to distinguish from MMOs driven by other mechanisms.

In addition to identifying features related to noise level and bifurcation or slowly varying control parameters, we identified  a number of additional behaviors that are related to the reset or refractory dynamics following a spike.

7. Reset and rate of increase of the control parameter in the simple IF model can be varied to capture the amplitude and ISI behavior of the different classes of time series in the physiological model. However, a simple IF model may not have the flexibility to capture different behaviors of the coherence measure.

8. For the dynamic voltage-dependent control parameters, an early escape to spiking can translate into a lower return value.  In that case the ISI density does not shift with larger noise but is just less concentrated.  This behavior distinguishes it from IF-type models with a fixed reset or models where the return mechanisms are independent of the spike.

9. For return mechanisms with limited damping of STOs following the spike, the coherence measure shows only a decrease in the coherence measure or no well-defined coherence over the ISI. 

One important element in these comparisons is that characteristics of the time series, captured by the suite of measures, can change in distinct ways when the noise level is varied. A direct  way of varying noise levels in neuroscience experiments is by injecting a noisy current. Recognizing  that the noise has its greatest impact in the ISI where the slow dynamics are prominent leads to the proposal that tunable extrinsic noise can be introduced through an applied current to mimic the effects of the intrinsic noise.  We show that this is indeed the case with an appropriate scaling of the noise in  the physiological model.

The distinctions between model features highlighted by these measures have been observed in an additional measure related to spike clusters, repeated spikes without STOs in the ISI~\cite{Kuske2009}. There it was shown that the dynamic return mechanism of NLMFN, where an earlier spike can result in a reset value farther from the HB, can result in more robust STOs in the ISI, consistent with the ISI density behavior of NLMFN.  The spike cluster frequency increases much more dramatically with noise for systems with SPHB-driven STOs, as compared with the case where STOs are CR-driven,  consistent with the amplitude and ISI density results observed here. Also, small perturbations in the reset value can translate into a large increase in spike cluster frequency in stochastic systems, particularly where the STOs are SPHB-driven for an underlying subcritical HB.

 Our analysis  in this paper is focused on a specific type of MMOs, namely a combination of small amplitude oscillations and spikes relevant for neural systems.  We have proposed measures that are focused on characteristics of MMOs that are particularly sensitive to the noise, due to the presence of multiple scales. This suggests that  understanding the presence of multiple time scales and noise-sensitive characteristics of the underlying bifurcation structures would provide a solid basis for identifying measures that characterize  MMOs in broader or more generic settings.




\appendix

\section{Abbreviations}

\begin{tabular}{ll}
STO & subthreshold oscillation \\
MMO & mixed mode oscillation \\
ISI & interspike interval \\
PSD & power spectral distribution \\
IF & integrate and fire \\
SC & stellate cell \\
FN & FitzHugh-Nagumo \\
MFN & modified FitzHugh-Nagumo \\
LMFN & linearly augmented modified FN \\
NLMFN & nonlinearly augmented modified FN \\
vdP & van der Pol \\
HB & Andronov-Hopf bifurcation \\
CR & coherence resonance \\
SPHB & slow passage through a Hopf bifurcation
\end{tabular}

\section{Equations for the seven dimensional biophysical model for the stellate cells}
\label{app:sc_eq}

The full seven dimensional system for the stellate cells (7DSC) as presented in Ref.~\onlinecite{Acker2003} consists of one differential equation for the transmembrane voltage and six for six gating variables. Throughout this article, we omit the units both in the equations and the parameters. Voltage and reversal potentials are in mV, time in ms, all gating variables are unitless, the membrane capacitance $C$ is in $\mathrm{\frac{\mu F}{cm^2}}$, $I_{\rm app}$ in $\mathrm{\frac{\mu A}{cm^2}}$ and the conductances in $\mathrm{\frac{mS}{cm^2}}$.

\begin{align}
\frac{\dd}{\dd t}{V} & = \frac{1}{C}\left[ I_{\rm app} - G_{\rm Na}m^3h(V-E_{\rm Na}) - G_{\rm K}n^4(V-E_{\rm K}) \right. \nonumber \\ 
 & \quad \left. - G_L(V-E_L) - G_h(0.65r_f+0.35r_s)(V-E_h) \right. \nonumber \\
 & \quad \left. - G_{p}p(V-E_{\rm Na}) \right] \\
\frac{\dd}{\dd t}{m} & = -0.1\frac{V+23}{\exp(-0.1(V+23))-1} (1-m) \nonumber \\
 & \quad - 4 m \exp\left(-(V+48)/18\right) \\
\frac{\dd}{\dd t}{h} & = 0.07\exp\left(-(V+37)/20\right) (1-h) \nonumber \\
 & \quad - h /(\exp(-0.1(V+7))+1) \\
\frac{\dd}{\dd t}{n} & = -0.01\frac{V+27}{\exp(-0.1(V+27))-1} (1-n) \nonumber \\
& \quad - 0.125n \exp\left(-(V+37)/80\right)\\
\frac{\dd}{\dd t}{p} & = \frac{1}{0.15}\left(\frac{1}{1+\exp\left(-\frac{V+38}{6.5}\right)}-p\right) \label{eq:sc_7D_p} \\
\frac{\dd}{\dd t}{r_f} & = \left[\frac{1}{1+\exp\left(\frac{V+79.2}{9.78}\right)}-r_f\right] \nonumber \\ 
 & \quad / \left[\frac{0.51}{\exp\left(\frac{V-1.7}{10}\right)+\exp\left(-\frac{V+340}{52}\right)}+1\right] \\
\frac{\dd}{\dd t}{r_s} & = \left[\frac{1}{1+\exp\left(\frac{V+71.3}{7.9}\right)}-r_s\right] \nonumber \\
 & \quad / \left[\frac{5.6}{\exp\left(\frac{V-1.7}{14}\right)+\exp\left(-\frac{V+260}{43}\right)}+1\right].
\label{eq:sc_rs_app}
\end{align}
The parameters used throughout this article are as follows:
\begin{align}
 & C=1; \; G_h = 1.5;\; G_p = 0.5;\; G_L = 0.5;\; G_{\rm K} = 11;\;  G_{\rm Na} = 52;  \nonumber \\
& E_h = -20;\; E_{\rm Na} = 55;\; E_L = -65;\; E_{\rm K} = -90 .
\end{align}
Equivalent to Ref.~\onlinecite{Rotstein2006} and what we did in the 3DSC model in Eq.~\ref{eq:sc_V}, we augment Eq.~\ref{eq:sc_7D_p} by the additive noise term $\sqrt{2D}\eta(t)$.

The following alternative equation for $r_s$ was introduced in Ref.~\onlinecite{Rotstein2006}:
\begin{align}
\frac{\dd}{\dd t}{r_s} =  & \left[\frac{1}{\left(1+\exp\left(\frac{V+2.83}{15.9}\right)\right)^{58}}-r_s\right] \nonumber \\
 & / \left[\frac{5.6}{\exp\left(\frac{V-1.7}{14}\right)+\exp\left(-\frac{V+260}{43}\right)}+1\right] .
\label{eq:sc_rs_58} 
\end{align}
The right hand sides of Eqs.~\ref{eq:sc_rs_app} (Eq.~\ref{eq:sc_rs}) and~\ref{eq:sc_rs_58} are very similar within the relevant parameter ranges of $r_s$ and $V$ with the only noticeable deviation at very small values of $V$ (between $-80$ and $-70$). The qualitative features of the reduced 3DSC model therefore are not altered. Here, we use Eq.~\ref{eq:sc_rs_app} throughout the paper, except for Subsec.~\ref{ssec:intra_sc_weak_noise}, where we replace Eq.~\ref{eq:sc_rs_app} by Eq.~\ref{eq:sc_rs_58} for the sake of comparability with the analysis in Ref.~\onlinecite{Wechselberger2009}. We refer to the respective footnote in Ref.~\onlinecite{Rotstein2006} for a discussion.


\section{Details for the measures used to characterize MMOs}
\label{app:measures}

\subsection{Amplitude trend before a spike}

First, the data (time series) was low-pass filtered by convolving it with a triangle function (e.g., Ref.~\onlinecite{Tucker}) of a length that is on the order of half a STO period. A simple algorithm then finds the spikes, chooses a window before each spike, removes the average and finds the local maxima in each window going backwards in time. The amplitude at each consecutive maximum before a spike is averaged over all ISIs in the time series. In this analysis the choice of the low pass filter (`triangle length') is crucial, to avoid reduction of amplitudes or miscounting of the order of the maxima resulting from the time scale of the filter being too short or long.

\subsection{Power spectral distribution}

For the spiking models, spikes were removed from the time series by deleting data points corresponding to a typical spike duration (including recovery) after the respective variable ($V$ or $u$) crossed a certain value. In the IF-type models, a similar but much shorter stretch of data was removed. The remaining data points were concatenated. Typically, PSDs computed from the full time series (including spikes and recovery) show significant power in broad frequency ranges that differ strongly between different models due to the different shapes of the spikes. Removing the spikes makes the power contribution of the STOs more clearly visible. To remove low frequency content in the PSDs, the average of the time series was removed. The routine \texttt{spctrm} from numerical recipes~\cite{NR} was used to compute an estimate of the PSD. The normalization is such that the total power in the PSD is equal to the mean squared amplitude of the time series. Our `standard' PSD for the MFN models is obtained from time series sampled at $\Delta t = 0.02$ with a Bartlett window of size 4096 data points with overlap~\cite{NR}. The `standard' PSD for the SC models is obtained from time series sampled at $\Delta t = 2.5$ with the same window function. For time series sampled at a different $\Delta t$, we scale the PSD accordingly. We scale the PSDs such that the frequencies are expressed in $1/T_{\rm STO}$ (see Subsec.~\ref{ssec:tools}).

\subsection{Coherence measure $\beta$}

To compute the coherence measure $\beta$ according to Eq.~\ref{eq:beta}, we obtain the relevant values of the PSD by fitting a non-normalized Cauchy-Lorentz distribution to the peak of the PSD corresponding to STOs.

The fit is generally good for intermediate and strong values of noise. For weaker noise values the PSDs are often multi-peaked making it difficult to obtain a meaningful coherence measure. The coherence measure $\beta$ has units of the PSD (squared unit of the considered variable when using $T_{\rm STO}$ as a time scale) and depends strongly on the sampling rate of the original time series and the windowsize of the PSD-estimator. It is therefore difficult to compare absolute values of $\beta$ between time series with different units, scales and sampling rates.


\begin{acknowledgments}
RK and PB received funding from an NSERC Discovery Grant. PB was partially funded by the Pacific Institute for the Mathematical Sciences. PB acknowledges support by the Indian Institute of Technology Madras (IIT-M) for a visit to the Department of Physics at the IIT-M. We thank Ozgur Yilmaz for helpful discussions.
\end{acknowledgments}

\vspace*{1cm}

\end{document}